\documentclass{jfm}
\usepackage{graphicx,etoolbox}
\makeatletter
\define@key{Gin}{noincl}[true]{%
  \lowercase{\Gin@boolkey{#1}}{draft}%
  \lowercase{\Gin@boolkey{#1}}{noincl}%
}
\newif\ifGin@noincl

\patchcmd{\Gin@setfile}{\hb@xt@}{\noinc@box{#3}}{}{}

\def\noinc@box#1#2#3{%
  \ifGin@noincl
    \do@noinc@box{#1}{#2}{#3}%
  \else
    \hb@xt@#2{#3}
  \fi
}

\def\do@noinc@box#1#2#3{%
  \hb@xt@\Gin@req@width{%
    \vrule\hss
    \vbox to \Gin@req@height{%
      \hrule \@width \Gin@req@width
      \vss
      \edef\@tempa{#1}%
      \expandafter\filename@parse\expandafter{\@tempa}%
      \rlap{%
        \kern3\p@
        \parbox{\dimexpr\Gin@req@width-6\p@}{
          \raggedright\footnotesize To be made
        }%
      }%
      \vss
      \hrule}%
    \hss\vrule}%
}
\makeatother
\usepackage{amsmath}
\usepackage{epstopdf, epsfig}
\usepackage{color}
\usepackage{chngcntr}
\usepackage{url}
\usepackage[colorlinks=true,citecolor=blue,linkcolor=blue,urlcolor=blue]{hyperref}

\usepackage{physics}

\usepackage{subcaption}
\usepackage[font={footnotesize}]{caption}

\newcommand{\pd}{\partial}

\newcommand{\rev}[1]{{\color{black}#1}}

\def\Re{\mbox{\it Re}}   
\def\Pr{\mbox{\it Pr}}  
\def\Ca{\mbox{\it Ca}}   

\renewcommand{\vec}[1]{\boldsymbol{\mathrm{#1}}} 

\providecommand\bnabla{\boldsymbol{\bnabla}}

\shorttitle{Nanoscale sheared droplet}
\shortauthor{U. L\={a}cis, M. Pellegrino, J. Sundin, et al.}

\title{Nanoscale sheared droplet: Volume-of-Fluid, phase-field and no-slip molecular dynamics}

\author{U\v{g}is L\={a}cis\aff{1,2}\thanks{\rev{These authors contributed equally to this work}},
  Michele Pellegrino\aff{3}\footnotemark[1],
  Johan Sundin\aff{1},
  Gustav Amberg\aff{1,4},
  St\'{e}phane Zaleski\aff{5,6},
  Berk Hess\aff{3}
 \and Shervin Bagheri\aff{1}
  \corresp{\email{shervin@mech.kth.se}}}
 
\affiliation{
\aff{1}FLOW Centre, Department of Engineering Mechanics KTH, 100 44 Stockholm, Sweden
\aff{2}FOTONIKA-LV, Institute of Atomic Physics and Spectroscopy, University of Latvia, LV-1586 Riga, Latvia
\aff{3}Swedish e-Science Research Centre, Science for Life Laboratory,
Department of Applied Physics KTH, 100 44 Stockholm, Sweden
\aff{4}S\"{o}dertorn University, Stockholm, Sweden
\aff{5}Sorbonne Universit\'{e} and CNRS, Institut Jean Le Rond $\partial$'Alembert, Paris, France
\aff{6}Institut Universitaire de France, Paris, France
}

\begin{document}

\maketitle

\begin{abstract}
The motion of the three-phase contact line between two immiscible fluids and a solid surface arises in a variety of wetting phenomena and technological applications.
One challenge in continuum theory is the effective representation of molecular phenomena close to the contact line.
Here, we characterize the molecular processes of the moving contact line to assess the accuracy of two different continuum two-phase models. Specifically, molecular dynamics (MD) simulations of a two-dimensional droplet between two moving plates are used to create reference data for different capillary numbers and contact angles. We use a simple-point-charge/extended (SPC/E) water model with particle-mesh Ewald electrostatics treatment. This model provides a very small slip and a more realistic representation of the molecular physics than Lennards-Jones models. The Cahn-Hilliard phase-field model and the Volume-of-Fluid model are calibrated against the drop displacement from MD reference data. It is demonstrated that the calibrated continuum models can accurately capture 
droplet displacement and  droplet breakup for different capillary numbers and contact angles. However, we also observe  differences between continuum and atomistic simulations in describing the transient and unsteady droplet behavior, in particular, close to dynamical wetting transitions.
The molecular dynamics of the sheared droplet provide insight of the line friction experienced by the advancing and receding contact lines and evidence of large-scale temporal ``stick-slip'' like oscillations. The presented results will serve as a stepping stone towards developing accurate continuum models for nanoscale hydrodynamics. 
%
\end{abstract}


\section{Introduction} \label{sec:intro}
The motion of a two-phase interface over a solid surface has turned out to be a challenging problem for continuum fluid mechanics (CFM) models. This is most evident for  models that assume a sharp interface between the phases and the no-slip velocity condition at the solid surface. Under these classical assumptions, one naturally ends up with an immobile line at which both fluid phases meet with the solid, so-called contact line.  Clearly, a fixed contact line is in contradiction with, for example, our observations of a spreading drop on a surface, or of a liquid imbibition into a porous medium. This theoretical issue was identified already half a century ago by~\cite{huh1971hydrodynamic} as a stress singularity at the contact line. Motivated by the importance of the moving contact line in applications such as printing~\citep{kumar2015liquid},  CO$_2$ storage and water management in fuel cells~\citep{singh2019capillary}, a number of approaches have been suggested for overcoming the stress singularity in Navier-Stokes based solvers. The extensive work on the topic~\citep{bonn:was,snoeijer2013moving,Sui2014} suggests that there exists no single continuum approach for describing the moving contact line. Instead, different models are suitable depending on the application and the representation of interfacial physics, and each model comes with its own sets of empirical parameters. Certain guidelines are required to determine these parameters. 

Despite the vast amount of theoretical developments that have been presented~\citep{voinov1976,cox1986dynamics,Shikhmurzaev1994,shikhmurzaev1997moving,kalliadasis2000steady,eggers2004hydrodynamic,flitton2004surface,wilson2006nonlocal,snoeijer2006free,pismen2008solvability,pismen2000disjoining,snoeijer2010asymptotic,chan2012theory,nold2018vicinity,chan2020cox}, a theoretical consensus is yet to be reached. To a large extent, progress is hindered by a lack of understanding the nanoscale physics of wetting phenomena~\citep{afkhami2020issueDresden,afkhami2021challenges}. For example, a  fundamental question relates to the nanoscopic contact angle. The typical choice up till now has been a constant equilibrium angle~\citep{kronbichler2017phase}. However, recent experimental~\citep{deng2016nanoscale} and numerical~\citep{fernandez2021taking} evidence suggest a contact line velocity-dependent dynamic contact angle even at nanoscale. The development of more accurate measurement techniques is ongoing~\citep{thormann2008force,eriksson2019direct}, however, as of yet, we lack a complete insight into interface shape and velocity field near moving contact lines in nanoscale. Atomistic simulations  have provided significant insight into the molecular physics at this scale~\citep{gentner2003forced,smith2016langevin,smith2018moving,perumanath2019droplet,lacis2020steadyDresden}, although most systems used so far have been based on idealized force models between the liquid and the substrate. 

In practice, there are few common approaches to numerically model moving contact lines in standard continuum methods. For sharp interface models, such as Volume-of-Fluid (VOF) and level-set (LS), the movement of the contact line is typically allowed by an explicit Navier-slip condition~\citep{navier1823memoire,spelt2005level} or by an implicit numerical slip at the contact line~\citep{renardy2001numerical,afkhami2009mesh}. For diffuse interface models, such as the Cahn-Hilliard phase-field (PF) model~\citep{jacqmin2000contact}, the contact line advances through diffusion even in the no-slip scenario. The Lattice-Boltzmann method (LBM) can be leveraged to solve different types of flow problems. The standard application of the LBM typically uses Shan-Chen~\citep{shan1993lattice} or Gunstensen et al.~\citep{gunstensen1991lattice} models. It is argued that in the Shan-Chen model the contact line moves through phase change~\citep{kamali2011contact}, similar as in LBM with thermodynamically consistent potentials~\citep{briant2004latticeP1}. The LBM method has been also used to solve phase-field equations~\citep{briant2004latticeP2}, consequently, the motion of the contact line then occurs through diffusion. In the current work, we focus particularly on two of these models, namely, geometric VOF and Cahn-Hilliard PF models.

For the VOF method, different components to model the contact line (static, dynamic angle, hysteresis window, Navier slip, etc.) have been proposed over time. A recent comparison between different options can be found in~\cite{legendre2015comparison}. They concluded that the models incorporating dynamic contact angle better represent the experiments of a spreading drop. For the PF model, there are guidelines to select the model parameters through calibration with experiments~\citep{yue2011wall}. A part of the calibration is choosing the diffuse interface thickness in a manner to satisfy the so-called sharp interface limit~\citep{yue2010sharp,magaletti2013sharp,xu2018sharp}. With this approach, a good agreement between PF simulations and capillary rise experiments has been demonstrated.

In general, however, the connection between the selected CFM models and molecular reality is not clear. It is not known how to choose the model parameters for an accurate representation of nanoscopic physics that determines the moving contact line speed. To address this, comparisons between molecular dynamics (MD) simulations and the chosen CFM models have been carried out~\citep{qian2003molecular,barclay2019cahn,mohand2019use}. The typical MD work considers Lennard-Jones types of surfaces with large slip lengths. Only more recently, water MD simulations have been carried out by some of us~\citep{johansson2018molecular,lacis2020steadyDresden} over surfaces with negligible slip. 

\begin{figure}
\centering
\includegraphics[width=1.0\linewidth]{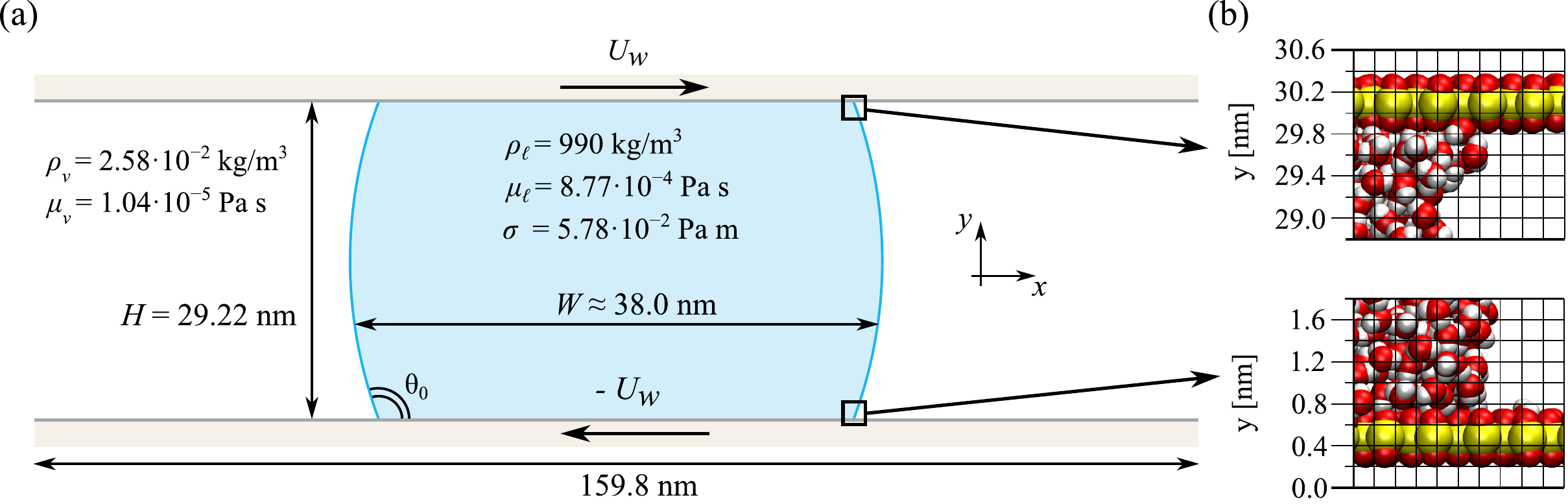}
\caption{Dimensions and properties of the sheared-droplet configuration considered in the current work (a). Close-up of the molecular system near the moving contact lines (b). Overlying solid mesh illustrates the binning boundaries for the collection of flow data from MD simulations.}
\label{fig:intro0}
\end{figure}

In this work, we generate benchmark data from MD simulations of wetting over no-slip surfaces. These conditions can be reproduced employing SPC/E water on a smooth, silica-like substrate~\citep{johansson2015water}. We choose a forced wetting set-up~\citep{blake2015forced} -- instead of a capillary-driven one~\citep{villanueva2006some} -- due to the more versatile control of the wetting process. In particular, we choose a sheared droplet configuration (see figure~\ref{fig:intro0}a), which is a well-studied canonical problem~\citep{jacqmin2004onset,sbragaglia2008wetting,gao2013wetting,wang2013shear} allowing simultaneous access to both receding and advancing contact lines. Furthermore, depending on the wall velocity $U_w$, the system either i) stabilizes at a steady state (if $U_w < U_{w,c}$) or ii) exhibits a non-trivial unsteady behaviour (if $U_w > U_{w,c}$). Here, $U_{w,c}$ is the critical wall velocity describing the boundary between i) and ii). Hence, the sheared drop configuration provides rich interfacial dynamics that are challenging to capture with CFM models.   

By assuming a negligible slippage, the space of input parameters is essentially reduced to wall velocities ($U_w$) and equilibrium contact angles between water and silica ($\theta_0$). We then adopt a two-step approach. In the first step, we calibrate the continuum simulations against MD for a given pair $(U_w,\theta_0)$. This yields the necessary PF and VOF parameters that best reproduce the steady droplet displacement measured from MD. In the second step, we fix the calibrated PF and VOF parameters and assess the predictive capability of the CFM  models for different $U_w$ by characterizing 
the interface shape and the drop displacement both in the stable and unstable regimes. 
This is a extension of the work presented by \cite{lacis2020steadyDresden}, who evaluated CFM model performance in matching the MD results for a single steady wall velocity ($U_w < U_{w,c}$) and a single equilibrium contact angle. 

The molecular dynamics of the sheared droplet reported herein provide insight of the friction experienced by the advancing and receding contact lines. We demonstrate asymmetric features of  advancing and receding lines and report evidence of large-scale temporal ``stick-slip'' like oscillations. These observations do not only enhance our physical understanding of moving contact lines, but also aid the development needed to increase the accuracy of continuum models.  

The paper is organized as follows. In \S\ref{sec:cont-mod} we describe the flow configuration, CFM models and demonstrate the effect of the unknown parameters. The reference MD simulations of the sheared droplet system are described in \S\ref{sec:md-sheared}. We calibrate the CFM models against MD in \S\ref{sec:CFM-MD-calib}. Predictions from CFM are evaluated against MD in \S\ref{sec:CFM-MD-pred}. Then, in \S\ref{sec:md-phys} we provide insights into the molecular physics of the sheared droplet system. Following that, in \S\ref{sec:discussion} limitations of CFM models, fluid slippage, friction and potential future modelling directions are discussed. We conclude the paper in \S\ref{sec:conclusions}. In appendices~\ref{app:vof}-\ref{app:wall-loc}, important physical and technical details are provided.

\section{Flow configuration and continuum models}\label{sec:cont-mod}
We consider a two-dimensional system that is periodic in
the streamwise direction and bounded in the vertical direction by two parallel horisontal plates located at $y=0$ and $y=H$.  A liquid drop of density $\rho_\ell$ and viscosity $\mu_\ell$ is sandwiched between the plates such that its maximum width is $W$. The drop is surrounded by water vapour with density $\rho_v$ and viscosity $\mu_v$. The surface tension between the phases is constant and denoted by $\sigma$. The numerical values of the geometry and fluid properties are reported in figure~\ref{fig:intro0}(a).

We study the response of the droplet to two parameters; i) $U_w$, the constant velocity of the upper and lower walls moving in opposite directions and; ii) the equilibrium contact angle $\theta_0$. The former is interchangeably discussed in its non-dimensional form, using the Capillary number, 
\begin{equation}
\Ca = \frac{2\,\mu_\ell\,U_w}{\sigma}.
\label{eq:capillary_number}
\end{equation}
The Capillary number, corresponding to critical wall velocity $U_{w,c}$, is $\Ca_c$.

For small $Ca$, the liquid slips on the solid, resulting in a steady deformed droplet shape as shown in figure \ref{fig:dx-def-VOF-lm-var}(a). Above a certain critical Capillary number, $Ca_c$, the liquid is deformed to such a degree that its interface to the surrounding vapour breaks, leaving behind multiple disconnected liquid droplets.  The second control parameter, $\theta_0$, is a measure of the surface's affinity to water. In hydrophilic conditions ($\theta_0<90^\circ$), the affinity is strong, resulting in a  larger droplet deformation compared to hydrophobic conditions ($\theta_0>90^\circ$).
%
Under dynamic conditions, the contact line can be different from the equilibrium one. For advancing contact lines (marked with A in figure  \ref{fig:dx-def-VOF-lm-var}a), the liquid displace the vapour, while for receding contact lines (R in figure \ref{fig:dx-def-VOF-lm-var}a), the vapour displace the liquid. Both these processes are largely determined by molecular interactions between the water and the substrate as depicted in figure~\ref{fig:intro0}(b). 

The drop deformation induced by the moving walls is characterized with measures defined in figure~\ref{fig:dx-def-VOF-lm-var}(a).  As a global measure, we introduce drop displacement  of left ($\Delta x_l$) and right ($\Delta x_r$) two-phase interface, respectively. For more detailed characterization, we also evaluate the interface angle $\theta (y)$ for steady configurations. It is obtained from the slope of each linear segment on the interface.


In the continuum setting, the flow and pressure  fields $\left(\vec{u}, p\right)$ are obtained by
solving the incompressible Navier--Stokes equations in the domain containing both phases. In two dimensions, the equations are
\begin{align}
\rho \left( x,y \right) \left[ \frac{\pd \vec{u}}{\pd t} + \left(\vec{u} \cdot \bnabla
 \right) \vec{u} \right] & = - \bnabla p + \bnabla \cdot \left[
 \mu \left( x,y \right) \left\{ \bnabla \vec{u} + \left( \bnabla \vec{u} \right)^T \right\}
 \right] + \vec{f}_\sigma, \label{eq:ns-1}\\
 \bnabla \cdot \vec{u} & = 0. \label{eq:ns-2}
\end{align}
Here, $\vec{f}_\sigma$ is the surface tension force, $\rho \left( x,y \right)$ and $\mu \left( x,y \right)$
are spatially dependent density and viscosity, respectively. 
The functions $\rho \left( x,y \right)$ and $\mu \left( x,y \right)$
take liquid and vapour values in the region occupied by each phase and
undergo a sharp transition at the boundary between the phases. In this transition
region, the volume force $\vec{f}_\sigma$ is applied to model the force induced by the surface tension.

Zero wall-normal velocity, $u_y = 0$, is imposed on the moving impermeable walls. For the tangential velocity component, we impose a
Navier-slip boundary condition
\begin{equation}
u_x = u_w + \ell_s \frac{\partial u_x}{\partial n}, \label{eq:ns-bc}
\end{equation}
where $\ell_s$ is the slip length, $u_w$ is the wall velocity ($U_w$ at the
top wall and $-U_w$ at the bottom wall) and $n$ is the wall-normal coordinate.
Although it is expected that $\ell_s = 0$ for the chosen liquid-solid combination, the implementation of the
CFM models allows for tests with non-zero value.

Equations (\ref{eq:ns-1}-\ref{eq:ns-2}) with corresponding boundary conditions
are shared between the Cahn-Hilliard PF model and the geometric VOF model. However,
the way surface tension and evolution of two-phase interface is treated differs.

\subsection{Geometric Volume-of-Fluid model} \label{sec:geom-vof}
In this model, a phase variable, $C(x,y)$, is defined as 0 in the gas and 1 in the liquid. The phase variable thus represents the liquid volume fraction in each cell. It satisfies the convection equation
\begin{equation}
    \frac{\pd C}{\pd t} = - \vec{u} \cdot \bnabla C. 
    \label{eq:vof-cov-diff}
\end{equation}
The surface tension force is applied by the continuous surface force (CSF) method \citep{brackbill1992continuum},
\begin{equation}
    \vec{f}_\sigma = -\sigma \kappa \bnabla C,
    \label{eq:continuous-surface-force}
\end{equation}
where $\kappa$ is the curvature of the interface. 

As a boundary condition for the phase variable, we impose a dynamic contact angle $\theta_\mathrm{num}$. The contact angle is the angle between the interface ($\hat{n}_i$) and wall ($\hat{n}$) normals, 
\begin{equation}
    \hat{n}_i\cdot \hat{n} = \cos(\theta_\mathrm{num}), \quad \text{ where } \quad  \hat{n}_i = -\frac{\nabla C}{|\nabla C|},
    \label{eq:cox}
\end{equation}
and the sign might vary depending on convention. 
The relationship for $\theta_\mathrm{num}$ is inspired by Cox's theory, as described and evaluated by~\cite{legendre2015comparison}.
It takes the form
\begin{equation}
    G(\theta_\mathrm{num}) = G(\theta_0) + \Ca_{cl}\ln\left(\frac{\Delta/2}{\lambda} \right),
    \label{eq:VOFContactAngle}
\end{equation}
where $G$ is a monotonically increasing function, $\lambda$ is a microscopic cut-off length scale and $\Delta$ is the wall-normal cell height. We confirm the grid convergence reported by~\cite{legendre2015comparison} in appendix~\ref{app:vof-num}. The capillary number $\Ca_{cl}$ is based on the velocity of the contact line with respect to the wall. It is estimated by linearly interpolating the velocity field at the first grid cell. For the most hydrophobic and hydrophilic configurations, 
only the constant wall velocity was used for increased robustness.

%

\begin{figure}
\centering
\includegraphics[width=0.95\linewidth]{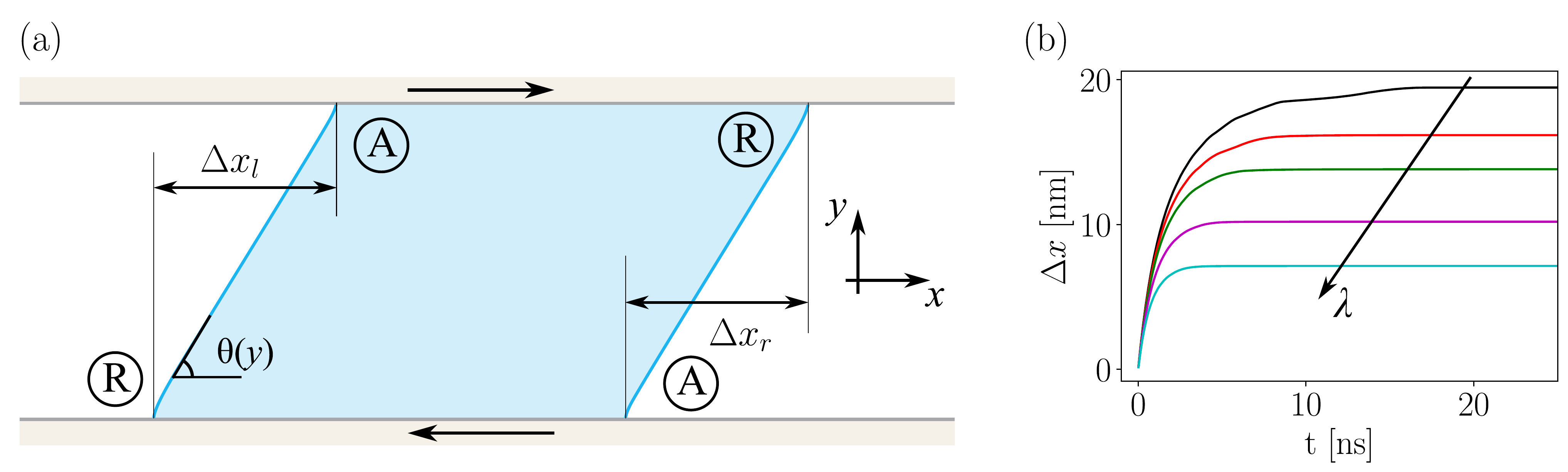}
\caption{Drop displacement measurement at the left $\Delta x_l$ and right $\Delta x_r$ sides (a) at selected time instance. Interface angle is defined with respect to the horisontal line and is sampled as a function of the vertical coordinate $\theta(y)$. Note that $\theta(y)$ is measured only for $\Ca < \Ca_{c}$, while $\Delta x$ is measured for all $U_w$. Variations of drop displacement $\Delta x = \left( \Delta x_l + \Delta x_r \right)/2$ over time in VOF simulations (b) for $\lambda = (0.47; 0.66; 0.94; 1.87; 3.74)$ nm (increasing with the arrow). Equilibrium contact angle is $\theta_0 = 95^\circ$ and Capillary number is $\Ca = 0.20 < \Ca_{c}$.}
\label{fig:dx-def-VOF-lm-var}
\end{figure}

The geometric VOF model is solved with an open-source finite volume code, called PArallel, Robust, Interface Simulator (PARIS) \citep{aniszewski2021parallel}. The full set of equations (\ref{eq:ns-1}--\ref{eq:ns-2},\ref{eq:vof-cov-diff}--\ref{eq:continuous-surface-force}) together with the boundary conditions (\ref{eq:ns-bc},\ref{eq:cox} and periodic inlet/outlet) are discretized on a regular cuboid grid with a staggered spatial representation. 
The cell spacing is constant in all directions. 
More details about the numerical method are given in appendix~\ref{app:vof-num}. In order to compute the curvature near the wall for very hydrophobic droplets ($\theta_0 = 127^\circ$), we use a high-order scheme to approximate derivatives (appendix~\ref{app:vof-127-cust} provides further details).


The only unknown free parameter for VOF simulations is the length scale $\lambda$. To test the effect of $\lambda$, we select $\theta_0 = 95^\circ$ and $\Ca = 0.20 < \Ca_{c}$. The  evolution of the drop displacement $\Delta x$ in time for $\lambda$ values $0.47$ nm, $0.66$ nm, $0.94$ nm, $1.87$ nm, and $3.74$ nm is shown in figure~\ref{fig:dx-def-VOF-lm-var}(b). Due to symmetry, $\Delta x_l = \Delta x_r = \Delta x$ in the CFM simulations, which we have verified numerically. We observe that larger $\lambda$  correspond to smaller steady $\Delta x$. As $\lambda$ is increased, the contact angle at the surface (\ref{eq:VOFContactAngle}) and thus the interface shape \eqref{eq:cox} near the wall is modified. The interface curvature is modified in such a way that the surface tension force across the interface opposes the friction force from the wall, which leads to a smaller $\Delta x$. The $\Delta x$ for the initial time agrees between all $\lambda$ values. In this short period, the contact line does not slip with respect to the substrate and the slope of $\Delta x (t)$ is equivalent to the wall separation velocity ($2 U_w$). 


%

\subsection{Cahn-Hilliard phase-field model} \label{sec:pf-mod-main}
We choose a model of a binary mixture with a classical fourth-order polynomial potential $\Psi$ (see eq.~\ref{app:eq:PF-potent} in appendix \ref{app:cahn-hil})~\citep{jacqmin2000contact,carlson2012thesis}. To describe the evolution of both phases, the PF model uses a phase variable $C(x,y)$ ranging from $1$ in the liquid to $-1$ in the vapour. At the interface, the function exhibits a smooth transition. The variable $C$ is governed by a convection-diffusion equation
\begin{equation}
    \frac{\pd C}{\pd t} = \bnabla \cdot \left[ M \bnabla
    \phi \right] - \vec{u} \cdot \bnabla C \, .
    \label{eq:cahn-hil-dim}
\end{equation}
Here, $M$ is the phase-field mobility and $\phi$ is the chemical potential. The latter is defined as
\begin{equation}
    \phi = \frac{2\sqrt{2}}{3} \frac{\sigma}{\epsilon} \Psi'\left(C \right) - \frac{2\sqrt{2}}{3} \sigma\,\epsilon \bnabla^2 C. \label{eq:cahn-hil-dim-2}
\end{equation}
The chemical potential (\ref{eq:cahn-hil-dim-2}) contains the surface tension ($\sigma$) and the interface thickness ($\epsilon$). The derivation of (\ref{eq:cahn-hil-dim}-\ref{eq:cahn-hil-dim-2}) is standard and can be found in \cite{carlson2012thesis} and \cite{jacqmin2000contact}. The surface tension $\sigma$ is a physical parameter and is set to a value corresponding to the water liquid-vapour interface. The constants $M$ and $\epsilon$, on other hand, are typically treated as numerical parameters. Having solved for function $C$, the surface tension force in (\ref{eq:ns-1}) is inserted as
\begin{equation}
    \vec{f}_\sigma = \phi\, \bnabla C.
\end{equation}

The Cahn-Hilliard equation (\ref{eq:cahn-hil-dim}) is a fourth-order partial differential equation and thus two boundary conditions are needed on solid walls. The lowest order boundary condition is
\begin{equation}
- \mu_f \epsilon \left( \frac{\pd C}{\pd t} + \vec{u} \cdot \bnabla C \right) =
\frac{2\sqrt{2}}{3} \sigma\,\epsilon \bnabla C \cdot \hat{n} - \sigma \cos \left( \theta_0 \right) g'\left( C \right),
\label{eq:wet-bc}
\end{equation}
where $\mu_f$ is contact-line friction and $g$ is switch function (\ref{app:eq:PF-swtch-func}). Equation (\ref{eq:wet-bc}) is also known as the non-equilibrium wetting condition and requires the equilibrium contact angle $\theta_0$ as an input. Setting a non-zero $\mu_f$ yields a dynamic contact angle different from $\theta_0$~\citep{jacqmin2000contact,qian2003molecular,carlson2011dissipation}. The second boundary condition on solid walls is $\bnabla \phi \cdot \hat{n} = 0$, which states that there is no diffusive flux through the walls.

The full system of fluid and phase-field equations (\ref{eq:ns-1}--\ref{eq:ns-2},\ref{eq:cahn-hil-dim}-\ref{eq:cahn-hil-dim-2}) together with the boundary conditions 
is discretized and solved using open-source code \textit{FreeFEM}~\citep{MR3043640}. Adaptive mesh is used near the two-phase interface to capture the variation of the function $C$. 
More details about the PF equations and simulations can be found in appendix~\ref{app:cahn-hil}.

\begin{figure}
\centering
\includegraphics[width=1.00\linewidth]{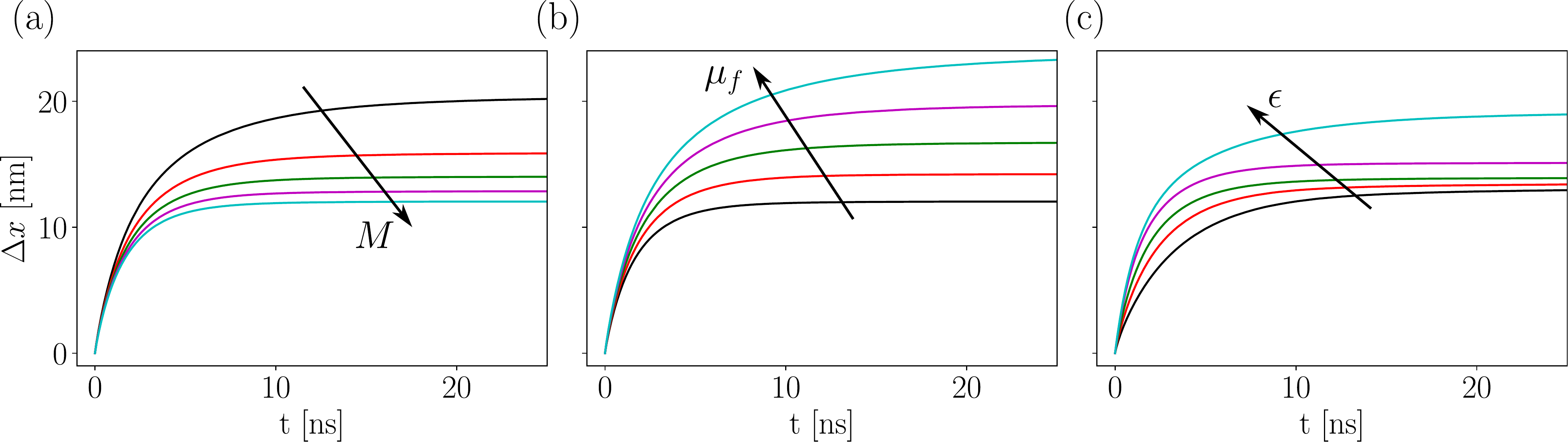}
\caption{Drop displacement $\Delta x$ over time in PF simulations. Varying PF mobility (a), contact line friction (b) and interface thickness (c). In (a), $\mu_f = 0$, $\epsilon = 0.7$ nm and PF mobility $M = (3.5; 7.0; 10.5; 14.0; 17.5) \times 10^{-16}$ m$^4$/(N s). In (b), $\epsilon = 0.7$ nm, $M = 1.75 \times 10^{-15}$ m$^4$/(N s) and contact line friction $\mu_f = \left(0.0; 0.5; 1.0; 1.5; 2.0\right)\,\mu_\ell$. In (c), $\mu_f = 0$, $M = 1.08 \times 10^{-15}$ m$^4$/(N s) and interface thickness $\epsilon = \left(0.18;0.35;0.70;1.40;2.80 \right)$ nm. Varying parameters are increasing along the black arrow in all panels. In all simulations, equilibrium contact angle $\theta_0 = 95^\circ$ and wall velocity $U_w = 6.67$ m/s.}
\label{fig:dx-PF-param-var}
\end{figure}

The PF model has three unknown free parameters; the phase-field mobility $M$, the contact-line friction $\mu_f$, and the interface thickness $\epsilon$. To provide an intuition about each parameter, we  vary them one at a time. As before, $\theta_0 = 95^\circ$ and $\Ca = 0.20$. 
Figure~\ref{fig:dx-PF-param-var}(a) shows $\Delta x$ for $\mu_f = 0$,  $\epsilon = 0.7$ nm and $M = (3.5; 7.0; 10.5; 14.0; 17.5) \times 10^{-16}$ m$^4$/(N s). We observe that the final steady $\Delta x$ value is reduced as $M$ is increased, i.e. smaller deformation with increased diffusion.
%
In figure~\ref{fig:dx-PF-param-var}(b), the displacement for $\epsilon = 0.7$ nm, $M = 1.75 \times 10^{-15}$ m$^4$/(N s) and $\mu_f = \left(0.0; 0.5; 1.0; 1.5; 2.0\right)\,\mu_\ell$ is shown. Here, the $\Delta x$ increases with $\mu_f$ and the contact-line friction has therefore an opposite effect compared to mobility. This competition has been previously clearly showcased by~\cite{yue2011wall}. 
%
Finally, the evolution of $\Delta x$ in time for $\epsilon = \left(0.18;0.35;0.70;1.40;2.80 \right)$ nm is shown in figure~\ref{fig:dx-PF-param-var}(c) with $\mu_f = 0$ and $M = 1.08 \times 10^{-15}$ m$^4$/(N s). It can be observed that -- in contrast to $M$ and $\mu_f$ variations -- notable differences are present for initial times. 
As $\epsilon$ is decreased (in the direction against the arrow), the steady $\Delta x$ value is converging. This is a signature of the sharp interface limit~\citep{yue2010sharp,xu2018sharp}. For enriched understanding, in appendix~\ref{app:pf-stream} we report streamlines from PF near receding contact line for similar parameter variations.
 
 \begin{figure}
    \centering
    \includegraphics[width=0.95\linewidth]{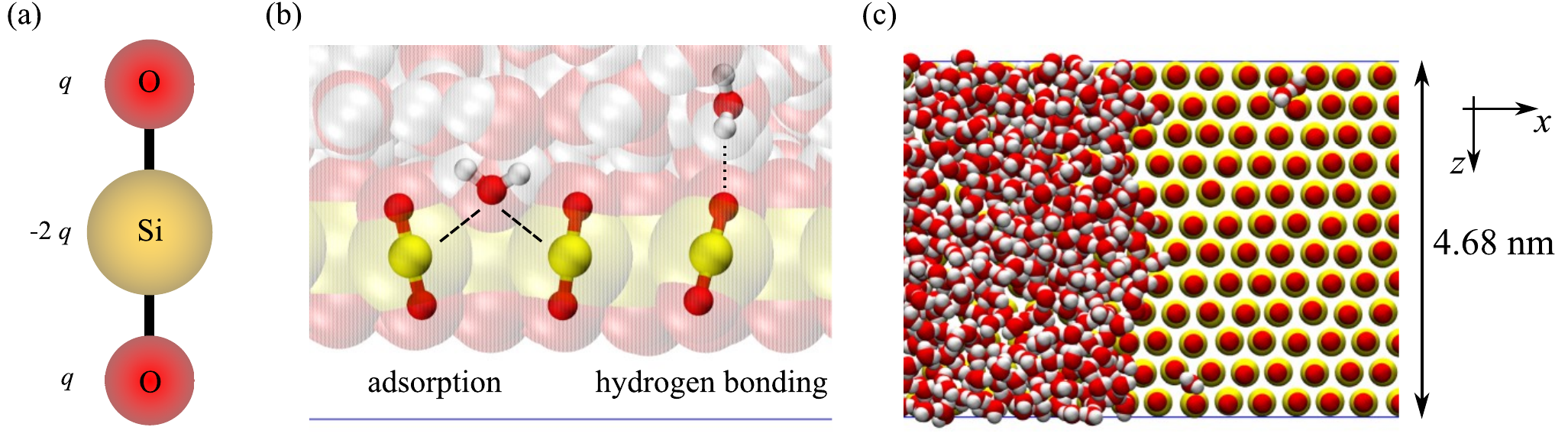}
    \caption{Molecular geometry of silica quadrupoles (a). Water molecules adsorbing (bonds sketched with dashed lines) and forming hydrogen bond (sketched with dotted line) with silica substrate (b). Top view of the contact line region (c), showing the lattice of silica quadrupoles and water up to $\approx$ 1.5 nm above the lower periodic boundary (blue line in panel b).}
    \label{fig:molecular_details_main}
\end{figure}

\section{Molecular dynamics simulations of the sheared droplet} \label{sec:md-sheared}
We describe the polar molecules of the water droplet using the SPC/E model. This is the simplest model allowing hydrogen bonds with the solid substrate. It also offers an accurate description of water bulk and interfacial properties and retains a relatively low computational cost. The bounding walls are formed as mono-layers of SiO$_2$ quadrupoles (figure~\ref{fig:molecular_details_main}a) that are restrained into a hexagonal lattice. A quasi 2D system with depth $4.68$ nm (figure~\ref{fig:molecular_details_main}c) is constructed. Albeit the composition of walls is structurally unrealistic, this simple surrogate configuration allows emulating the two fundamental electrostatic interactions characteristic to hydrophilic substrates \citep{johansson2018molecular}. The first is the hydrogen bond between water and silica (dotted line in figure~\ref{fig:molecular_details_main}b). The second interaction is the adsorption of water molecules on the substrate. The adsorption occurs due to the attraction between water oxygen and silicon atoms (dashed line in figure~\ref{fig:molecular_details_main}b). The strong electrostatic interaction is responsible for a very small hydrodynamic slip at the wall. Note that the current MD configuration does not include any chemical reactions that would occur at a real crystalline or amorphous silica surface. 

The strength of the water-substrate interaction can be tuned by adjusting the charge distribution in SiO$_2$ (figure~\ref{fig:molecular_details_main}a). Different interaction strengths will result in different equilibrium contact angles $\theta_0$. We simulate the system via atomistic molecular dynamics in the NVT ensemble, using well-established force fields and thermostats. Details regarding the physical and numerical simulation setup can be found in appendix~\ref{app:md-det-num}.

\subsection{Equilibration runs}
First, we measure the equilibrium properties of the water drop between two static plates and generate thermodynamically consistent initial conditions. This is done through so-called equilibration runs
(see appendix~\ref{app:md-equil}). During the run-time of the MD simulation, we collect
flow data in regular 0.2 nm $\times$ 0.2 nm bins for a time interval of 12.5 ps.
This yields instantaneous density $\rho^i\left(x,y,t\right)$ and flow velocity $u^i_x\left(x,y,t\right)$, $u^i_y\left(x,y,t\right)$ data as functions of space and time. After the initial transient, $\rho^i\left(x,y,t\right)$ is averaged over time to reduce noise in the liquid-vapour interface shape. The interface shape is extracted based on this averaged density, i.e.~$\rho\left(x,y\right)$. The interface position is determined by seeking the location where the liquid density transitions from bulk density to very small vapour density. Example of density distributions are shown in figure~\ref{fig:wall-allQ-summary}(e,f). More details on interface extraction are provided in appendix~\ref{app:md-itf-extr-th0}.

To determine the equilibrium contact angle, we extrapolate the interface angle towards a hydrodynamic wall position. We assume that the wall position is at the centre of the bin coloured in red (figure~\ref{fig:wall-allQ-summary}a-d). The $q$ values are tuned to yield $\theta_0 = 127^\circ$, $95^\circ$, $69^\circ$ and $38^\circ$. This allows us to investigate hydrophobic, neutral and  hydrophilic wetting conditions. In parallel to $\theta_0$ extraction, we also identify the first reliable bin for interface shape measurement. This bin is shown with green in figure~\ref{fig:wall-allQ-summary}(a-d). The interface angle computed from points closer to the wall exhibit extreme deviation from continuum description (figure~\ref{fig:q2-molEff-eqCA}d).
Consequently, for comparisons with CFM predictions (presented later), we always extract the interface shape from MD simulations neglecting the unreliable data points.

\begin{figure}
\centering
\includegraphics[width=0.90\linewidth]{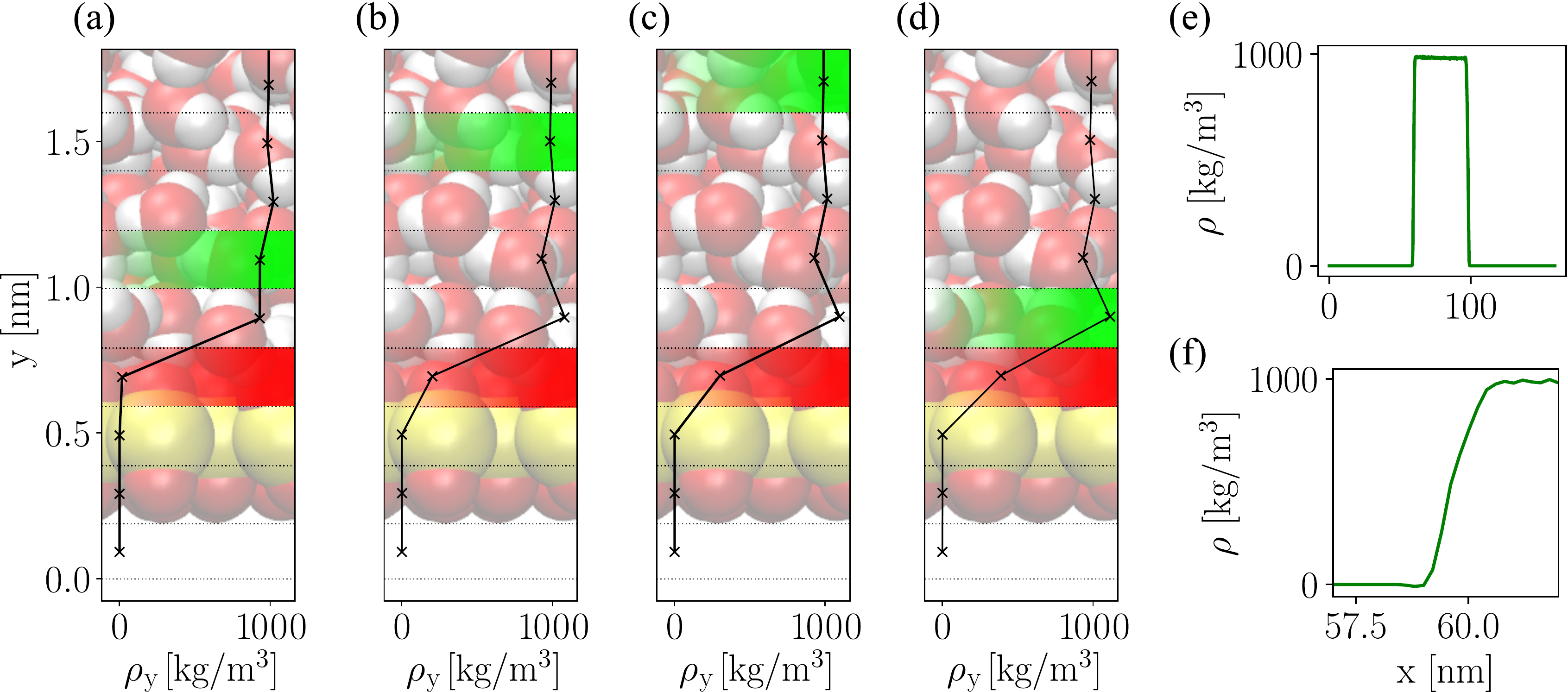}
\caption{Liquid water density variation near
the bottom wall for equilibrium angles $\theta_0 = $
$127^\circ$ (a), $95^\circ$ (b), $69^\circ$ (c) and $38^\circ$ (d).
Dashed
black lines illustrate the boundaries of the bins. The bin filled with
red shows the selected location of the solid wall, while the bin filled with
green shows the first reliable interface measurement.
Water density distribution along the green bin for $\theta_0 = 95^\circ$
configuration over the 
full span of $x$ coordinate (e) and near the
left liquid-vapor interface (f). 
}
\label{fig:wall-allQ-summary}
\end{figure}

In previous investigations, density variations have been observed for Lennard-Jones (LJ) liquids near solid walls and two-phase interfaces~\citep{bugel2011hybrid,stephan2018vapor}. We determine the extent of $\rho$ oscillations near the wall in our MD system. The water liquid density distribution along the height of the channel is computed as
\begin{equation}
\rho_y (y) = \int\limits_{x_l}^{x_r} \rho \left(x,y\right)\,
\mathit{dx}.
\end{equation}
Here, the boundaries for integration $x_l$ and $x_r$ are selected for each $\theta_0$ to fall within the liquid phase for all $y$ coordinates. The close-up near the bottom wall of $\rho_y (y)$ is shown in figure~\ref{fig:wall-allQ-summary}(a-d). Oscillations in $\rho$ have smaller amplitude and occur over smaller distances than typically observed in LJ systems.
The small layering is an outcome of the combination of the SPC/E water model and SiO$_2$ surface model. For the same value of $\theta_0$, the layering would propagate further into the liquid phase if an LJ surface would be used instead of SiO$_2$.

\begin{figure}
\centering
\includegraphics[width=1.0\linewidth]{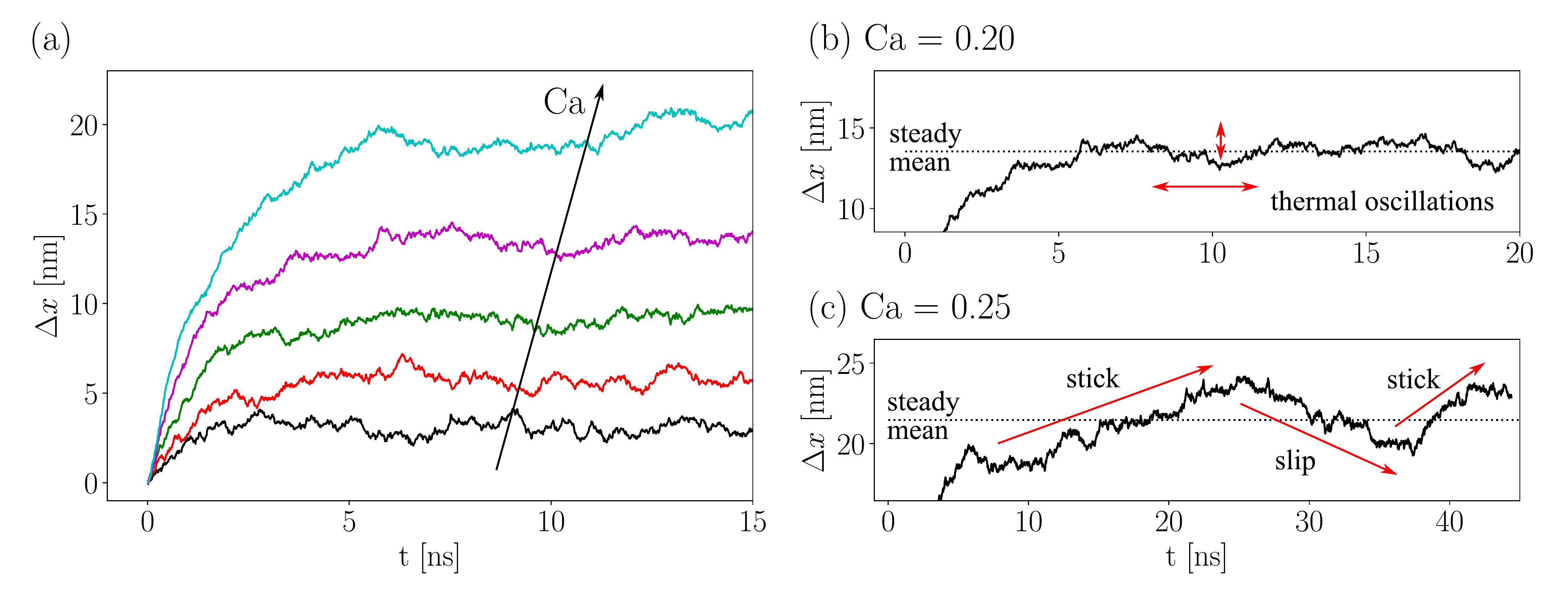}
\caption{Time evolution of $\Delta x$ from MD with $\theta_0 = 95^\circ$ for simulations reaching steady state. In (a), $\Ca$ increases along the black line starting at $\Ca = 0.05$ with increments of $0.05$. Window of $\pm  5$ nm around the steady mean value for $\Ca = 0.20$ (b) and $\Ca = 0.25$ (c). With red arrows we indicate thermal oscillations and motion similar to stick-slip.}
\label{fig:all-q2-dx-tevol}
\end{figure}

\subsection{Dynamic configuration}
Having obtained the initial state from the equilibrium simulations, we turn to the dynamic configuration. 
Since the configuration is symmetric in a continuum sense, the final $\Delta x$ is obtained as an average between the left ($\Delta x_l)$ and right ($\Delta x_r)$ interface. We determine the 
drop displacement using the first reliable bins near the top and bottom walls (figure~\ref{fig:wall-allQ-summary}, green bins). Since there is no interface data near the hydrodynamic wall, the final steady drop displacement is obtained by extrapolating the interface shape. 
We use polynomial extrapolation as detailed in appendix~\ref{app:MD-poly-fit}.

Figure~\ref{fig:all-q2-dx-tevol}(a) shows $\Delta x (t)$ for $\theta_0 = 95^\circ$ for different capillary numbers, starting at $\Ca = 0.05$ and then increased incrementally by $\Delta \Ca = 0.05$. For all $\Ca$ numbers up to and including $\Ca = 0.25$, we observed a stable configuration. 
%
%
%
The obtained steady drop displacements for $\theta_0 = 95^\circ$ are summarized in second row of table~\ref{tab:MD-run-summary}. The table also reports displacement for the other equilibrium contact angles. Different $\Ca$ values are gradually tested until at least 3 simulations in a stable regime are gathered. 
For all $\theta_0$, as expected, we observe that $\Delta x$ increases with $\Ca$.

The largest stable $\Ca = 0.25$ at $\theta_0=95^\circ$ exhibits oscillations similar to the so-called ``stick-slip'' behaviour~\citep{orejon2011stick,varma2021inertial}. For a prolonged time, the contact line shows more resistance towards movement (i.e., stick). This period is followed by another in which the contact line exhibits less resistance towards movement (i.e., slip). We show a zoomed view of $\Delta x (t)$ for $\Ca = 0.25$ in figure~\ref{fig:all-q2-dx-tevol}(c), where the stick-slip behaviour is identified with red arrows. Note that the partial stick-slip effect for $(\Ca, \theta_0) = (0.25,90^\circ)$ is distinct from the oscillations observed for $\Ca = 0.20$. We show the enlarged view of $\Delta x$ versus time for $\Ca = 0.20$ in figure~\ref{fig:all-q2-dx-tevol}(b), where the red arrows depict the oscillation magnitude and time scale. We observe that the oscillations at $\Ca = 0.20$ are much smaller in magnitude and span smaller time scales compared to stick-slip-like motion (figure~\ref{fig:all-q2-dx-tevol}c). More discussion on the physics behind these oscillations is provided in \S\ref{sec:md-stick-slip}. Similar oscillations with increased magnitude are observed also for $\theta_0 = 127^\circ$ at $\Ca = 0.90$ and $\theta_0 = 38^\circ$ at $\Ca = 0.02$.  Simulations exhibiting  stick-slip oscillations are denoted with $^*$ in table~\ref{tab:MD-run-summary}.

\begin{table}
\begin{center}
\begin{tabular}{p{9mm}p{8mm}p{8mm}p{8mm}p{8mm}p{8mm}p{8mm}p{8mm}p{8mm}p{8mm}p{8mm}p{8mm}p{8mm}}
$\theta_0$ & \multicolumn{2}{c}{sim. 1} & \multicolumn{2}{c}{sim. 2} & \multicolumn{2}{c}{sim. 3} & \multicolumn{2}{c}{sim. 4} & \multicolumn{2}{c}{sim. 5} & \multicolumn{2}{c}{sim. 6} \\ \\
 & $\Ca$ & $\Delta x$ & $\Ca$ & $\Delta x$ & $\Ca$ & $\Delta x$ & $\Ca$ & $\Delta x$ & $\Ca$ & $\Delta x$ & $\Ca$ & $\Delta x$ \\ \\
$127^\circ$ & $0.150$ & $3.01$ & $0.300$ & $5.13$ & $0.600$ & $11.7^\dagger$ & $0.900$ & $24.9^*$ & $1.080$ & unst. \\
$95^\circ$ & $0.050$ & $3.40$ & $0.100$ & $5.93$ & $0.150$ & $9.11$ & $0.200$ & $13.9^\dagger$ & $0.250$ & $21.8^*$ & $0.300$ & unst. \\
$69^\circ$ & $0.030$ & $4.09$ & $0.050$ & $6.45$ & $0.060$ & $7.73$ & $0.080$ & $11.6$ & $0.100$ & $17.0^\dagger$ & $0.150$ & unst. \\
$38^\circ$ & $0.010$ & $2.73$ & $0.015$ & $5.31^\dagger$ & $0.020$ & $7.35^*$ & $0.030$ & unst. & $0.050$ & unst.
\end{tabular}
\end{center}
\caption{Overview of MD simulations carried out in this work. For each simulation, $\Ca$ and steady $\Delta x$ (in nm) are given. If simulation is unstable, ``unst.'' is reported instead of $\Delta x$ value. Stick-slip like behaviour is indicated with $^*$, while calibration simulation is denoted by $^\dagger$.}
\label{tab:MD-run-summary}
\end{table}

\begin{figure}
\centering
\includegraphics[width=1.0\linewidth]{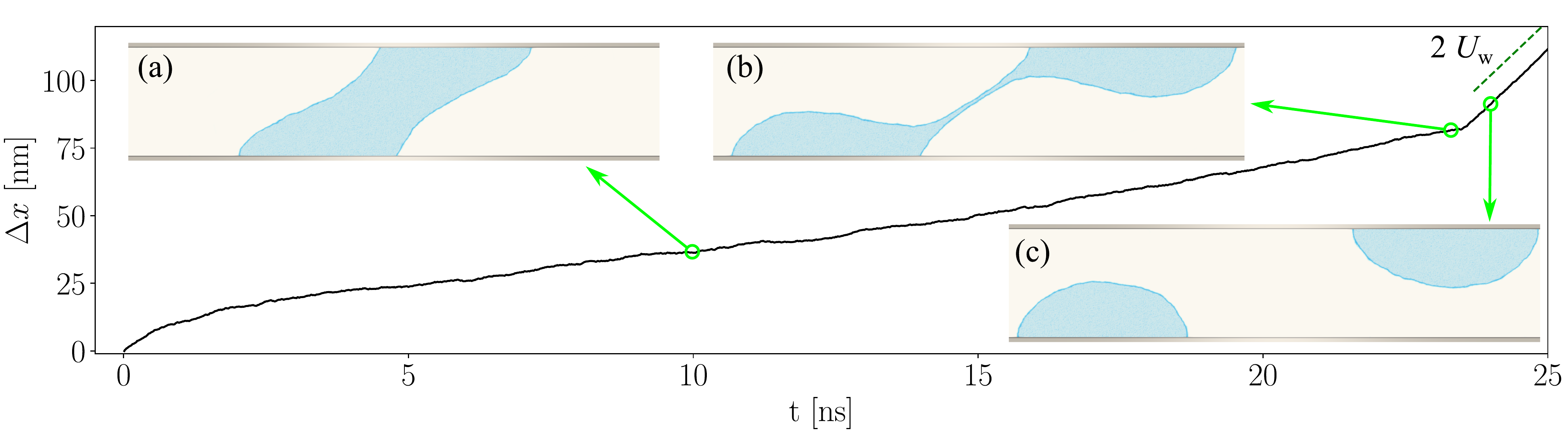}
\caption{Drop displacement in MD with $\theta_0 = 95^\circ$ at $\Ca = 0.30$. In insets (a-c), we show the drop shape at 3 selected time instances. The selected time is shown with green circle with arrow pointing to the inset. Green dashed line corresponds to the relative wall speed $2 U_w$.}
\label{fig:MD-illust-unsted}
\end{figure}

\subsection{Droplet break-up}
As the $\Ca$ number is increased further,  $\Delta x(t)$  measured from MD does not stabilise around some finite value. Instead, $\Delta x(t)$ continuously grows. An example of $\Delta x(t)$ at $(\Ca,\theta_0) = (0.30,95^\circ)$ is shown in figure~\ref{fig:MD-illust-unsted}.  
At $10$ ns, the drop is only moderately deformed (figure~\ref{fig:MD-illust-unsted}a). At $23$ ns (figure~\ref{fig:MD-illust-unsted}b), there are  two drops at the top and bottom walls connected by thin thread of liquid water. Then, at around $23.5$ ns break-up occurs and at $24$ ns we observe two completely separated drops (figure~\ref{fig:MD-illust-unsted}c). Note that at the break-up instant, the slope of $\Delta x(t)$ changes distinctively. This is due to the absence of the surface tension force that was resisting the displacement. For the two separate drops, there is no competition between the friction at the top and bottom contact lines. Instead, the friction at the contact lines now ensures that the two drops follow the wall velocity and separate with speed $2 U_w$. Hence, the critical Capillary number $\Ca_c$ lies in between $\Ca = 0.25$ and $\Ca = 0.30$. In the sheared droplet configuration, the exact value of $\Ca_{c}$ is determined by the most unstable contact line (either advancing or receding). For investigations of individual contact lines, other types of experiments should be performed, such as plunging/withdrawn plate~\cite{eggers2005cl} or hydrodynamic assist~\citep{afkhami2018transition,liu2019dynamic,fullana2020dynamic}.

\begin{figure}
\centering
\includegraphics[width=0.667\linewidth]{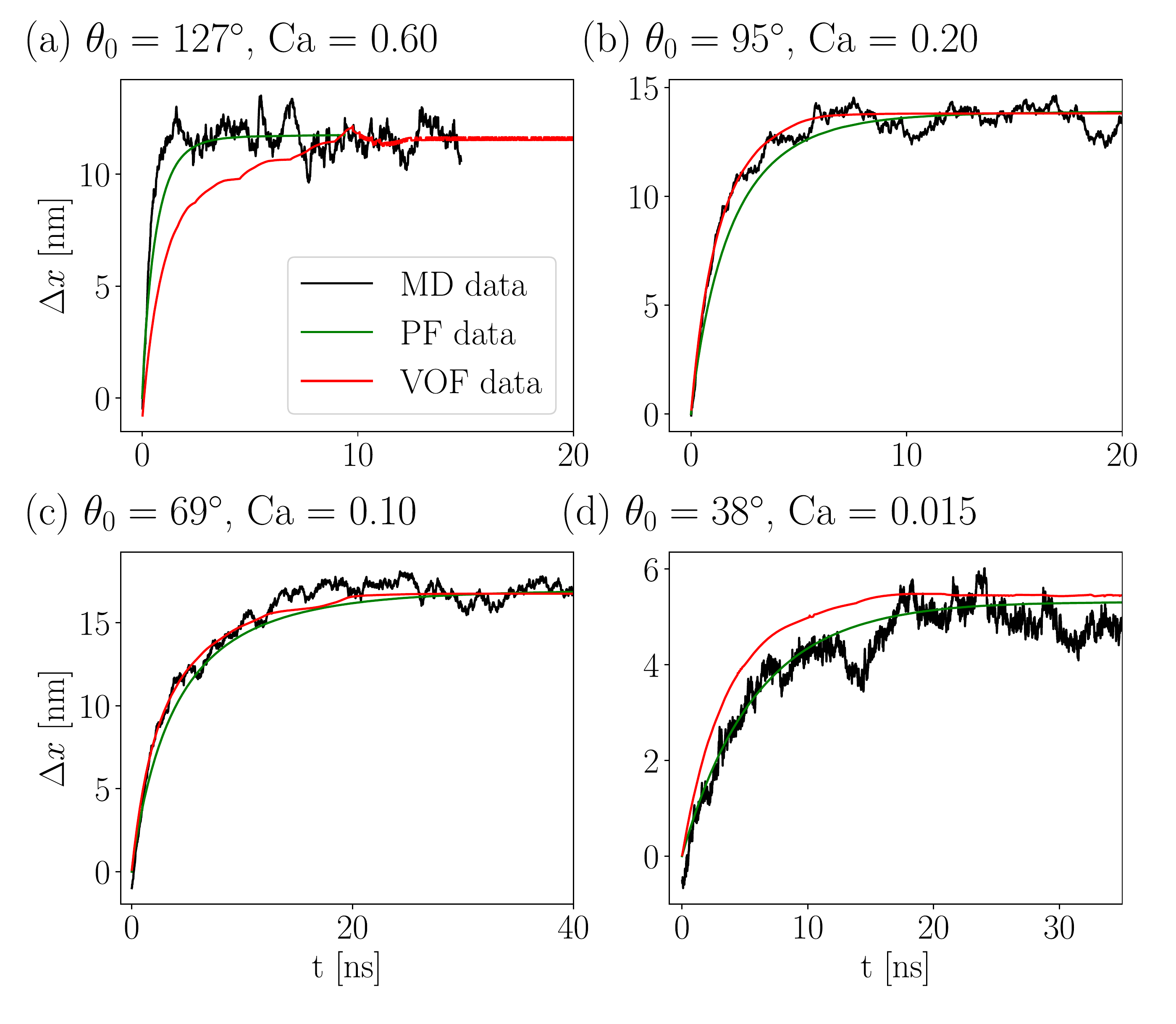}
\caption{Time evolution of $\Delta x$ in MD, PF and VOF for all calibration configurations.}
\label{fig:all-calib-dx-tevol-wCFM}
\end{figure}

\section{Calibrating CFM models against MD} \label{sec:CFM-MD-calib}
The aim of this section is to identify the free parameters in PF and VOF such that the displacement $\Delta x$ obtained from the continuum models match the displacement obtained from MD simulations.   The four calibration pairs $(Ca,\theta_0)$ are (marked with $^\dagger$ in table~\ref{tab:MD-run-summary})  $(0.60,127^\circ)$, $(0.20,95^\circ)$, $(0.10,69^\circ)$ and $(0.015,38^\circ)$. For each $\theta_0$, we have chosen the largest steady $\Ca$ number available from MD simulations. If the chosen steady $\Ca$ number leads to large stick-slip like oscillations of $\Delta x$ (such as observed in figure~\ref{fig:all-q2-dx-tevol}c), we select the previous (smaller) steady $\Ca$ number. The unsteady simulations are not suitable for calibration, because $\Delta x(t)$ grows in time  (figure~\ref{fig:MD-illust-unsted}).
%

Before calibration, we have to make sure that the same system -- in terms of geometry and fluid properties -- is represented in CFM and MD. The dimensions and physical properties  are reported in figure~\ref{fig:intro0}(a). The bulk liquid density is obtained by taking an average of $\rho_y$ over the height of the channel
\begin{equation}
    \rho_\ell = \int\limits_{y_b}^{y_t} \rho_y \left(y\right)\,
    \mathit{dy} \approx 990\,\mathrm{kg/m}^3.
\end{equation}
This density is valid for all equilibrium angles. Viscosity and surface tension of liquid SPC/E water are taken from previous work \citep{lacis2020steadyDresden}. The viscosity measurement of the vapour SPC/E phase from MD is impractical. Therefore, both viscosity and density of the vapour are determined from engineering tables (appendix~\ref{app:vap-prop}).
For VOF simulations, we increase the vapour density to $\rho_v^{VOF} = 9.9$ kg/m$^3$. This ensures numerical stability of the simulations, while keeping the influence of the vapour inertia negligible.

The size of the channel in the $x$-direction (distance between periodic boundary conditions) is matched with the distance between left and right periodic boundary conditions in MD system. The channel height is based on the chosen hydrodynamic wall position at the centre of the red bin (figure~\ref{fig:wall-allQ-summary}a-d), which results in $H = 29.22$ nm. This position is the subject of an investigation in itself~\citep{herrero2019shear} and further discussed in appendix~\ref{app:wall-loc}.  

\rev{T}he slip length \eqref{eq:ns-bc} used in CFM models \rev{in principle can be directly related to MD simulations. Previous work~\citep{huang2008water} has demonstrated that the slip length of MD system follows a quasi-universal relationship with respect to equilibrium contact angle $\theta_0$. However, for the chosen MD system (SPC/E water on SiO$_2$ surrogate wall), accurate slip length quantification has not yet been done. To obtain an indication about validity of $\ell_s = 0$ for the selected MD system, we compare streamwise velocity near the wall between PF and MD (appendix~\ref{app:wall-loc}). Through this comparison, } we found that $\ell_s = 0$ holds for $\theta_0 = 38^\circ - 95^\circ$, while for $\theta_0 = 127^\circ$ the appropriate choice is $\ell_s = 0.44$ nm. \rev{These values are used both in VOF and PF as input parameters without any additional fitting, see sixth and ninth columns of table~\ref{tab:calibration}. Note that $\ell_s$ in VOF is fixed and distinct from $\lambda$ -- the length scale in condition (\ref{eq:VOFContactAngle}), which is used for calibration.}



\begin{table}
\begin{center}
\begin{tabular}{p{6mm}p{8mm}|p{9mm} | p{8mm}p{8mm}p{7mm}p{8mm} | p{7mm}p{7mm}p{14mm}p{10mm}p{10mm}}
$\theta_0$ & $\Ca$ &  $\Delta x$ & $\Delta/2$ & $\lambda$ & $\rev{\ell_s}$ & $\Delta x$ & $\epsilon$ & $\rev{\ell_s}$ & $M \times 10^{16}$ & $\mu_f/\mu_\ell$ &  $\Delta x$\\
 & &  [nm] &  [nm] & [nm] &  $\mbox{\rev{[nm]}}$ & [nm]&  [nm] & $\mbox{\rev{[nm]}}$ &  [m$^4$/N$\cdot$s] & &  [nm]\\
   &&&&&&&&&\\
$127^\circ$ & $0.60$ & $11.73$  & $0.457$ &  $4.325$ & $\rev{0.44}$ & $11.53$ & $0.7$ & $\rev{0.44}$ & $235.5$ & $0$ & $11.74$\\
$95^\circ$ & $0.20$ & $13.89$  & $0.457$ & $0.935$ & $\rev{0.00}$ & $13.81$ & $0.7$ & $\rev{0.00}$ & $10.80$ & $0$ & $13.89$ \\
$69^\circ$ & $0.10$ & $16.98$  & $0.457$ & $0.313$ & $\rev{0.00}$ & $16.75$ & $0.7$ &  $\rev{0.00}$ & $3.500$ & $2.361$ & $16.99$\\
$38^\circ$ & $0.015$ & $5.31$  & $0.457$ & $0.146$ & $\rev{0.00}$ &$5.38$ & $0.7$ & $\rev{0.00}$ & $3.500$ & $11.84$ & $5.31$ 
\end{tabular}
\end{center}
\caption{Results of CFM calibration against MD. For each $(\Ca,\theta_0)$ pair, we report MD reference data (steady displacement $\Delta x$) in third column. In following \rev{four} columns, we report VOF parameters \rev{(eq. \ref{eq:ns-bc} and \ref{eq:VOFContactAngle})} and resulting steady displacements. In final \rev{five} columns, PF parameters \rev{(eq. \ref{eq:ns-bc}, \ref{eq:cahn-hil-dim} and  \ref{eq:wet-bc})} and resulting displacements are given.}
\label{tab:calibration}
\end{table}

\subsection{VOF calibration} \label{sec:vof-calib} 
Figure \ref{fig:all-calib-dx-tevol-wCFM} compares $\Delta x$ obtained from VOF (red) and MD (black) for the four calibration configurations.
We have adjusted the parameter $\lambda$ in (\ref{eq:VOFContactAngle}) to find the best fit of $\Delta x$ as $t\rightarrow\infty$. We observe that the droplet displacement from VOF stabilizes at values that agree well the atomistic simulations with the most challenging calibration configuration being $\theta_0=127^\circ$ (figure \ref{fig:all-calib-dx-tevol-wCFM}a).
The $\lambda$ values that reproduce the displacement obtained from MD are listed in sixth column of table~\ref{tab:calibration}.
%
For the hydrophobic configurations, we obtained $\lambda = 4.325$ nm for $\theta_0 = 127^\circ$ and $\lambda = 0.935$ nm for $\theta_0 = 95^\circ$.
%
%
We observe that for $\theta_0 = 95^\circ$, $\lambda$ is roughly 4 times smaller than for $\theta_0 = 127^\circ$. Naively, $\lambda$ can be regarded as a slip-related length scale near the contact line. Larger $\lambda$ for $\theta_0 = 127^\circ$ then suggests smaller friction, in line with the findings from PF calibration (\S\ref{sec:PF-calib}).


For $\theta_0 = 69^\circ$ and $38^\circ$ we found even smaller $\lambda$ values, $0.313$ nm and $0.146$ nm, respectively. This continuous behaviour of decaying $\lambda$ for smaller $\theta_0$, qualitatively, follows the contact line friction argument presented later (\S\ref{sec:PF-calib}). It is also interesting to note that the obtained $\lambda$ values from this calibration procedure are of order $0.1$ nm - $4$ nm. This is similar to what is used by~\cite{legendre2015comparison}, where they set $\lambda = 1$ nm in macroscopic simulations.
%
The mesh spacing in the VOF simulations is $\Delta/2 = 0.457$ nm. 
For $\theta_0 = 38^\circ$ and $69^\circ$, we have $\lambda < \Delta/2$ (table~\ref{tab:calibration}).
%
On the other hand, for $\theta_0 = 95^\circ$ and $127^\circ$, we have $\lambda > \Delta/2$ (table~\ref{tab:calibration}). Here the sign of the logarithm in (\ref{eq:VOFContactAngle}) is negative.
This is typically not the case when applying Cox inspired relationships such as (\ref{eq:VOFContactAngle}). Therefore imposed $\theta_{num}$ for $\theta_0 = 95^\circ$ and $127^\circ$ should be viewed as numerical parameter to tweak the curvature near the wall.


\subsection{PF calibration} \label{sec:PF-calib}
The droplet displacement $\Delta x$ produced by PF (green) and MD (black) are compared in figure \ref{fig:all-calib-dx-tevol-wCFM} for the calibration configurations. The interface thickness, mobility and line friction that provide the best match with the displacement obtained from MD are listed in table~\ref{tab:calibration}. For PF, different combinations of parameters may provide a good fit; therefore it is prudent to have physically motivated guidelines.

\subsubsection{PF calibration proposed in the literature} \label{sec:PF-calib-Yue}
For the Cahn-Hilliard PF model, there is a standard calibration procedure proposed by \cite{yue2010sharp,yue2011wall}. The sequential steps are; i) choosing the interface thickness $\epsilon$ that is suitable to describe the physical problem; ii) setting the PF mobility $M$ according to the sharp interface limit~\citep{yue2010sharp} and; iii) calibrating the contact line friction $\mu_f$ against experiments.

The interface thickness has to be smaller than the important physical length scales in the chosen system, which for the sheared droplet configuration are the water drop height $H = 29.22$ nm and width $W \approx 38$ nm (figure~\ref{fig:intro0}a). Based on this, the interface thickness is set to $\epsilon = 0.7$ nm.
According to the sharp interface limit~\citep{yue2010sharp}, the criterion for choosing $M$ is
\begin{equation}
M > 1/16\, \epsilon^2 / \sqrt{\mu_v\, \mu_\ell} . \label{eq:sh-itf-lim-M}
\end{equation}
Inserting the chosen values for interface thickness, vapour viscosity and liquid viscosity, we obtain $M >  3.21 \times 10^{-16}$ m$^4$/(N s). In the MD simulations, we do not observe any physical effect that would hint towards a large $M$ value. Therefore, for all $\theta_0$ in this section, we select $M = 3.5 \times 10^{-16}$ m$^4$/(N s), which is close to the lower limit. 

 \begin{table}
 \begin{center}
 \begin{tabular}{p{8mm}p{12mm}p{18mm}p{10mm}p{10mm}p{10mm}p{10mm}p{10mm}p{10mm}}
 \multicolumn{3}{l}{ \mbox{ } } & \multicolumn{6}{c}{ $\mu_f / \mu_\ell$ in PF } \\
 $\theta_0$ & $\Ca$ & $\Delta x$ [MD] & $0$ &  $0.1$ & $0.2$ & $0.3$ & $2.36$ & $11.84$ \\ \\
 $127^\circ$ & $0.60$ & $11.73$ nm & unst. & unst. & unst. & unst. & unst. & unst. \\
 $95^\circ$ & $0.20$ & $13.89$ nm & $20.36$ & $21.21$ & $22.18$ & $23.26$ & unst. & unst.  \\
 $69^\circ$ & $0.10$ & $16.98$ nm & $11.00$ & $11.23$ & $11.43$ & $11.64$ & $16.99$ & unst. \\
 $38^\circ$ & $0.015$ & $5.31$ nm & $2.54$ & $2.56$ & $2.58$ & $2.60$ & $3.08$ & $5.31$
 \end{tabular}
 \end{center}
 \caption{Calibrating PF with MD by changing $\mu_f$. Strategy proposed by~\cite{yue2011wall}. In third column, we show the reference $\Delta x$ from MD. To the right, we show steady $\Delta x$ (in nm) obtained from PF for specified $\mu_f / \mu_\ell$. If no steady state exists, we write ``unst.''. In all simulations, $M = 3.5 \times 10^{-16}$ m$^4$/(N s) and $\epsilon = 0.7$ nm.}
 \label{tab:try-Yue-match}
 \end{table}


The last step is to carry out simulations with different $\mu_f$ until a steady state is attained that matches the displacement obtained from MD. Table~\ref{tab:try-Yue-match} summarizes the displacement obtained for each calibration couple $(\Ca,\theta_0)$ at $\mu_f = 0, 0.1, 0.2, 0.3, 2.36$ and $11.84$.
The fourth column of table~\ref{tab:try-Yue-match} shows the displacements for $\mu_f = 0$. We observe that no steady solution has been obtained for $\theta_0 = 127^\circ$.  The steady $\Delta x$ obtained for $\theta_0=95^\circ$ overestimates the MD value (third column of table~\ref{tab:try-Yue-match}) by roughly $50\%$. Increasing $\mu_f$ only deteriorates the agreement. Therefore, we conclude that the matching procedure proposed by~\cite{yue2011wall} is not adequate to calibrate the PF model for the chosen nanoscale configuration at $\theta_0 = 127^\circ$ and $95^\circ$.



For $\theta_0 = 69^\circ$ and $38^\circ$, on the other hand, $\mu_f = 0$ results in an underestimated drop displacement (table~\ref{tab:try-Yue-match}) compared to MD. 
We observe that the required PF contact line friction for $\theta_0 = 38^\circ$ ($\mu_f = 11.84\,\mu_\ell$) is roughly five times larger than the one for $\theta_0 = 69^\circ$ ($\mu_f = 2.36\,\mu_\ell$). 
More hydrophilic $\theta_0$ entails a stronger affinity between the liquid and the wall. Stronger affinity, in turn, can yield larger friction, which is consistent with the obtained $\mu_f$ values.

\subsubsection{Calibrating PF by adjusting mobility for $\theta_0 = 127^\circ$ and $95^\circ$} \label{sec:match-PF-MD-M}


A way to reduce friction near the contact line is to increase $M$ (figure~\ref{fig:dx-PF-param-var}a) and thus allow for more diffusion. Therefore, we impose $\mu_f = 0$ and increase $M$ for $\theta_0 = 127^\circ$ and $95^\circ$ until the steady $\Delta x$ from the PF agrees with MD.
The required PF mobility values are reported in table~\ref{tab:calibration}. We observe that for $\theta_0 = 95^\circ$, a three times larger $M$ is required compared to the $M$ set by the sharp interface limit (see tenth column of table~\ref{tab:calibration}). Whereas for $\theta_0 = 127^\circ$ the $M$ value has to be increased by a factor of hundred.

Note that this calibration procedure is compatible with the sharp interface limit. The condition (\ref{eq:sh-itf-lim-M}) can be rewritten in terms of interface thickness as $\epsilon < 4\,M^{1/2}\,\mu_v^{1/4}\,\mu_\ell^{1/4}$. As we increased $M$, the interface thickness $\epsilon = 0.7$ nm was kept constant, and therefore the condition is satisfied.
This calibration procedure produces however a noticeable diffusion near the contact line
(figure~\ref{fig:streamL-PF-param-var}c), which is not observed in the MD simulations.

\begin{figure}
\centering
\includegraphics[width=1.0\linewidth]{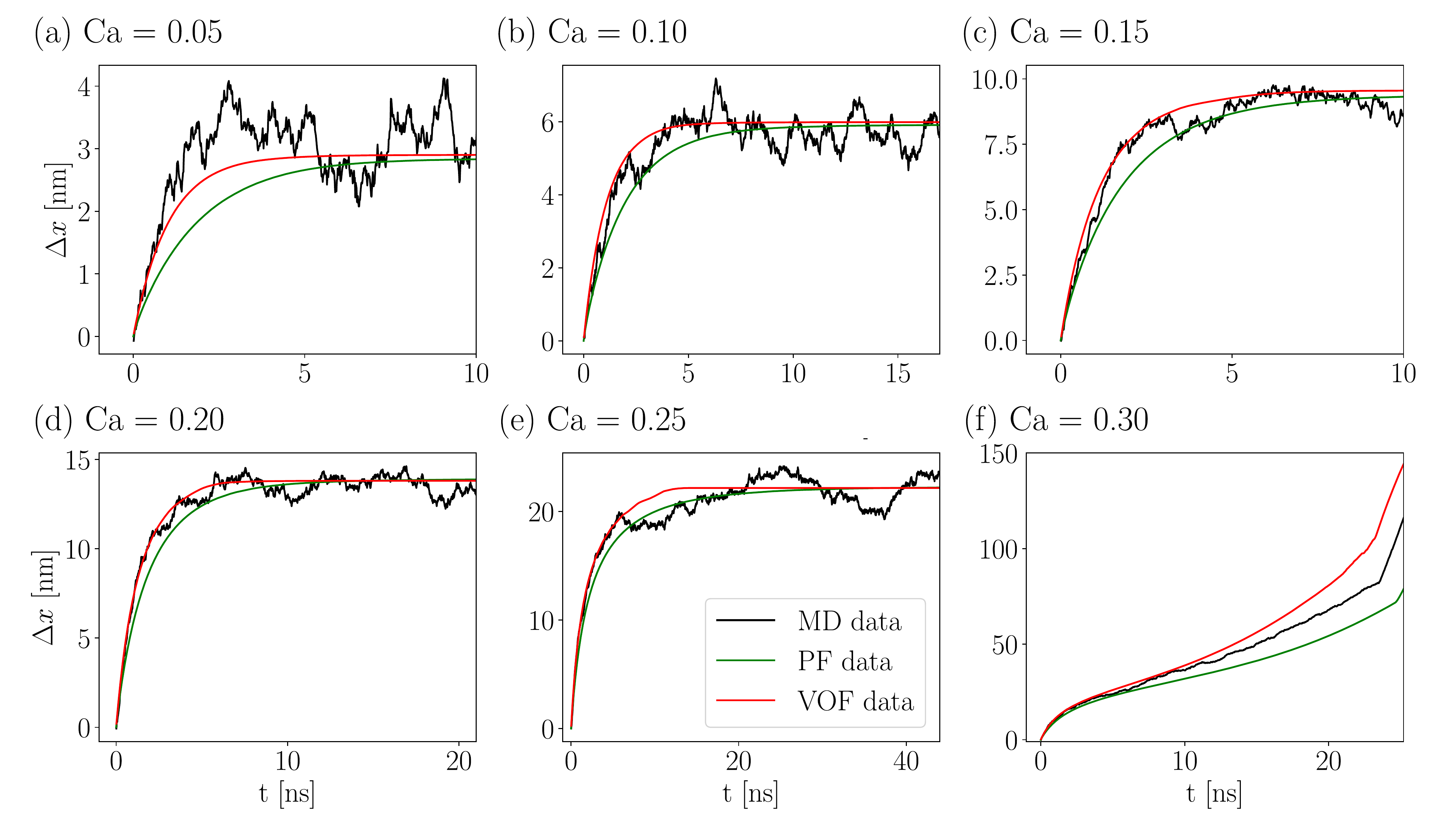}
\caption{Time evolution of $\Delta x$ in MD, PF and VOF for $\Ca \in (0.05, 0.30)$. Equilibrium
contact angle $\theta_0 = 95^\circ$. The calibrated function $\Delta x (t)$ from the PF and the VOF is shown in figure~\ref{fig:all-q2-dx-tevol-wCFM}(d) with green and red lines, respectively. The 
\emph{a priori} 
measurements of $\Delta x (t)$ from MD (\S\ref{sec:md-sheared}) are shown with black line. This colour code is retained for all comparisons that follow. 
}
\label{fig:all-q2-dx-tevol-wCFM}
\end{figure}

\section{Predictions from PF and VOF models} \label{sec:CFM-MD-pred}
We have shown that continuum systems can be tuned to match the final steady droplet displacement computed from MD simulations. It is also necessary to understand how well the CFM models capture other key features of the system, including the interface shape and the time-dependent transient behavior of the droplet. Moreover, an important practical aspect is the accuracy of the CFM when it comes to predicting the droplet behavior away from calibration conditions. 
The aim of this section is therefore to quantitatively characterize the sheared droplet system for a range of capillary numbers. 
Specifically, in this section, we fix the parameters for PF and VOF to values reported in table~\ref{tab:calibration}.


\subsection{Time evolution of drop displacement} \label{sec:pred-dx-vs-t}
Figure \ref{fig:all-q2-dx-tevol-wCFM} shows the droplet displacement as a function of time for a fixed $\theta_0=95^\circ$ but different $\Ca$. Figure~\ref{fig:all-q2-dx-tevol-wCFM}(d) is identical to figure~\ref{fig:all-calib-dx-tevol-wCFM}(b) and corresponds to the conditions for which the system was calibrated for, i.e. $\Ca=0.20$.  
We observe that both PF and VOF capture the transient dynamics  very well overall. The PF model is slightly slower (predicts smaller $\Delta x$ at the same time instant) than VOF and MD. 

\begin{figure}
\centering
\includegraphics[width=1.0\linewidth]{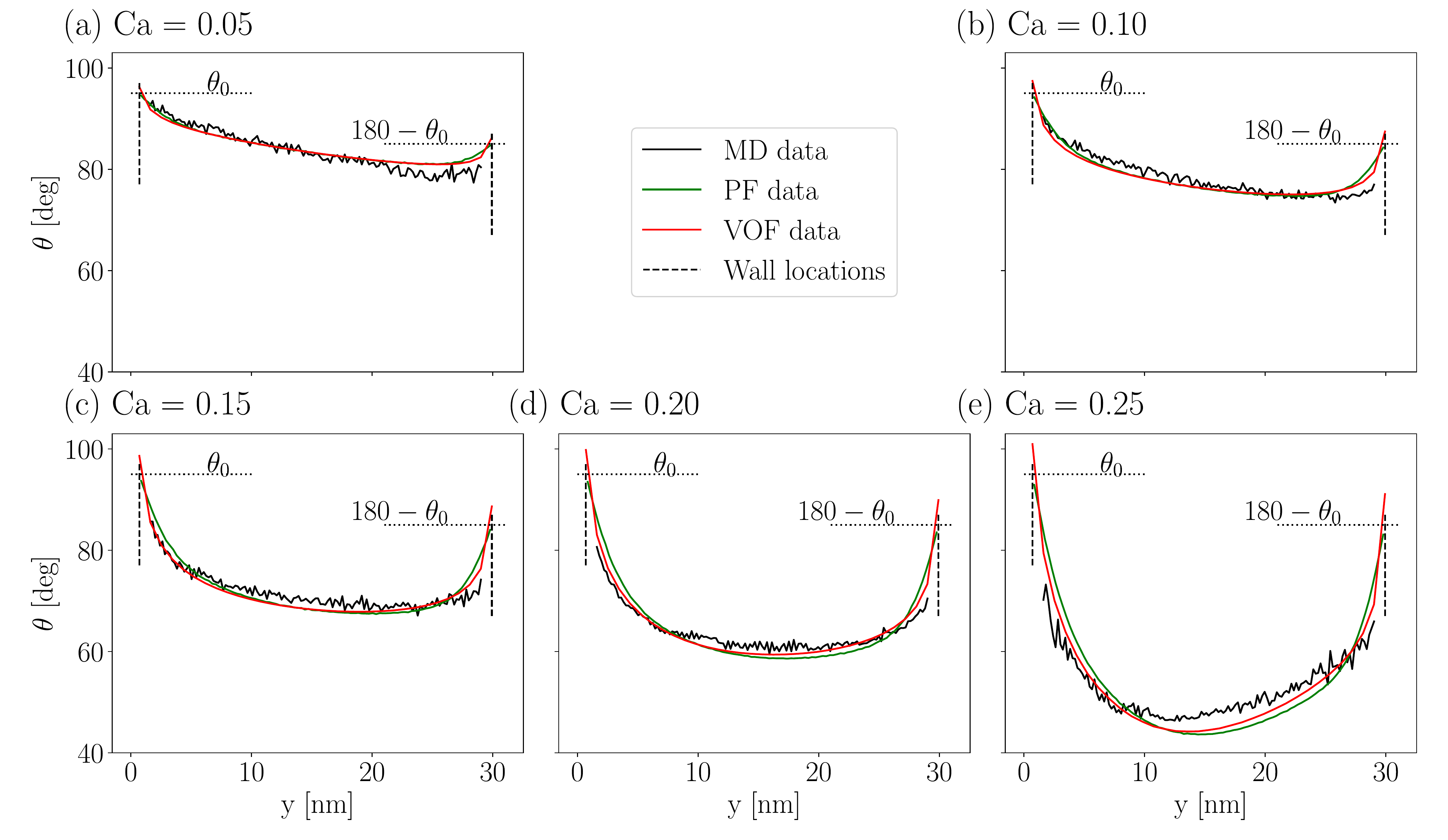}
\caption{Steady interface shape from MD, PF, and VOF simulations, $\theta_0 = 95^\circ$. Equilibrium angles and solid wall locations are
shown with black dotted and dashed lines, respectively.}
\label{fig:q2-allStedCa-angDist}
\end{figure}

Figure~\ref{fig:all-q2-dx-tevol-wCFM}(a-c,e) shows the transient behavior of CFM models in off-calibrations conditions for $\Ca = 0.05$, $0.10$, $0.15$ and $0.25$. For moderate Capillary numbers $\Ca \geq 0.10$, we observe qualitatively similar behaviour as for $\Ca = 0.20$. The CFM models predict the transient and steady $\Delta x$ values rather accurately. The PF, however, is always slightly slower compared to the VOF model. For $\Ca = 0.05$, the steady $\Delta x$ value is slightly lower in CFM models than in MD. The PF model is slower in the transient compared to the VOF, and the agreement with MD is arguably worse. 

To conclude $\theta_0 = 95^\circ$ investigations, we consider the CFM model predictions for the unsteady configuration with $\Ca = 0.30$ (figure~\ref{fig:all-q2-dx-tevol-wCFM}f). The results from PF, VOF, and MD are indistinguishable for the first $5$ ns showing excellent predictive capability. For later times VOF simulation over-predicts and PF simulation under-predicts the $\Delta x$ observed from MD. This is in line with observations in steady situations (figure~\ref{fig:all-q2-dx-tevol-wCFM}a-e). Both VOF and PF exhibit the rapid change of slope at around $23.5$ ns, corresponding to the drop break-up (discussed in \S\ref{sec:md-sheared}).

We have carried out the same investigation for $\theta_0=38^\circ, 95^\circ$ and $127^\circ$. Qualitatively similar results to those observed in figure~\ref{fig:all-q2-dx-tevol-wCFM} are obtained, see Supplementary Figures 1-3. 


\subsection{Interface shape}
For interface shape comparisons, the data is presented as the variation of the angle along the interface $\theta \left( y \right)$ (figure~\ref{fig:dx-def-VOF-lm-var}a). 
Figure~\ref{fig:q2-allStedCa-angDist} compares $\theta(y)$ obtained from MD, PF, and VOF for $\theta_0=95^\circ$ and different $\Ca$.
The equilibrium angles at the bottom and top walls are shown with a black dotted line while the  hydrodynamic wall positions are presented with black dashed lines. 
%
%
The interface shapes in both calibration ($\Ca=0.20$) and in off-calibration conditions ($\Ca = 0.05$, $0.10$, $0.15$ and $0.25$) are similar between the models. We observe that both PF and VOF exhibit more pronounced differences from MD at the top wall (near advancing contact line). At $\Ca = 0.25$ (figure~\ref{fig:q2-allStedCa-angDist}e) the CFM model predictions are notably different from MD data compared to the other capillary numbers. The loss of accuracy could arise from the large $\Delta x$ oscillations in MD at $\Ca = 0.25$ (\S\ref{sec:md-sheared} and \S\ref{sec:md-stick-slip}) that are not captured with the CFM models. 


The MD data in Figure~\ref{fig:q2-allStedCa-angDist} does not extend all the way to the wall, so that the exact value of the dynamic contact angle is not known. At $\theta_0=95^\circ$, the dynamic contact angle in PF simulations is equal to the equilibrium angle since $\mu_f = 0$ (table~\ref{tab:calibration}). For VOF, the dynamic contact angle is different than $\theta_0$. However, the angle is larger (smaller) than the equilibrium angle for the receding (advancing) contact line. This is opposite to the understanding arising from analysis of uncompensated Young stress. The source of this effect is the value of $\lambda = 0.935$ nm $> 0.467$ nm $ = \Delta / 2$ (table~\ref{tab:calibration}). 

We have repeated this comparison of interface shapes at different $\Ca$ numbers between MD, PF, and VOF for equilibrium contact angles $\theta_0 = 127^\circ$, $69^\circ$, and $38^\circ$. The results are reported in Supplementary Figures 4-6 and are similar to what is presented above.

\subsection{Steady drop displacement} \label{sec:CFM-pred-dx-vs-Ca}

\begin{figure}
\centering
\includegraphics[width=1.0\linewidth]{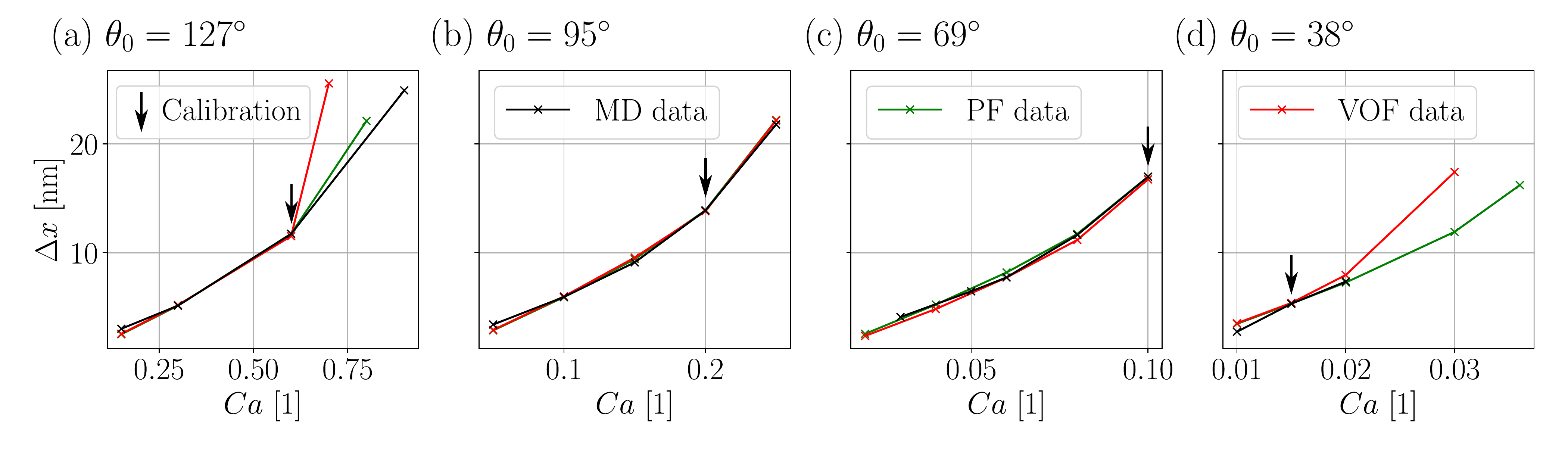}
\caption{Steady $\Delta x$ as a function of $\Ca$ number for all $\theta_0$.}
\label{fig:q24-dx-vs-Ca}
\end{figure}

Figure~\ref{fig:q24-dx-vs-Ca} shows the steady displacement as function of $Ca$ for different $\theta_0$.  
As expected, the MD, PF and VOF points collapse for the calibration configurations (marked with vertical arrows). 
%
%
For $\theta_0 = 127^\circ$ configuration, the PF and VOF model overestimates the steady displacement for large $\Ca$. A possible reason behind the discrepancy between PF, VOF and MD could be the slip length $\ell_s$. It has been determined for $\Ca = 0.60$ and kept constant for runs with different $\Ca$. In general, the slip length can increase for larger wall shear stress \citep{thompson1997general}. This, consequently, would reduce $\Delta x$ and possibly move PF and VOF predictions closer to MD results.
%
For $\theta_0 = 38^\circ$, the agreement between PF, VOF, and MD diverge
as we increase $\Ca > 0.02$. A reason of the disagreement could be the increased contact line friction (\S\ref{sec:md-clf-meas}) at receding contact line, from which liquid film is formed (\S\ref{sec:drop-break} and figure~\ref{fig:q4-Ca005-break-showcase}). This asymmetry is not taken into account in the CFM simulations.
%

\subsection{Critical Capillary number}
\begin{figure}
\centering
\includegraphics[width=0.49\linewidth]{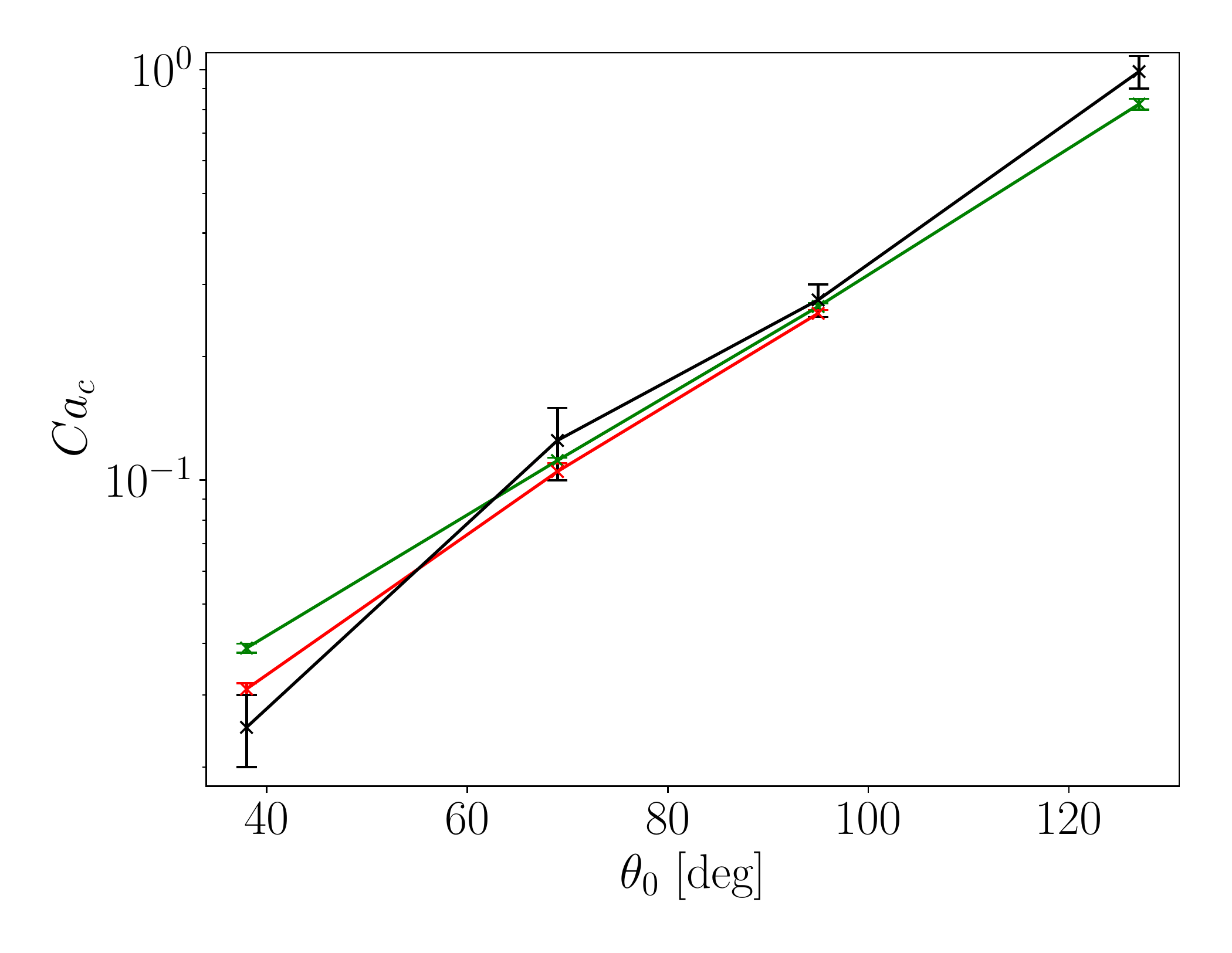}\vspace*{-15pt}
\caption{Critical Capillary number $\Ca_c$ as a function of $\theta_0$ }
\label{fig:MD-PF-VOF-stab-tresh}
\end{figure}

Figure~\ref{fig:MD-PF-VOF-stab-tresh} shows the critical Capillary number, $\Ca_c$, as a function of $\theta_0$ from the three methods. 
To determine $\Ca_c$ for a given $\theta_0$, we take the mean of the largest steady ($\Ca_s$) and smallest unsteady ($\Ca_u$) Capillary number that was simulated, i.e.
\begin{equation}
\Ca_c = \frac{\Ca_u + \Ca_s}{2} \pm \frac{\Ca_u - \Ca_s}{2}.
\end{equation}
The uncertainty is determined as half of the difference between these $\Ca$ numbers. 
To reduce $\Delta Ca_c$ in CFM, we sample the $\Ca$ space with smaller intervals. 
%
It can be observed from figure~\ref{fig:MD-PF-VOF-stab-tresh} that $\Ca_c$ increases for larger $\theta_0$, which has also been reported in previous numerical \citep{sbragaglia2008wetting} and analytical \citep{hocking2001meniscus,eggers2004hydrodynamic} investigations.

We did not manage to run any VOF simulations for $\theta_0 = 127^\circ$ and $\Ca \geq 0.80$.
Due to a large $\lambda$ (table~\ref{tab:calibration}), the dynamic angle for $\Ca \geq 0.80$ falls outside of physically admissible range $\left(0^\circ, 180^\circ\right)$. Consequently, in figure~\ref{fig:MD-PF-VOF-stab-tresh} we compare only PF prediction with the MD data at $\theta_0 = 127^\circ$. The PF model slightly under-predicts the $\Ca_c$ value. This discrepancy could again be due to the uncertainty in the slip length $\ell_s$.
%

Figure~\ref{fig:MD-PF-VOF-stab-tresh} shows that predictions of $\Ca_c$ for $95^\circ$ and $69^\circ$  agree with MD results within the accuracy bounds of the MD.
%
For $\theta_0=38^\circ$, however, discrepancies are observed, where $\Ca_c$ computed from both CFM models are larger compared to the MD results. 
Note that the differences at $38^\circ$ are enhanced due to the logarithmic scale of $\Ca_c$ axis in figure~\ref{fig:MD-PF-VOF-stab-tresh}.

\subsection{Above the critical Capillary number} \label{sec:drop-break}

In this section, we investigate the accuracy of the CFM models for predicting the unsteady breakage of the droplet. This constitutes the most challenging test of the CFM models in off-calibration conditions. Two configurations are selected with $\Ca > \Ca_c$, namely,  $\theta_0 = 95^{\circ}$ at $\Ca = 0.30$ and $\theta_0 = 38^{\circ}$ at $\Ca = 0.05$.

\begin{figure}
\centering
\includegraphics[width=0.90\linewidth]{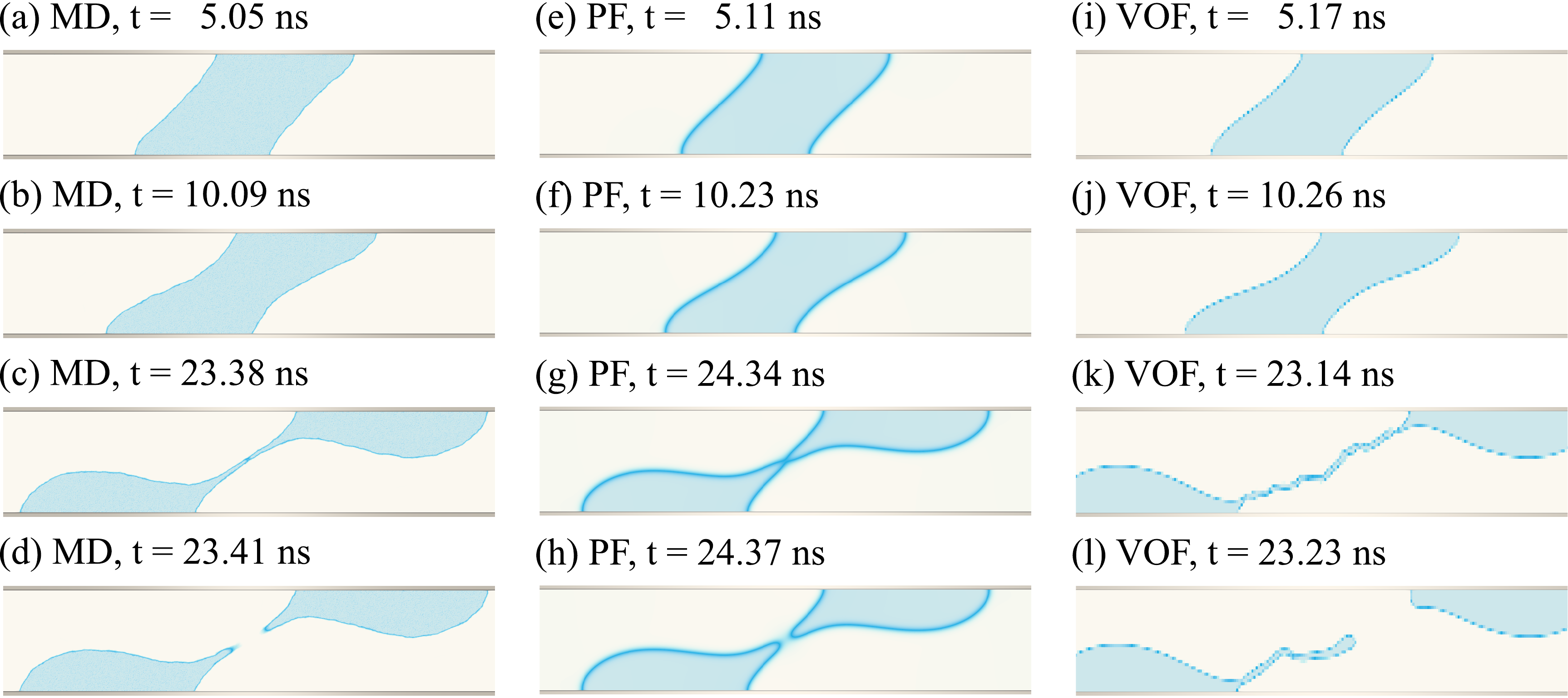}
\caption{Snapshots of interface shape evolution over time from MD (a-d), PF (e-h) and VOF (i-l). Equilibrium contact angle $\theta_0 = 95^\circ$ and $\Ca = 0.30$.}
\label{fig:q2-Ca030-break-showcase}
\end{figure}

Figure~\ref{fig:q2-Ca030-break-showcase} shows the drop shape at four time instances for $(Ca,\theta_0)=(0.30,95^{\circ})$.  We observe that the three models provide similar deformed states at $t \approx 5$ ns and $t \approx 10$ ns, see top two rows in figure~\ref{fig:q2-Ca030-break-showcase}. 
The time instance just before and right after the break-up is shown in third and fourth rows in figure~\ref{fig:q2-Ca030-break-showcase}. 
The thread connecting the lower and upper drop is very thin in MD and VOF simulations, compared to the PF simulation. In addition, the thread in VOF  is comparably  long and exhibits grid-to-grid like oscillations. There are also pronounced differences in the neck of each satellite drop -- the region in which the thread transitions to the drop shape. The PF simulations show the thickest neck (figure~\ref{fig:q2-Ca030-break-showcase}g), followed by MD with slightly thinner neck (figure~\ref{fig:q2-Ca030-break-showcase}c) and VOF with very small drop neck (figure~\ref{fig:q2-Ca030-break-showcase}k). The time instant at which the break-up occurs is remarkably similar ($t^{MD}_b = 23.41$ ns for MD, $t^{PF}_b = 24.37$ ns for PF and $t^{VF}_b = 23.23$ ns for VOF). For a complete time-dependent animation of MD, PF, and VOF simulations side by side, see Supplementary Movie 1.

\begin{figure}
\centering
\includegraphics[width=1.00\linewidth]{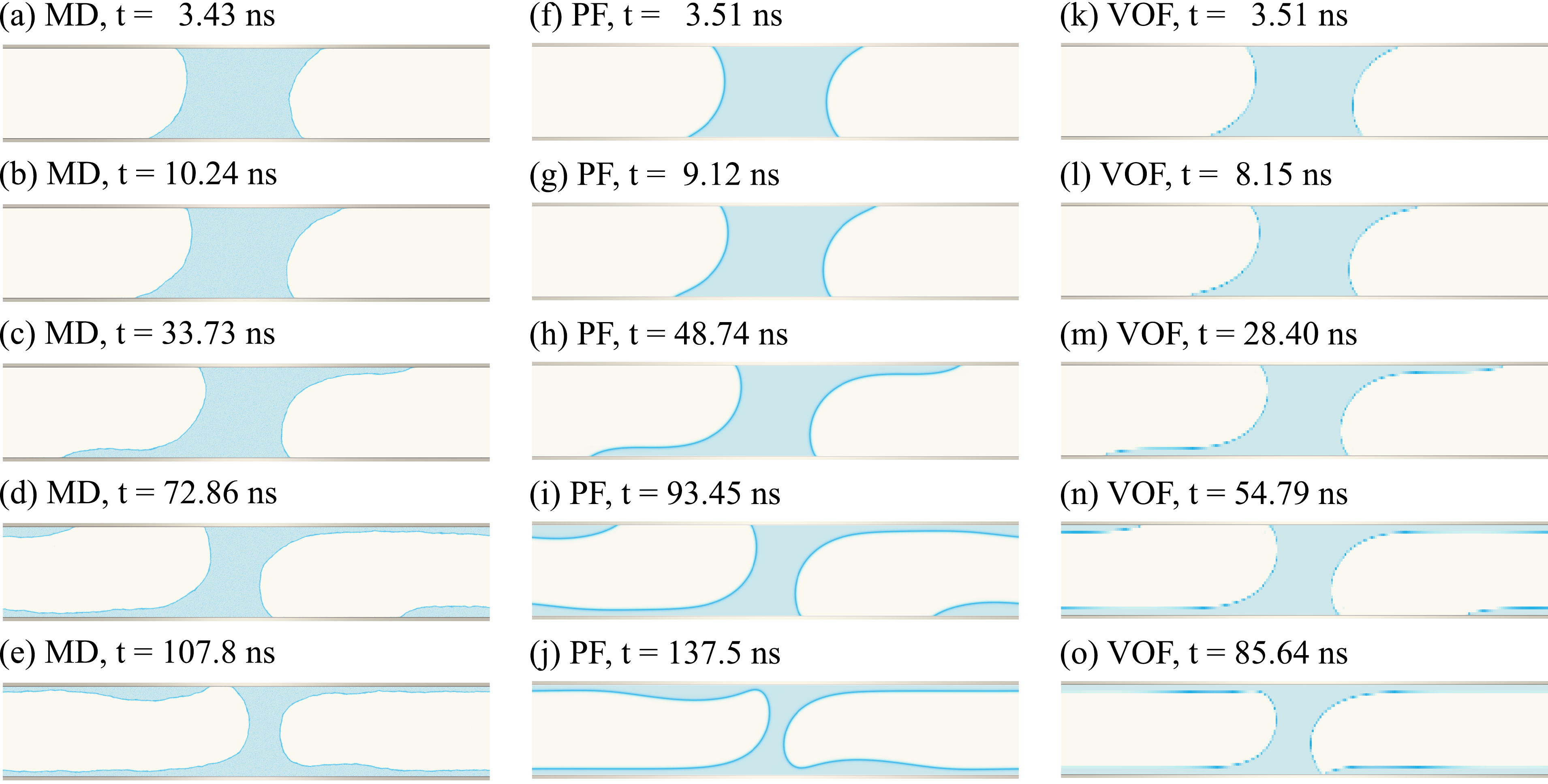}
\caption{Snapshots of interface shape evolution over time from MD (a-e), PF (f-j), and VOF (k-o).
Equilibrium contact angle $\theta_0 = 38^\circ$ and Capillary number $\Ca = 0.05$. For PF, we have $\epsilon = 0.35$ nm.}
\label{fig:q4-Ca005-break-showcase}
\end{figure}

We turn our attention to $\theta_0 = 38^{\circ}$ and  $\Ca = 0.05$. To avoid a premature break-up of the liquid bridge in the PF simulation, we use $\epsilon = 0.35$ nm for this particular simulation since
in this unsteady example, a third length scale -- width of the liquid bridge in the deformed state -- becomes important. 
Droplet shapes at five time instances from MD, PF, and VOF are shown in figure~\ref{fig:q4-Ca005-break-showcase}. We observe that initially (top two rows in figure~\ref{fig:q4-Ca005-break-showcase}) the drop is deforming only slightly. After a certain time (figure~\ref{fig:q4-Ca005-break-showcase}c,h,m) a liquid film is generated at the receding contact lines of the drop. Very similar drop shapes are observed in MD, PF, and VOF at different time instances.  The VOF takes roughly $5$ ns less to form a film similar to one observed in MD. For the PF, one has to wait for around $15$ ns more than in MD. As time progresses, the liquid film becomes longer (figure~\ref{fig:q4-Ca005-break-showcase}d,i,n), and the liquid bridge becomes thinner. We also note that the tip of the film forms a thicker drop-like region. Finally, the liquid film from the periodic image merges with the liquid bridge (figure~\ref{fig:q4-Ca005-break-showcase}e,j,o). For MD and VOF simulations, the coalescence occurs only near one of the walls. For PF, on the other hand, a symmetric configuration is obtained with fully wetted top and bottom walls. Animations of the simulations can be found in supplementary movie 2.


\section{Molecular physics of the sheared droplet} \label{sec:md-phys}
In this section, we present molecular phenomena of the sheared droplet that are particularly challenging to model in a continuum model.  First, we present contact line friction measurements directly from MD and asymmetry between advancing and receding lines for hydrophilic and hydrophobic configurations. Second, we discuss the nature of the stick-slip like oscillations. Finally, we show the inevitable three-dimensional nature of the drop break-up.

\subsection{Contact line friction measurements from MD} \label{sec:md-clf-meas}

To extract the $\mu_f$ from MD, we use the formula proposed by~\cite{yue2011wall},
\begin{equation}    \label{eq:wall_energy_relaxation}
    \Bigg[\frac{\sqrt{2}}{3}\frac{\mu_f}{\mu_\ell}\sin\theta\Bigg]Ca = \cos\theta_0 - \cos\theta .
\end{equation}
Here, $\theta$ is the dynamic contact angle at the wall. This expression is an approximation of the wetting boundary condition (\ref{eq:wet-bc}) in case of no-slip and small Capillary number. \rev{Note that there are alternative approaches to determine contact line friction, for example, based on equilibrium simulations, as proposed by~\cite{fernandez2019contact,fernandez2020hidden}. For this work, however, we have determined that non-equilibrium approach based on fitting (\ref{eq:wall_energy_relaxation}) to MD data is the most efficient approach.}


\begin{figure}
    \centering
    \includegraphics[scale=0.225,trim={5cm 2cm 5cm 2cm},clip]{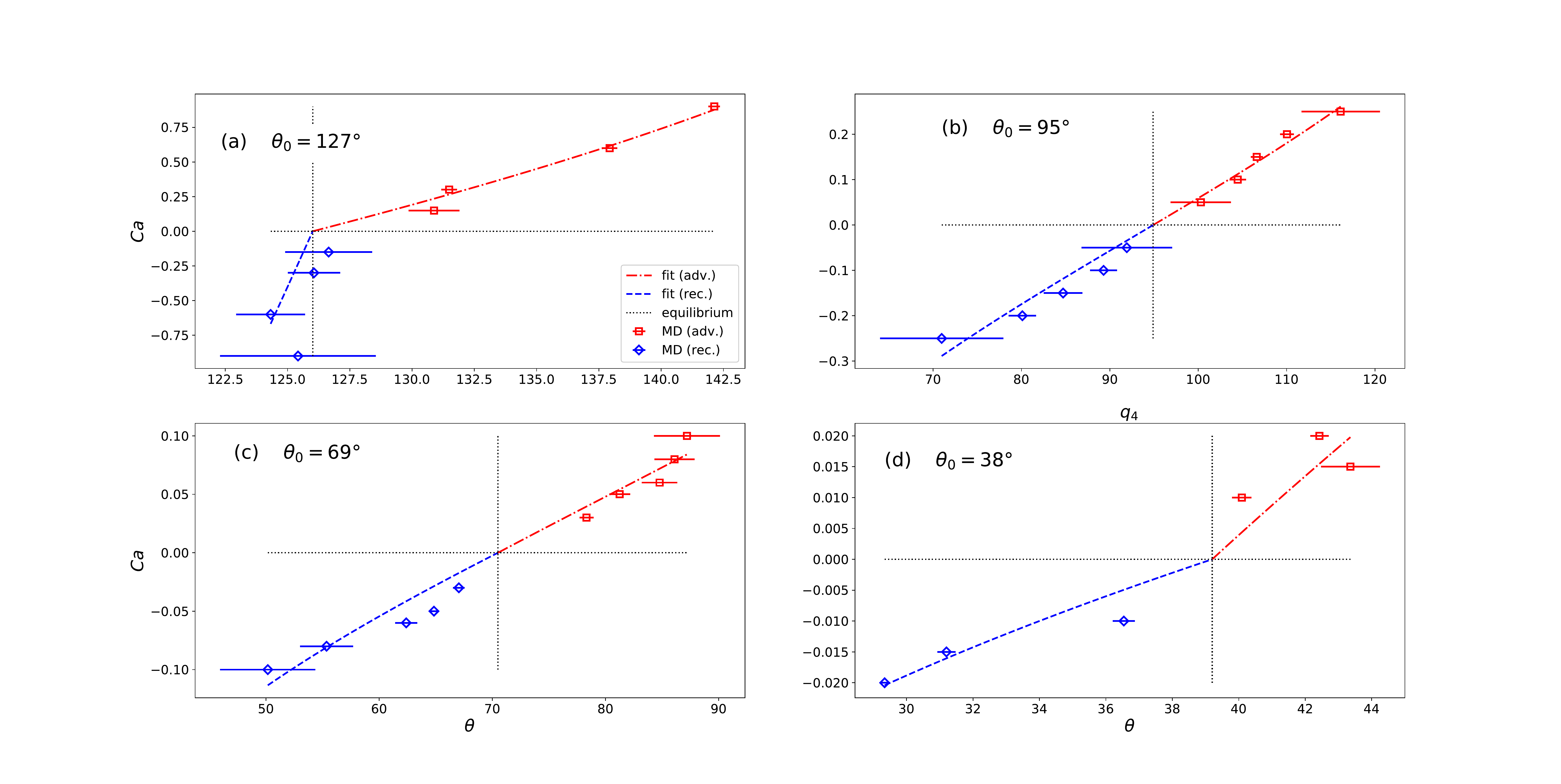}
    \caption{Measured dynamic contact angles $\theta$ for different $\Ca$ along with uncertainty interval for advancing (red squares) and receding (blue rhombus) contact lines. The dashed lines are the best least-squares fit of the expression (\ref{eq:wall_energy_relaxation}). Equilibrium contact angles $\theta_0 = 127^\circ$ (a), $95^\circ$ (b), $69^\circ$ (c) and $38^\circ$ (d).}
    \label{fig:line_friction_direct_md}
\end{figure}

To determine the $\mu_f$ from (\ref{eq:wall_energy_relaxation}), the dynamic contact angle $\theta$ has to first be determined for each $\Ca$ number. We extract the dynamic contact angle above the first reliable bin (green bins in figure~\ref{fig:wall-allQ-summary}a-d). To reduce the noise, we use polynomial interpolation of the interface shape to read the dynamic contact angle at the chosen location. For consistency, we also re-evaluate the $\theta_0$ at the same distance from the wall. The obtained dynamic contact angles for all $\theta_0$ values are gathered in figure~\ref{fig:line_friction_direct_md}. For each $\theta$, we also display error bars, obtained by adding $\pm 2$ to the polynomial order of the interpolation. As expected, when $\Ca$ number increases so does the deviation of $\theta$ from $\theta_0$.

We use least-squares fit to match (\ref{eq:wall_energy_relaxation}) to $\theta$ measurements. The fit provides $\mu_f/\mu_\ell$ values and error intervals. We fit the measurements taken at the top-left/bottom-right and top-right/bottom-left contact lines separately. This allows the observation of an asymmetric line friction. The obtained best-fit lines for all equilibrium contact lines are shown in figure~\ref{fig:line_friction_direct_md}. The  contact-line friction are listed in table~\ref{tab:PD-MD-muf} along with previously reported $\mu_f$ from calibration of PF simulations.

\begin{table}
\begin{center}
\begin{tabular}{p{38mm}p{18mm}p{18mm}p{18mm}p{18mm}}
$\theta_0$ & $127^\circ$ & $95^\circ$ & $69^\circ$ & $38^\circ$ \\ \\
PF, $\mu_f / \mu_\ell$ (adv \& rec) & $0.00$ & $0.00$ & $2.361$ & $11.84$ \\
MD $\mu_f / \mu_\ell$ (adv) & $0.79\pm0.019$ & $3.21\pm0.059$ & $7.20\pm0.24$ & $7.48\pm1.04$ \\
MD $\mu_f / \mu_\ell$ (rec)& $0.093\pm0.030$ & $3.20\pm0.13$ & $7.48\pm0.38$ & $20.5\pm1.64$
\end{tabular}
\end{center}
\caption{Comparison of contact line friction used in PF simulations and
values obtained directly from MD results. For PF, the same contact line friction
is used for advancing and receding contact lines, while in MD the contact 
line friction can be different.}
\label{tab:PD-MD-muf}
\end{table}

We observe that larger line friction parameters are measured for smaller $\theta_0$. This observation is consistent with the molecular-kinetic-theory (MKT). It states that the line friction scales with the work of adhesion needed to desorb water molecules from the substrate, which in turn increases as the surface becomes more hydrophilic~\citep{blake1969haynes}. Interestingly, we observe a difference between advancing and receding line friction for $\theta_0 = 127^\circ$ and $38^\circ$. Future and dedicated work is required to investigate this asymmetry in depth. In particular, to determine if the asymmetry depends on the position of the hydrodynamic wall. 
The observation of asymmetry is not specific to (\ref{eq:wall_energy_relaxation}). A linearised MKT model -- where line friction is defined in a slightly different way -- would not change the conclusions.

The line friction obtained by fitting expression (\ref{eq:wall_energy_relaxation}) directly to MD data are larger than the ones obtained through calibration of PF against MD (table~\ref{tab:PD-MD-muf}). This fact seems to entail some missing physical effects in the PF model. 
Ideally -- in a potentially more advanced PF model --  one would employ  $\mu_f$ obtained from MD directly in the PF boundary condition (\ref{eq:wet-bc}). 

\subsection{Stick-slip-like oscillations} \label{sec:md-stick-slip}

 \begin{figure}
    \centering
    \includegraphics[width=1.0\linewidth]{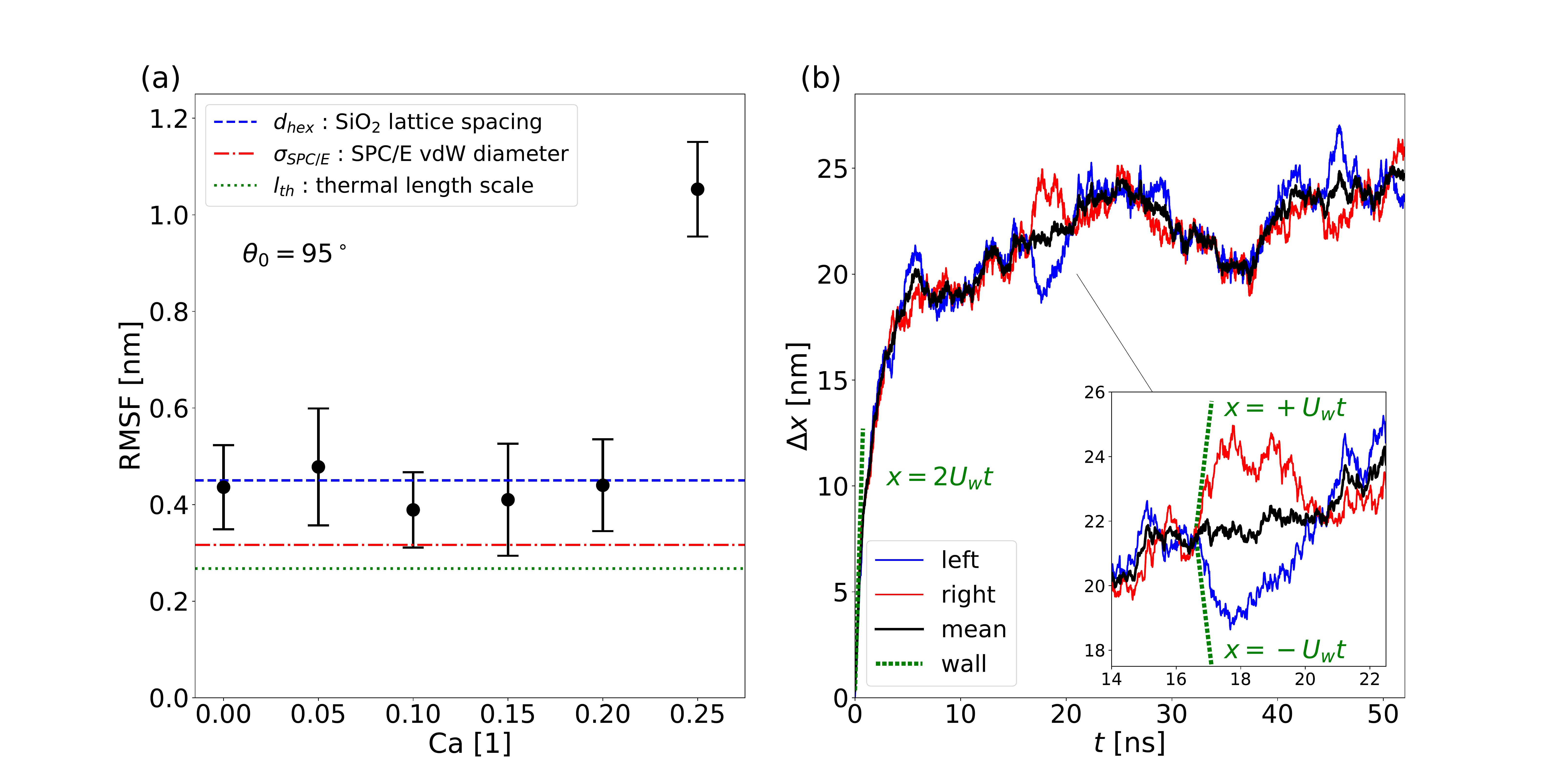}
    \vspace*{-10pt}
     \caption{\rev{(a) Root-mean-square fluctuations of all four contact lines against Capillary number for $\theta_0=95^\circ$. The error bars show a conservative estimate of the standard error. 
     The spacing in the hexagonal silica lattice (blue dashed line), the van der Waals diameter of the SPC/E water molecules (red dot-dashed line) and the length scale of thermal fluctuations (green dotted line) is shown.} (b) Drop displacement over time for $\theta_0=95^\circ$ and $Ca=0.25$. The mean ($\Delta x$), left ($\Delta x_l$) and right ($\Delta x_r$) displacement is shown. Close-up on a time interval showing pinning/depinning is shown in the inset. The green dotted lines show the slope corresponding to wall velocity: at the beginning of the simulation both advancing and receding contact lines stick to the wall and thus move apart from each other with velocity $2U_w$; during stick-slip events, the receding contact line sticks to the wall while the advancing maintain a steady motion, thus the displacement matches the velocity of a single wall $U_w$.}
     \label{fig:stick_slip_details}
\end{figure}


\rev{To differentiate between fluctuations caused by molecular-scale motion and large stick-slip-like oscillations we define three reference length scales. The characteristic length scale of interface fluctuations far from contact lines can be estimated by balancing the thermal energy $k_B T$ with the energy due to surface tension $\sigma l_{th}^2$, giving $l_{th}=\sqrt{k_B T/\sigma}\simeq 0.27$ nm. The other two scales are the van der Walls diamater of SPC/E water molecules $\sigma_{SPC/E} \simeq 0.32$ nm and the hexagonal lattice spacing of the substrate $d_{hex}=0.45$ nm. We then compute the root-mean-square for fluctuations of the contact line displacement RMSF$=(\expval{\Delta x^2}-\expval{\Delta x}^2)^{1/2}$ for each of the four contact lines, both at equilibrium and under shear conditions. For $\theta_0 = 95^\circ$, we show the obtained RMSF values as function of $\Ca$ number in figure~\ref{fig:stick_slip_details}(a).

We observe that for $Ca<0.25$ the RMSF is comparable with $d_{hex}$, hinting that the observed fluctuations for moderate Capillary numbers are caused by the same process producing fluctuations at the equilibrium, that is the local thermal-induced pinning (de-pinning) of the contact line on (from) adsorption sites close to the average contact line position. It is worth noticing that the scale of these fluctuations is larger than the one expected on the interface far from contact lines. This observation is confirmed by examining the RMSF across the whole interface at equilibrium (see Supplementary Figures~7-10). Note that for some contact angles, the fluctuations in equilibrium simulations are smaller compared to $d_{hex}$.}

We examine closer the MD simulation with $\theta_0=95^\circ$ at $Ca=0.25$ (figure~\ref{fig:all-q2-dx-tevol}a)\rev{, which shows a much larger contact line RMSF, incompatible with lattice spacing driven fluctuations.} The speed of the drop displacement during the stick-slip like motion is much smaller than $U_w$ (figure~\ref{fig:stick_slip_details}b). We conclude that the contact line does not entirely pin when resisting motion and only partially slips when complying with it. Moreover, for most of the time evolution, $\Delta x_l$ and $\Delta x_r$ are synchronized in an oscillatory motion. However, there are a few time intervals (between 17 ns and 20 ns; 45 ns and 47 ns) where indeed complete stick-slip occurs. In these intervals, the advancing and receding speeds match the magnitude of wall velocity, see inset of figure~\ref{fig:stick_slip_details}. 

It appears that a local (pinning/depinning) and a global (oscillations) processes co-exit. Pinning/de-pinning can be explained by the fact that $\Ca$ is close to the critical Capillary number, $\Ca_c$. The physical interpretation of the global oscillations is more delicate. In our modelling approach, we have implicitly assumed that the contact line motion is over-damped. This means that there should be a (possibly nonlinear) direct relation between force and speed, in which the proportionality constant is the contact line friction. This may not be true for large wall velocities. In such a case, the effects of the neighbouring flow inertia come into play. In this scenario, the coupling between positions and forces is more complex. The stochastic forcing produced by thermal fluctuations of the microscopic contact angle is no more completely dissipated. Instead, these may excite oscillation modes of the whole interface. Stick-slip has been theorized for flat surfaces and homogeneous fluids under some flow conditions~\citep{hocking2001meniscus,eggers2005cl,varma2021inertial}. However, to the best of the author's knowledge, it has not been directly observed.

The selected standard CFM models are not capable of describing stick-slip like oscillations. The Navier-Stokes equations, underlying the PF and VOF models, are inherently deterministic, where all the thermal oscillations are averaged out. To model  stick-slip like processes  in a continuum model, a possible approach could be to introduce random fluctuations on the imposed $\theta_0$. The distribution of the contact angle oscillations has been previously identified~\citep{smith2016langevin}. Fluctuations of the contact angle would in turn induce oscillations in $\Delta x (t)$.

\subsection{Three-dimensional nature of drop break-up}

In figure~\ref{fig:q2-Ca030-break-showcase}(d,h), we have observed that the interface shape obtained from MD and PF  display  more diffuse regions at the tip of the broken thread. It is tempting to conclude that PF correctly captures the MD behaviour. However, the MD snapshots in figure~\ref{fig:q2-Ca030-break-showcase} have been averaged in spanwise $z-$direction, while PF simulation is a pure 2D result.

\begin{figure}
    \centering
    \includegraphics[width=0.90\linewidth]{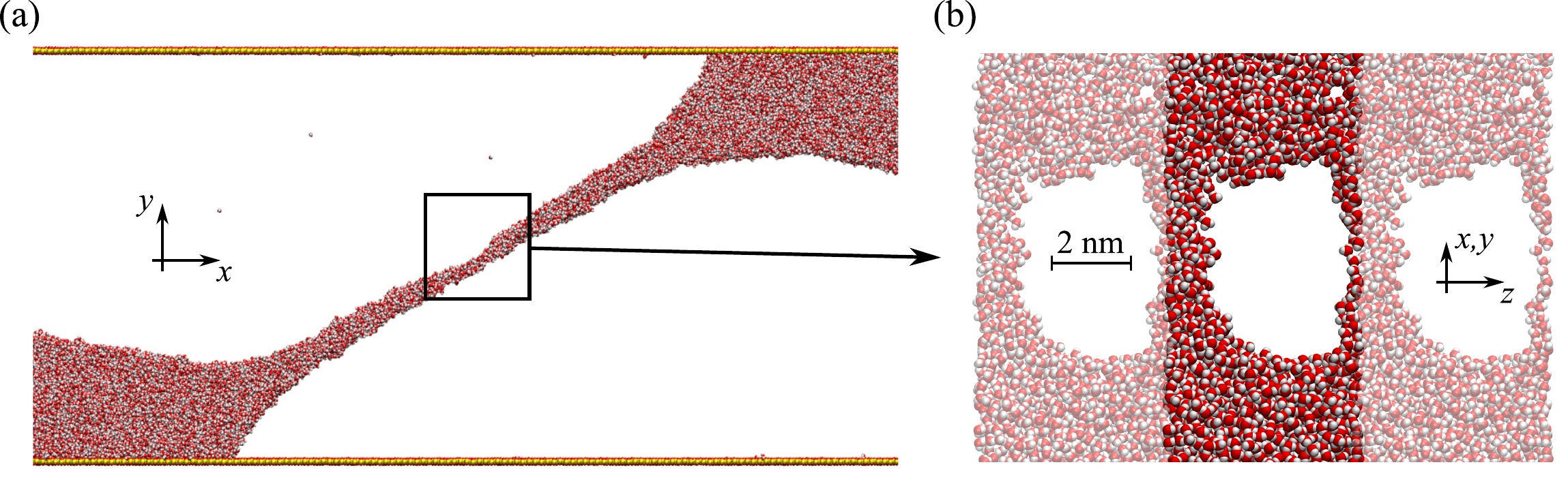}
    \vspace*{-5pt}
     \caption{Detailed view of the molecular system upon breakage for $\theta_0=95^\circ$ and $Ca=0.4$, observed at $t=9.45$ ns. Overview from the side (a) and close-up on the neck region (b). Two transparent periodic images are added at the sides. The close-up (b) is obtained by positioning the camera orthogonal to the thin thread (a).}
     \label{fig:breakage_necking}
\end{figure}

To obtain a better physical understanding of the exact break-up process, we investigate the molecular field of the MD system exactly at the break-up. In figure~\ref{fig:breakage_necking} we show water molecule locations for $\theta_0 = 95^\circ$ at $\Ca = 0.4$ shortly before the break-up. Observing the drop from the side (figure~\ref{fig:breakage_necking}a) it seems that the thread is not interrupted. In reality, the thin thread develops three-dimensional holes (figure~\ref{fig:breakage_necking}b) before disconnecting completely. Averaging in the $z$ direction results in a lower density (than the bulk) and thus giving the impression of a diffuse interface (figure~\ref{fig:q2-Ca030-break-showcase}e).

It is also worth noticing that the formation of these three-dimensional threads occurs very quickly. The time it takes for the threads to form is around $50$ ps before the actual break-up. The break-up itself occurs after a few nanoseconds from the instant the two-dimensional neck starts to form. We can thus infer that for the chosen combination of a molecular system and domain size there exists a time scale separation between 2D and 3D breakage dynamics. This in turn suggests that the selected MD system dimensions are reasonable to produce 2D results and can be directly compared with 2D CFM simulations.

\section{Discussion} \label{sec:discussion}
Based on the data presented in the previous sections, we discuss the accuracy, limitations and future development directions of CFM models when it comes to modeling the molecular physics in a nanoscale channel.

\subsection{Challenges of the chosen CFM models} \label{sec:limit-large-th}

For $\theta_0 = 127^\circ$, we observed differences in the steady drop displacement for $\Ca>0.6$,  see figure~\ref{fig:q24-dx-vs-Ca}(a). In general, very high contact angles  are challenging to model using the particular CFM parameters  we have chosen for calibration.
%
For VOF, it is important to consider condition \eqref{eq:VOFContactAngle} as a numerical model that allows adjusting the interface curvature at the contact line. For $\theta_0 = 127^\circ$ this was achieved by increasing $\lambda$ to values larger than the grid size, i.e. $\lambda > \Delta/2$ (table~\ref{tab:calibration}). This leads to a negative sign of logarithm in condition (\ref{eq:VOFContactAngle}). This means that for the bottom-left receding contact line (figure~\ref{fig:dx-def-VOF-lm-var}a), where $\Ca_{cl}$ is negative, we have $G(\theta_{num}) > G(\theta_0)$ and thus $\theta_{num} > \theta_0$. In other words, the imposed numerical contact angle is increased such that the curvature at the receding contact line results in a sufficiently large positive forcing in the $x$-direction on the fluid. The imposed numerical contact angle does therefore not approximate the true dynamical contact angle of the system. For PF, the mobility parameter $M$ is calibrated (\S\ref{sec:match-PF-MD-M}) to increase contact line velocity for high $\theta_0$. Here, mobility is considered as numerical tuning parameter to match the droplet displacement and does not correspond to the actual molecular diffusion at the interface (see appendix~\ref{app:pf-stream} and figure~\ref{fig:streamL-PF-param-var}a-c). 

Very  low contact angles impose other challenges for CFM.  For $\theta_0 = 38^\circ$, we observed that steady drop displacement  diverged (figure \ref{fig:q24-dx-vs-Ca}d) between the three models. Indeed, the prediction of critical capillary number showed  deviations between the methods (figure \ref{fig:MD-PF-VOF-stab-tresh}). 
In addition, the time it takes for the drop to evolve in different shapes (figure~\ref{fig:q4-Ca005-break-showcase}) is different in all simulation methods. This is despite the fact that the $\lambda$ obtained through VOF calibration (\S\ref{sec:vof-calib}) is reasonable ($\lambda < \Delta / 2$) and the standard PF calibration procedure (\S\ref{sec:PF-calib-Yue}) works well.
By investigating MD data directly, asymmetric contact line friction was observed (table~\ref{tab:PD-MD-muf}). For $\theta_0 = 38^\circ$ we observed significantly larger $\mu_f$ for receding contact line compared to one for the advancing contact line. In the CFM models, on the other hand, the contact line properties ($\mu_f$ for PF and $\lambda$ for VOF) were the same for both advancing and receding sides. Since the receding contact line becomes unstable first (figure~\ref{fig:q4-Ca005-break-showcase}), this is the likely reason of the CFM model inaccuracy.

Finally, there are limitations of the chosen CFM at very low and high $\Ca$ numbers.
At high $\Ca$ numbers (close to $\Ca_c$) we observed enhanced oscillations of $\Delta x (t)$ (figures~\ref{fig:all-q2-dx-tevol}c and \ref{fig:stick_slip_details}, stars in table~\ref{tab:MD-run-summary}). More detailed analysis of these oscillations were presented in \S\ref{sec:md-stick-slip} for $\theta_0 = 95^\circ$. The CFM models does not include the intrinsic oscillations present in the molecular reality. Consequently, the agreement between the CFM and MD in drop time evolution (figure~\ref{fig:all-q2-dx-tevol-wCFM}e and Supplementary Figure~6) is degraded. This could also be a potential source of increased discrepancy between interface shapes from CFM models and MD (figure~\ref{fig:q2-allStedCa-angDist}e). 
%
%
%
Also for low $\Ca$ numbers (for example, $\Ca = 0.05$ in figure~\ref{fig:all-q2-dx-tevol-wCFM}a), we have observed relatively large difference between CFM models and MD results. The underlying cause for this inaccuracy is the relatively small drop displacement. As seen in figure~\ref{fig:all-q2-dx-tevol}(a), the MD oscillations have roughly the same magnitude for all $\Ca$ numbers.
Consequently, the signal to noise ratio in MD is much larger for smaller $\Ca$ numbers and the large oscillations can give impression of larger inaccuracy of CFM models.

\subsection{Fluid slippage and contact line friction}

In this work, the nanoscale molecular system has a negligible hydrodynamic slip. It was observed that the slip length $\ell_s$ exhibits only small variations to $\theta_0$. Below bulk liquid, the slip length had to be $0.44$ nm for $\theta_0 = 127^\circ$ and $0$ nm for all other $\theta_0$.  As discussed in \S\ref{sec:limit-large-th}, for low-friction configurations $\theta_0>90^\circ$, it was necessary to adjust $\lambda$ and $M$ to further enhance contact line movement relative to the wall. This serves as a hint that the friction near the contact line is much smaller compared to what is modelled through bulk slip length $\ell_s$. In general however, other physical mechanisms that are intrinsic to the contact line are expected to co-exist with molecular slippage. 

We have  extracted contact line friction directly from MD (table~\ref{tab:PD-MD-muf}). For hydrophilic substrates, a particularly high friction value was obtained. This is consistent with the formation of a microscopic water film for a relatively small capillary number (i.e. $\theta_{rec}\sim0$ for $Ca\ll1$). On the other hand, the interpretation for the almost vanishing receding friction on the $\theta_0 = 127^\circ$ surface is less obvious. 
To the best of the author's knowledge, nowhere before an \textit{asymmetric} behaviour has been reported. It is tempting to explain the asymmetry by stating that hydrophilic surfaces are easier to wet rather than de-wet and vice-versa for hydrophobic surfaces. This puts the classical view -- that hydrophilic substrates are high-friction surfaces and that hydrophobic substrates are low-friction surfaces -- under doubt. Indeed, it is not clear whether the value of contact line friction can be predicted from $\theta_0$ alone~\citep{liu2019hysteresis,wang2019review}. Reasoning with the frame of mind of molecular-kinetic-theory instead, line friction asymmetry hints toward some complex physics modulating adsorption/desorption of molecules at the contact line. Fluid/surface interface energy alone is not sufficient to describe asymmetry between adsorption and desorption. More in-depth examination of sub-continuum fluid displacement near contact lines will be required to arrive at a physical understanding of this process. When the asymmetry is understood, it is straightforward to impose different $\mu_f$ values in PF and different $\lambda$ values in VOF for advancing and receding contact lines.

\subsection{Potential modelling directions}


It has been previously proposed that a better way to model the moving contact line is to use the so-called generalized Navier boundary condition (GNBC) as first hinted by~\cite{blake1993dynangleChapter}. This approach has been later on evaluated against MD simulations~\citep{qian2003molecular,QIAN:2006hl,mohand2019use} and good agreement has been found. However, recently~\cite{lacis2020steadyDresden} have tried to match GNBC with MD simulations exhibiting negligible slip but were not successful in demonstrating any advantage of GNBC against no-slip and Navier-slip boundary conditions.

In this work, we also use substrate with negligible hydrodynamic slip ($\ell_s = 0.44$ nm for $\theta_0 = 127^\circ$ and $\ell_s = 0.0$ nm for $\theta_0 = 38^\circ$ -- $95^\circ$). Previously~\citep{lacis2020steadyDresden}, for $\theta_0 = 95^\circ$ we extracted $\ell_s = 0.17$ nm, which is close to what we have in the current work. However, the most important overlap between the current and our previous~\citep{lacis2020steadyDresden} work is that the slip length imposed at the solid wall is constant over the surface. This is the main reason why the GNBC for the selected system \citep{lacis2020steadyDresden} did not exhibit any advantage. Very small slip length corresponds to very large friction at the contact line. Consequently, the addition of uncompensated Young's stress does not lead to significantly modified flow near the contact line.

Nevertheless, there are no solid arguments as to why the effective friction exactly at the contact line should be the same as below the bulk liquid. It could very well be that the effective slippage (friction) exactly at the contact line must be prescribed larger (smaller) compared to below the bulk liquid. This has the potential for improving both PF and VOF ability to match the MD results. An alternative approach to prescribing larger friction in VOF simulations would be to use so-called staggered slip or negative slip~\citep{hartmann2021breakup}. \rev{Another effect, which was not considered in this work, is so called disjoining pressure~\citep{pismen2000disjoining}. Including it would also allow for direct modification of contact line motion. Furthermore, more detailed studies of local rheological effects (including orientation parameter of water) could provide insight of detailed mechanisms governing film formation and contact line friction.} An accurate description of Navier slip related friction near the contact line could improve results attainable also using the GNBC condition. This is because the friction parameter is an important input in the GNBC. Infinite friction parameter renders GNBC ineffective, whereas gradually reducing friction parameter (increasing slip) would amplify the GNBC effect on the velocity near the contact line.

Another aspect is that the MD results hint towards asymmetric properties of advancing and receding contact lines. Consequently, contact line friction and possibly local slip length could be different. This enlarges the parameter space enormously. Fundamental investigations into the possible cause of asymmetry between the adsorption and desorption process would be welcome. These could potentially shed more light on what input should be given to CFM models to match the molecular reality. Alternatively some kind of hybrid methods that allow the matching between MD and Navier-Stokes solvers~\citep{hadjiconstantinou1999hybrid,zhang2017multiscale,borg2018multiscale} could be used. These approaches would alleviate the need to understand the asymmetric properties of the contact line and provide direct coupling between MD and CFM solvers.

Finally, there are other CFM models available that could be benchmarked against the molecular data produced through this work. For example, there exist different PF models, such as van der Walls~\citep{laurila2012thermohydrodynamics} or Cahn--Allen~\citep{eggleston2001phase}. The level-set~\citep{tornberg2000finite} model or Lattice Boltzman~\citep{chen2014critical} method are other potential candidates for the simulation of multiphase flows\rev{; Latva-Kokko and Rothman~\citep{latvakokko2007latticeboltzmann} showed for instance how a no-slip LB model is able to automatically capture the speed-dependent dynamic contact angle and the interface-local slip length}. The number of freely adjustable parameters differs between all alternatives. However, the issue of the appropriate velocity boundary condition will be shared between all of models \rev{based on single continuum description}. \rev{A good recent classification of multiphase models can be found in work by \cite{soligo2021turbulent}. It has to be recognised that the hybrid models~\citep{zhang2017multiscale,liu2021multiscale} alleviate the need to understand the fundamental mechanisms near the moving contact line and provide a way to couple MD and CFM directly. This is another alternative that should be considered for efficient simulations of multiphase systems.} \rev{I}f the stick-slip like oscillations of the contact line are important, some other means of continuum modelling can be pursued. For example, the fluctuating hydrodynamic interfaces model proposed by~\cite{flekkoyrothman1996,smith2016langevin} could be evaluated as a suitable choice. \rev{Lastly, we mention the possibility of regularizing the contact line stress singularity via the Brinkman's model for porous surfaces~\citep{devauchelle2007porous}.}


Despite the missing physical mechanisms in the PF and the VOF, we have demonstrated that sufficiently accurate predictions of interface shape, drop displacement and critical Capillary number can be obtained. The only prerequisite is that the PF and VOF simulations have to be calibrated with the MD once for given $\theta_0$.

\section{Conclusions} \label{sec:conclusions}

We have calibrated a standard Cahn-Hilliard phase-field model, as well as a standard geometric Volume-of-Fluid model with Cox-like wetting condition, against molecular dynamics simulations of water over a no-slip substrate. The no-slip behaviour in the MD system is an outcome of electro-static bonds between polar water molecules and polar wall molecules. Two parameters (mobility $M$ and contact line friction $\mu_f$) were adjusted in PF simulations and one parameter (Cox cut-off length scale $\lambda$) in VOF simulations. Four different equilibrium contact angles ($\theta_0 = 127^\circ$, $95^\circ$, $69^\circ$ and $38^\circ$) were investigated. For each $\theta_0$, the largest steady stick-slip free simulation was selected for calibration. The PF and VOF models were calibrated to match the steady $\Delta x$ -- single scalar macroscopic measurement -- observed in MD. Using the calibrated parameters, a series of simulations were carried out for other solid wall velocities. We demonstrated that CFM simulations can sufficiently accurately predict the drop displacement without any additional adjustments. The critical Capillary number predictions from CFM models also displayed good agreement with MD data.

In addition, we have showcased predictions for two unsteady sheared droplet configurations of $\theta_0 = 95^\circ$ and $\theta_0 = 38^\circ$. The CFM models predicted all qualitative features of the MD simulations. For $\theta_0 = 95^\circ$, drop displacement and break-up in half were predicted. For $\theta_0 = 38^\circ$, thin film deposition and coalescence with the periodic image were predicted. Despite the quantitative differences in the time instances of the similar shapes, the CFM predictions exhibited good agreement with MD results.

We identified molecular physics that to the best of the authors' knowledge have not been previously reported. We extracted line friction directly from MD and compared it with the PF calibration results. The MD results showed the same trend as obtained with PF. In addition, the resulting contact line friction $\mu_f$ values were asymmetric between advancing and receding contact lines for $\theta_0 = 127^\circ$ and $38^\circ$.

Finally, we have discussed the variations of PF and VOF parameters for matching the MD results for all $\theta_0$ values. We identified that the currently chosen CFM models seem to be lacking a way to describe reduced friction near the contact line for $\theta_0 \geq 95^\circ$. A possible future direction to remedy this shortcoming would be to introduce larger slippage (lower friction) locally near the contact line. In a continuum setting, this could correspond to having a spatially varying slip. In addition, we anticipate that the asymmetric behaviour of the advancing and  the receding contact lines is the source of the inaccuracies in PF and VOF when predicting the $\theta_0 = 38^\circ$ results.

We have demonstrated that by calibrating the CFM once for each $\theta_0$ by targeting a single global measure $\Delta x$ it is possible to obtain many accurate predictions of interface shape, $\Delta x$ as a function of $\Ca$ and also prediction of $\Ca_c$. This is despite the fact, that the selected MD configuration exhibits practically negligible slippage. The selected CFM models seem to perform very well for close to neutral and slightly hydrophobic configurations ($\theta_0 = 95^\circ$ and $69^\circ$). On the other hand, more deviations were observed for hydrophobic ($\theta_0 = 127^\circ$) and hydrophilic ($\theta_0 = 38^\circ$) configurations. These accuracy limits have to be kept in mind, if these CFM models are applied in similar conditions.

The results of this study continue to enrich our understanding of connections between continuum mechanics simulations and molecular reality. We believe that this work provides important insights into PF and VOF models, associated open questions, and the required calibration procedures. Properly calibrated, both PF and VOF can serve as useful tools for investigations of technological applications.



\section*{Supplementary data}

Supplementary figures, movies and data files for easier figure reproductions are available at ...

\section*{Acknowledgments}
Numerical simulations were performed on resources provided by the Swedish National Infrastructure for Computing (SNIC) at PDC, NSC, and HPC2N.
%
The authors thank Tomas Fullana and Dr. Petter Johansson for fruitful discussions.
The developers of the PARIS Simulator code are acknowledged for continuous advancement and support of the tool. S.Z. thanks the dean of the faculty of sciences of Sorbonne Universit\'{e}, St\'{e}phane R\'{e}gnier, for the grant of a sabbatical leave that made his contribution to this work possible. 

\section*{Funding}

U.L., S.B., J.S., B.H. and M.P. were supported by 
Swedish Research Council (INTERFACE centre and grant nr. VR-2014-5680).
U.L. was supported by the ERDF (project No. 1.1.1.1/20/A/070).
S.Z. acknowledges funding from ERC (advanced grant TRUFLOW).

\section*{Declaration of Interests}
The authors report no conflict of interest.

\section*{Data availability statement}
The dataset containing all the flow field maps used for the calibration of CFM methods is openly available under Creative Commons Attribution 4.0 International license on Zenodo \citep{md-dataset}, at \hyperlink{http://doi.org/10.5281/zenodo.5997091}{http://doi.org/10.5281/zenodo.5997091}. Data files and scripts to plot figures in the main paper are provided in Supplementary Material.

\section*{Author ORCID}

\noindent U. L\={a}cis, \href{https://orcid.org/0000-0003-3094-0848}{orcid.org/0000-0003-3094-0848}\\
M. Pellegrino, \href{https://orcid.org/0000-0002-2603-8440}{orcid.org/0000-0002-2603-8440}\\
J. Sundin, \href{https://orcid.org/0000-0001-5673-5178}{orcid.org/0000-0001-5673-5178}\\
G. Amberg, \href{https://orcid.org/0000-0003-3336-1462}{orcid.org/0000-0003-3336-1462}\\
S. Zaleski, \href{https://orcid.org/0000-0003-2004-9090}{orcid.org/0000-0003-2004-9090}\\
B. Hess, \href{https://orcid.org/0000-0002-7498-7763}{orcid.org/0000-0002-7498-7763}\\
S. Bagheri, \href{https://orcid.org/0000-0002-8209-1449}{orcid.org/0000-0002-8209-1449}\\

\section*{Author contributions}
G.A., S.B., B.H., and S.Z. conceived the original idea. U.L. performed the PF simulations, carried out MD post-processing and constructed PF/VOF/MD comparisons. M.P. configured the MD system, carried out MD simulations, improved flow field extraction from MD data and extracted contact line friction from MD. J.S. performed the VOF simulations and implemented required customisations with feedback from S.Z. All authors analysed data. U.L., M.P., J.S. and S.B. wrote the paper with feedback from all co-authors.

\appendix
\counterwithin{figure}{section}
\counterwithin{table}{section}

\section{Details of the geometric Volume-of-Fluid model} \label{app:vof}

In this appendix, we provide additional details of the VOF model. First, we describe the numerical implementation of the solver. After that, customisation for the $\theta_0 = 127^\circ$ configuration is explained.

\subsection{Numerical implementation} \label{app:vof-num}

All VOF simulations used a resolution of $N_x = 256$ ($\Delta_x = 0.624$ nm) cells in the streamwise and $N_y = 32$ ($\Delta = \Delta_y = 0.913$ nm) in the wall-normal direction. PARIS solves the general three-dimensional equations. The two-dimensional behaviour was obtained using a thin domain, two cells wide in the spanwise direction (with periodic boundary conditions). We performed the simulations using a first-order time scheme, and the pressure was computed using the HYPRE library. Momentum was advected with a second-order central difference scheme. Equation~(\ref{eq:vof-cov-diff}) was solved using the built-in implementation of the algorithm by \citet{weymouth10conservative}, computing the fluxes of $C$ on the faces of each cell and updating $C$ accordingly. The equation for the dynamic contact angle (\ref{eq:VOFContactAngle}) was solved using the implementation reported by~\cite{sundin2021roughness}. Boundary conditions were implemented through a ghost layer (a layer of cells outside the computational domain where numerical quantities can be imposed).

The curvature of the interface was calculated using height functions. Height functions give the distance to the interface from a reference plane aligned with the grid. The values of the height function in specific cells, called heights, are computed by integrating $C$. 
Wall-parallel (wall-normal) heights provide interface $x$-coordinates ($y$-coordinates)  for equidistant $y$-locations ($x$-locations) corresponding to the simulation mesh. The dynamic contact angle was imposed by prescribing wall-parallel heights in the ghost layer \citep{afkhami2008height}. The value of $C$ in the ghost layer was also set according to the dynamic contact angle to give a consistent interface normal for the flux computations. 
In the simulations, we imposed a density ratio of 0.01 to make the simulations stable. We consider this sufficient to reproduce the main features of the water liquid-vapour system.

\begin{figure}
    \centering
    \begin{subfigure}{0.48\textwidth}
        \centering
        \includegraphics[width=5cm]{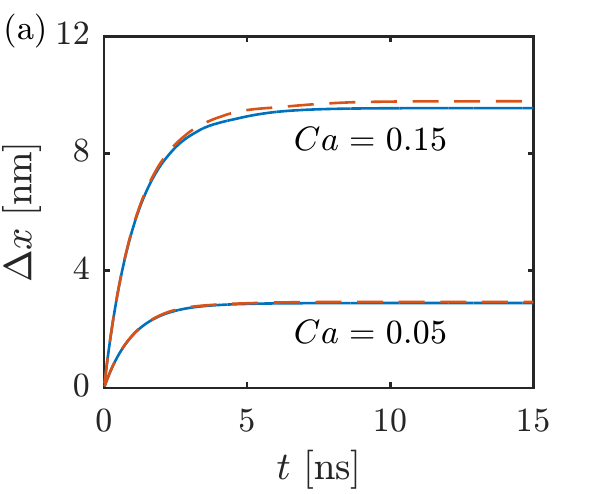}
    \end{subfigure}
    \begin{subfigure}{0.48\textwidth}
        \centering
        \includegraphics[width=5cm]{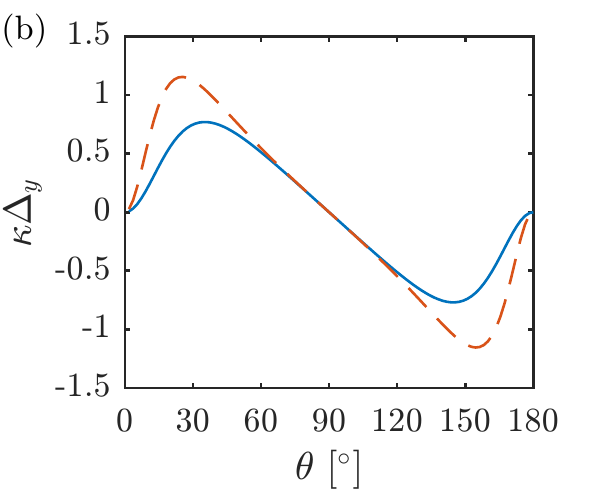}
    \end{subfigure}
    \caption{ (a) VOF refinement study, showing drop displacement for $N_x = 256$ and $N_y = 32$ cells (solid lines) and $N_x = 512$ and $N_y = 64$ cells (dashed lines) for $Ca = 0.05$ and $0.15$. (b) Curvatures estimated by the analytical relations \eqref{eq:vof-curvature-estimation-1} and \eqref{eq:vof-curvature-estimation-2} (solid and dashed lines, respectively).}
    \label{fig:VOF-refinement-study}
\end{figure}

To evaluate the grid independence of the simulations, we performed a grid refinement study. A refined grid with $N_x = 512$ and $N_y = 64$ was used for $\theta_0 = 95^\circ$, with $Ca = 0.05$ and $0.15$. The time series of the drop displacements are presented in figure~\ref{fig:VOF-refinement-study}(a). For $Ca = 0.05$, the drop displacement changed by $1.3\%$ and for $Ca = 0.15$ by $2.4\%$. Accordingly, the simulations seem more sensitive to the grid for higher capillary numbers. The difference could appear because of the refinement of the velocity field and because the condition for the dynamic contact angle (\ref{eq:VOFContactAngle}) does not completely remove the grid dependence \citep{legendre2015comparison}. Nevertheless, we consider the observed convergence sufficient.

\subsection{Customisation for $\theta_0 = 127^\circ$} \label{app:vof-127-cust}


Using wall-parallel heights in the ghost layer limits the maximum absolute value of the curvature at the contact lines. This curvature corresponds to a maximum forcing on the fluid ($|\vec{f}_\sigma|$, equation \ref{eq:continuous-surface-force}). The expression for the curvature is 
\begin{equation}
    \kappa = \frac{d^2 h/d y^2}{(1 + (d h/ d y)^2)^{3/2}},
    \label{eq:curvature-by-heights}
\end{equation}
where $h = h(y)$ is the height function giving the $x$-location of the interface. The sign of $\kappa$ also depends on the interface orientation; this discussion is not included for brevity (see \citet{aniszewski2021parallel}). The derivatives are computed with central finite differences. We denote the value of the heights $h_0$, $h_1$, and $h_2$ in the ghost, first and second cell layer above the wall, respectively. From the definition of the angle, $h_0 = h_1 + \Delta_y/\tan(\theta)$ (the sign in front of $\tan(\theta)$ depends on the interface orientation). We assume that $h_2 \approx h_1$. The curvature in the first cell layer becomes
\begin{equation}
    \kappa_\mathrm{num} = \frac{(h_2 - 2h_1 + h_0)/\Delta_y^2}{(1 + (h_2 - h_0)^2/(2\Delta_y)^2)^{3/2} } \approx \frac{1}{\Delta_y}\frac{1/\tan(\theta)}{(1 + 1/(4\tan^2(\theta)))^{3/2}}, 
    \label{eq:vof-curvature-estimation-1}
\end{equation}
shown in figure~\ref{fig:VOF-refinement-study}(b) (solid line). The approximation of $\kappa_\mathrm{num}$ is zero for $\theta = 0^\circ$, $ 90^\circ$, and $180^\circ$. As shown in the figure, $|\kappa_\mathrm{num}|$ has one maximum in each of the intervals $0^\circ < \theta < 90^\circ$ and  $90^\circ < \theta < 180^\circ$. 
For $\theta_0 = 127^\circ$, a minimum separation velocity was achieved for a steady-state receding contact line angle smaller than $180^\circ$, as expected from equation~\eqref{eq:vof-curvature-estimation-1}. However, this minimum was not low enough to match the target displacement from MD.

To match the MD results, we allowed the usage of the height of the third cell layer, $h_3$. The finite-difference scheme of the second derivative was left unchanged. The order of the first derivative, on the other hand, was increased, resulting in 
\begin{equation}
    \kappa_\mathrm{num} = \frac{(h_2 - 2h_1 + h_0)/\Delta_y^2}{(1 + (-h_3 + 6h_2 - 3h_1 - 2h_0)^2/(6\Delta_y)^2)^{3/2} } \approx \frac{1}{\Delta_y}\frac{1/\tan(\theta)}{(1 + 1/(9\tan^2(\theta)))^{3/2}}, 
    \label{eq:vof-curvature-estimation-2}
\end{equation}
where we assumed $h_3 \approx h_2 \approx h_1$. 
This expression results in significantly higher curvatures for extreme angles (figure~\ref{fig:VOF-refinement-study}b, dashed line). We were then able to match the MD result.

Another possible remedy would be to impose the angle by wall-normal heights. However, in many instances for $\theta_0 = 127^\circ$, not enough wall-normal heights could be computed at receding nor advancing contact lines for curvature calculations. The current solution is, therefore, more robust. Wall-normal heights could be a viable solution if resolution significantly is increased.

\section{Details of Cahn-Hilliard phase-field model} \label{app:cahn-hil}

In this appendix, we provide additional details of the Cahn-Hilliard PF model used in this work. The model is briefly introduced in \S\ref{sec:pf-mod-main}.
The standard double-well
potential is
\begin{equation}
    \Psi \left(C\right) = \left( C+1 \right)^2\left( C-1 \right)^2/4. \label{app:eq:PF-potent}
\end{equation}
The wetting boundary condition (\ref{eq:wet-bc}) contains the so-called switch function, defined as
\begin{equation}
    g\left(C\right) = 0.5 - 0.75 C + 0.25 C^3. \label{app:eq:PF-swtch-func}
\end{equation}
This expression serves as a window function that ensures the wetting boundary condition is applied only at the contact line. Furthermore, the cubic expression (\ref{app:eq:PF-swtch-func}) is not empirical. Instead, it is derived as the equilibrium solution of PF equations based on the selected potential (\ref{app:eq:PF-potent}) and hyperbolic tangent variation of $C$ across the interface.

The density and viscosity in the momentum equation (\ref{eq:ns-1}) are spatially dependent. In the PF model, we define these fluid parameters through linear combination based on phase-field variable $C$ as
\refstepcounter{equation}\label{eq:pf-def-den-visc}
\begin{equation}
\rho\left(C\right) = \rho_\ell \frac{C+1}{2} - \rho_v
\frac{C-1}{2} \ \ \ \ \mbox{and} \ \ \ \
\mu\left(C\right) = \mu_\ell \frac{C+1}{2} - \mu_v
\frac{C-1}{2}. \tag{\theequation a,b}
\end{equation}
Recall that $\rho_\ell$ and $\mu_\ell$ are the density and the viscosity of the liquid component,
and $\rho_v$ and $\mu_v$ are the density and the viscosity of the vapour
component. 

The introduced governing equations are linearised, written into the weak form, and solved using an open-source finite-element software \textit{FreeFEM}~\citep{MR3043640}, which allows easy specification of finite-element weak form. Linear elements were used for the phase-field variables, while the fluid flow was resolved using Taylor-Hood elements (quadratic for velocity and linear for pressure).
Mesh resolution was refined and results were checked for a few selected simulation cases.
The production resolution selected and used for most simulations is $\Delta s_1 = 3.65$ nm far from the interface, and down to $\Delta s_2 = 0.24$ nm within the interface region. The constant time step was used through the simulation as $\Delta t = 0.002$ dimensionless time units. For very small $\Ca$, $\Delta t$ was reduced to ensure numerical stability. For simulations with smaller $\epsilon$, the mesh was refined at the interface to maintain roughly the same amount of elements over the interface and the time step was reduced to ensure numerical stability. Exact numerical code used to produce the PF results is available freely from Github repository~\citep{githubMCL2020UgisShervin}. The data to reproduce the figures in the main paper is available in Supplementary File 1.

\section{Details of molecular dynamics simulations} \label{app:md-det-main}

In this appendix, we provide the necessary details for the reader to understand the simulation procedure, interface extraction, and determination of equilibrium angle.


\subsection{Numerical implementation} \label{app:md-det-num}


The thin quasi-2D liquid meniscus (figure~\ref{fig:intro0}) is composed of 172933 water molecules. The parameters for the SPC/E model are taken from the OPLS-AA force field. The parametrization of SiO$_2$ quadrupoles is summarized in table \ref{tab:MD-silica-char}. Silicon atoms are treated as virtual sites without mass. Oxygen atoms are restrained to absolute coordinates by a spring of constant $\kappa_O$. The usage of position restraints grants the substrate some flexibility to re-configure and accommodate water adsorption and desorption. All covalent bonds and angles are treated as rigid constraints. Non-bonded parameters for the interactions between different species are generated via the geometric combination rule. The time-marching step is the same for equilibrium and non-equilibrium runs, $\delta t=2$ fs. We use the leap-frog time marching to update atomistic coordinates. All simulations have been pre-processed and run with GROMACS 2020 \citep{gromacs2015}.

To obtain $\theta_0$ values stated in \S\ref{sec:md-sheared}, charge values $\left( q_1, q_2, q_3, q_4\right) = \left( 0.40, 0.60, 0.67, 0.74\right)\,e$ were required. Here, $e$ is the elementary electron charge. Ideal purely-repulsive Lennards-Jones (LJ) walls are placed beyond the silica surfaces at the location of periodic boundary condition. The LJ walls decouple periodic images along $y$. The starting configuration for production runs is obtained by letting the droplet relax to its equilibrium shape.
\begin{table}
\begin{center}
\begin{tabular}{llll}
    Oxygen mass & \quad$m_O$ & \quad$9.95140$ & u \\
    L-J well depth (silicon) & \quad$\varepsilon_{Si}$ & \quad$0.2$ & kJ mol$^{-1}$ \\
    L-J char. distance (silicon)  & \quad$\sigma_{Si}$ & \quad$0.45$ & nm   \\
    L-J well depth (oxygen) & \quad$\varepsilon_{O}$ & \quad$0.65019$ & kJ mol$^{-1}$ \\
    L-J char. distance (oxygen) & \quad$\sigma_{O}$ & \quad$0.316557$ & nm \\
    Si-O bond distance & \quad$d_{so}$ & \quad$0.151$ & nm \\
    Hexagonal lattice spacing & \quad$d_{hex}$ & \quad$0.45$ & nm \\
    Restraint force constant & \quad$\kappa_O$ & \quad$10^5$ & kJ mol$^{-1}$ nm$^{-2}$
\end{tabular}
\end{center}
\caption{Parametrization of the force field of silica quadrupoles (electrostatics excluded).}
\label{tab:MD-silica-char}
\end{table}


\begin{figure}
    \centering
    \includegraphics[width=0.8\linewidth]{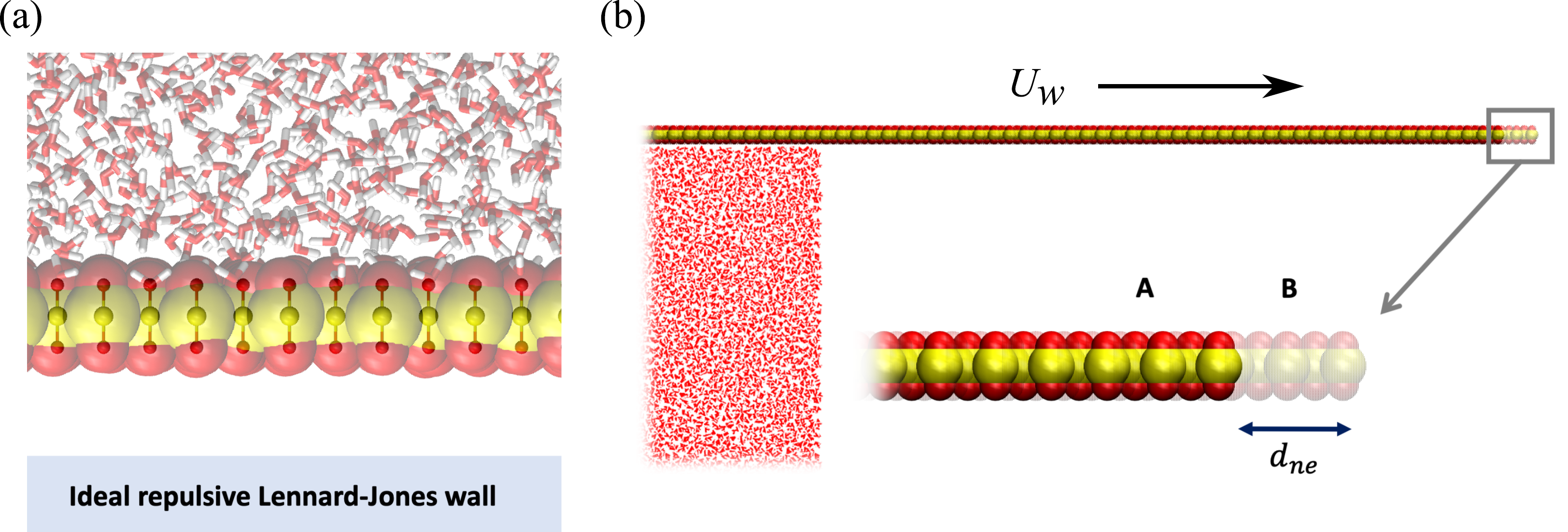}
    \caption{Position restraints and wall treatment. In (a), references for restrained positions are represented as CPK sticks, while actual atoms of SiO$_2$ molecules are shown as transparent VDW spheres. Auxiliary interpolation procedure for non-equilibrium simulations (b) to produce an effective wall velocity $U_w$.}
    \label{fig:md-details}
\end{figure}

The desired shear rate is produced by interpolating position restraints between two reference configurations (figure~\ref{fig:md-details}c). The configuration A is  the equilibrium one. In configuration B, horizontal coordinates of the silica layer have been offset by $+d_{ne}$ on the top and by $-d_{ne}$ on the bottom walls. The effective wall velocity is quantified as
\begin{equation}
    U_w = \frac{\delta x}{\delta t} = d_{ne}\frac{\delta\lambda}{\delta t}\, .
\end{equation}
Here, $\delta\lambda$ is the increment of an auxiliary variable $\lambda\in[0,\infty)$ applied at each time step. The $\lambda=0$ corresponds to configuration A and $\lambda=1$ to configuration B. The desired wall velocity is obtained by setting the interpolation increment to $\delta\lambda=U_w\delta t/d_{ne}$.

All simulations are performed in the NVT ensemble at $T=300K$ and fixing the extent of simulation box to $(L_x, L_y, L_z) = (159.75,30.634,4.6765)$ nm. Periodic boundary conditions are employed along the direction of flow homogeneity $z$ and along the shear direction $x$, while periodic image interactions along the vertical direction $y$ are avoided by placing ideal Lennard-Jones walls at $y=0$ and $y=L_y$ (figure~\ref{fig:md-details}a). Bussi-Donadio-Parrinello thermostat (GROMACS `v-rescale') is applied to both water and silica, with coupling time $0.1$ ps for equilibration runs and $10$ ps for non-equilibrium runs.

When performing non-equilibrium simulations of liquids one has to bear in mind that most standard choices for thermal coupling either lead to local flow hindering or to artificial cooling where the flow velocity is larger. We estimated the maximum local temperature deviation for the probed range of capillary numbers and we concluded that it has only a marginal effect on steady regimes when surfaces are hydrophilic or moderately hydrophobic. The estimate can be obtained as follows. Imagine a fluid composed of spherical particles. Then temperature and kinetic energy (per particle) can be related \textit{at equilibrium} via the equipartition theorem
\begin{equation}
    E_{kpp} = \frac{1}{2}m\big(\expval{c_x^2}+\expval{c_y^2}+\expval{c_z^2}\big) = \frac{3}{2}k_B T \; ,
\end{equation}
where $\boldsymbol{c}$ is the particle's peculiar velocity. For a steady flow and in case of no-slip, molecules close to a solid wall have a deterministic velocity component $U_w$ in the $x$ direction. An equilibrium thermostat that is oblivious to hydrodynamics will attempt to re-scale the kinetic energy per particle in order to match the prescribed temperature $T_0$, defined as
\begin{equation}
    \frac{3}{2}k_B T_0 = \frac{1}{2}m\big(u^2+\expval{c_x^2}+\expval{c_y^2}+\expval{c_z^2}\big) \, .
\end{equation}
The difference between imposed and effective temperature is
\begin{equation}
    T_0-T = \frac{mu^2}{3k_B} \simeq \frac{m(\gamma/\mu)^2}{12k_B}Ca^2 = \Theta Ca^2 > 0 \; ,
\end{equation}
where $\Theta$ is a characteristic temperature differential that tunes how the system is cooled down in the function of the imposed capillary number. For our molecular model, we estimate $\Theta\simeq0.78$ K, which entails that to cool down the near-wall molecules by $1K$ one needs to prescribe at least $Ca\simeq1.132$. This rough calculation does not account for the rotational degrees of freedom of water and thus can be regarded as a conservative estimate.

There exist several techniques for correctly thermalizing flow simulations. One among the simplest consists in only coupling the solid substrate and letting the liquid thermalize due to heat exchange. Trying this approach we noticed that in the configurations with $\theta_0 \geq 95^\circ$ (which are also problematic due to typically larger $\Ca$) heat transfer between silica and water is not large enough to effectively render the system isothermal. Other techniques would employ either a profile-biased thermostat~\citep{bernerdi2010pbt} or a dissipative-particle-dynamics thermostat~\citep{soddemann2003dpdt,goga2012dpd}. However, these thermostats are not currently implemented in GROMACS.

Hydrodynamic fields (density, velocity, and temperature) are directly measured from MD trajectories. Each quantity is averaged in space on a grid with spacing $(h_x, h_y) \simeq (0.20,0.20)$ nm (figure~\ref{fig:intro0}b), and over time by aggregating all measurements in consecutive windows of $12.5$ ps. Averaged and binned variables are saved to file ``on-the-fly'' concurrently with the simulation, thus vastly reducing the output size. Saving all atomistic trajectories would not be feasible. Consequently, the output of MD simulations is a range of data files containing so-called frames, corresponding to the sequence in time of the partially averaged MD data. Each frame contains the instantaneous field outputs $\rho^i\left(x,y,t\right)$, $u^i_x\left(x,y,t\right)$ and $u^i_y\left(x,y,t\right)$ as defined in \S\ref{sec:md-sheared}. Post-processing to obtain averages over time intervals of several ns has been performed with in-house codes based on the freely available repository (\href{https://github.com/MicPellegrino/densmap.git}{https://github.com/MicPellegrino/densmap.git}). The exact scripts are available upon reasonable request.

\subsection{Equilibration runs and centre of mass correction} \label{app:md-equil}

To determine that the signature of the initialisation is fully disappeared, we first visually inspect the time series of the relaxing contact angle. Based on the decay of the signal we have determined conservative cut-off times for different $q$ values. Then we check \textit{a-posteriori} the cross-correlation between the signals at each contact line. This is done to ensure that any transient relaxation dynamic has disappeared and that the size of the molecular system is large enough to effectively localize contact line motion. After the cut-off time, we have continued the runs to collect a sufficient amount of statistics for the measurements, at least $4$ ns for all $q$ values.

From MD simulations of the equilibrium configuration, we obtain many sequential frames of hydrodynamic variables. To measure the geometrical features, such as the local interface curvature, we average all frames in the equilibrium state by shifting the centre of mass (COM) of the liquid droplet in each frame to the centre of the domain. The reason behind this procedure lies in the fact that COM correction is turned off in the MD simulations themselves. Run-time COM correction can potentially hinder relaxation in equilibrium simulations and create velocity measurement artefacts in non-equilibrium ones. This averaging procedure is employed before interface extraction for both equilibration and sheared MD runs.

\subsection{Interface extraction and $\theta_0$ measurement} \label{app:md-itf-extr-th0}

\begin{figure}
\centering
\includegraphics[width=0.95\linewidth]{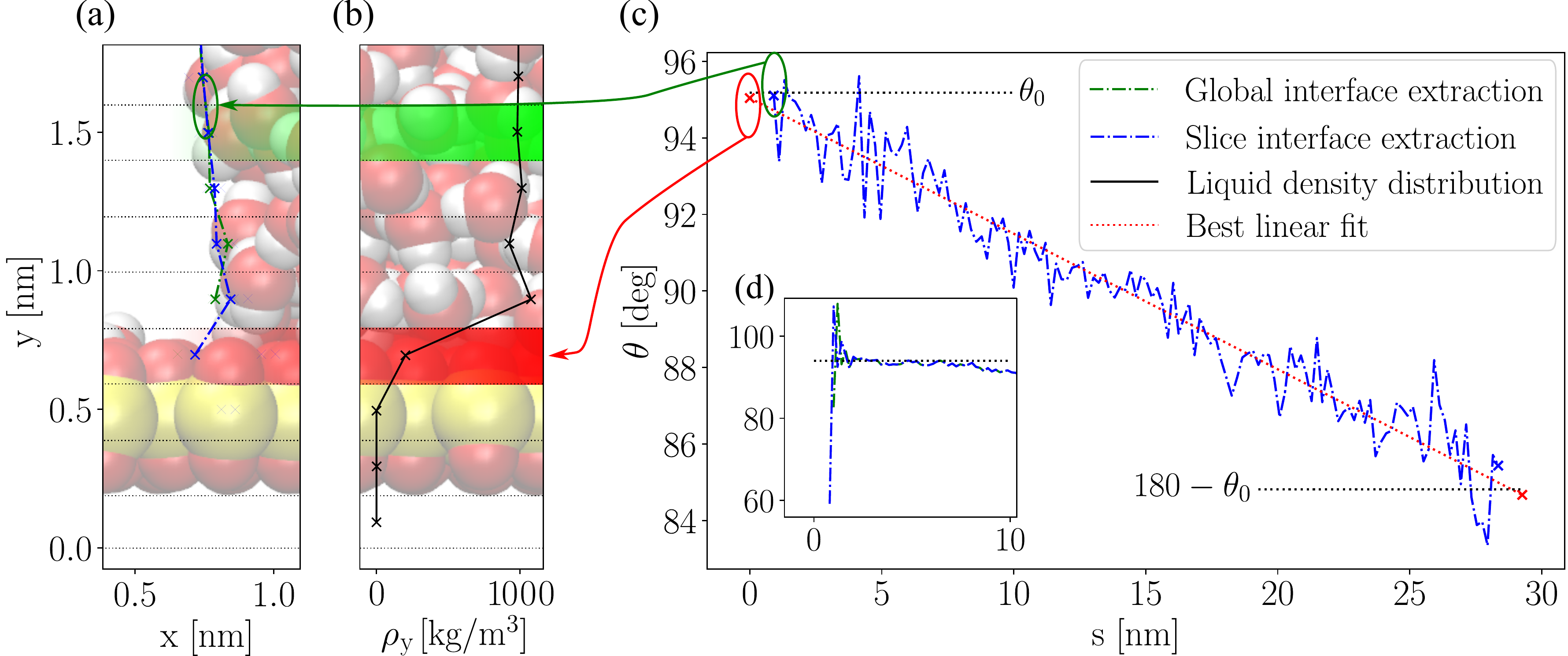}
\caption{Equilibration MD run yielding $\theta_0 = 95^\circ$. Close-up near the bottom wall of extracted interface shape (a) and liquid density variation (b). Interface angle variation (c) along the curvilinear coordinate $s$ excluding the interface measurements closest to the walls. Inset (d) shows the interface angle along the vertical coordinate $y$ with all interface points included. Dashed lines in panels (c) and (d) correspond to the measured equilibrium contact angle $\theta_0$.}
\label{fig:q2-molEff-eqCA}
\end{figure}

In this appendix, we describe the extraction of the interface shape from the water density distribution $\rho(x,y)$. The distribution for equilibration run with $\theta_0 = 95^\circ$ along the bin with vertical coordinate $y \approx 1.5$ nm is shown in figure~\ref{fig:wall-allQ-summary}(e,f). We consider two criteria to define the exact $x_{ni}$ coordinate of the interface, i.e.
\refstepcounter{equation}\label{eq:itf-def}
\begin{equation}
\rho\left(x_{ni}, y_n\right) = 0.5\,\rho_\ell \qquad \mbox{and} \qquad
\rho\left(x_{ni}, y_n\right) = 0.5\,\rho_y \left(y_n\right). \tag{\theequation a,b}
\end{equation}
These are based on global liquid density (\ref{eq:itf-def}a) and slice liquid density (\ref{eq:itf-def}b) at an $n$-th vertical bin with coordinate $y_n$. If the sought density value is not located in any single bin, linear interpolation is used to find the exact coordinate. The interface extraction according to (\ref{eq:itf-def}a) we call ``global interface extraction'' and the obtained interface we call ``global interface''. This is the approach typically used in literature. However, the density layering (figure~\ref{fig:q2-molEff-eqCA}b) exhibits itself in the interface shape near the surface (figure~\ref{fig:q2-molEff-eqCA}a, green line). The interface extracted according to (\ref{eq:itf-def}b) we call ``slice interface''. The $x_{ni}$ obtained from (\ref{eq:itf-def}b) show reduced layering influence (figure~\ref{fig:q2-molEff-eqCA}a, blue line). Further away from the wall results from (\ref{eq:itf-def}a,b) agree. From (\ref{eq:itf-def}b) it is also possible to define the interface point for adsorbed water layer (figure~\ref{fig:q2-molEff-eqCA}a, $y = 0.7$ nm). However, this is only a visual evaluation of both extraction procedures and it is not yet clear if any one of those is advantageous when comparing MD with CFM.

As the next step, we measure the equilibrium contact angle $\theta_0$. We use both interface shapes extracted according to (\ref{eq:itf-def}). From the interface shape, we compute the interface angle $\theta(y)$ along the height of the interface, as defined in figure~\ref{fig:dx-def-VOF-lm-var}(a). The angle is obtained from the slope of the interface segments (encircled with green in figure~\ref{fig:q2-molEff-eqCA}a). Therefore angle measurements are located at the boundaries of the MD bins. The Young-Laplace equation for constant surface tension is
\begin{equation}
\Delta p = - \sigma\, \gamma, \label{eq:p-jump}
\end{equation}
where $\Delta p$ is the pressure jump across the interface and $\gamma$ is the local curvature of the interface. In equilibrium, we expect the pressure in the whole droplet to be constant and the pressure in the vapour phase to be negligible. According to the Young-Laplace equation (\ref{eq:p-jump}), this results in constant curvature along the interface. We can express the interface angle as
\begin{equation}
\theta(s) = c_1\,s + c_2, \label{eq:eq-lin-fit}
\end{equation}
where $c_1$ and $c_2$ are constants determined by boundary conditions and $s$ is curvilinear coordinate along the interface. For convenient comparison, we transfer $\theta(y)$ to $\theta(s)$. Since the theoretical function is known, we fit the obtained MD results with (\ref{eq:eq-lin-fit}). We observe that both global and slice interfaces rapidly deviate from linear relationships near the wall (figure~\ref{fig:q2-molEff-eqCA}d). The same effect is observed in figure~\ref{fig:q2-molEff-eqCA}(a), where the interface segments closest to the wall exhibit significantly different angles compared to segments $\approx 1.5$ nm above the wall. Therefore we conclude that several interface points closest to the wall can not be used for a reliable comparison with continuum description.


Next, we determine how many MD points near the wall need to be excluded. We do this by gradually removing the closest interface points near the wall. The procedure is carried out until the standard error of the linear fit (\ref{eq:eq-lin-fit}) reaches its minimum. This process is applied to both global and slice interfaces. We mostly observe that a larger number of points have to be removed from the global interface shape to obtain the minimum in the standard error. Therefore for producing MD results in this work, we choose to focus only on the slice interface. We postulate that the first remaining point on the interface is the first reliable interface measurement for comparison with CFM.  The obtained $\theta\left(s\right)$ from slice interface for the remaining interface points is presented in figure~\ref{fig:q2-molEff-eqCA}(c) with a blue line. The corresponding best linear fit is given with a red dotted line. To obtain the equilibrium angle $\theta_0$, we use the hydrodynamic wall position assumed in the main paper. The wall position is shown in red in figure~\ref{fig:q2-molEff-eqCA}(b). We extrapolate the linear fit to the assumed positions. The equilibrium angle is computed as an average of extrapolated values at the top and bottom walls (figure~\ref{fig:q2-molEff-eqCA}c, red crosses). However, the total arc length of the interface depends on the contact angle. Therefore the equilibrium angle $\theta_0$ is iteratively obtained in the following procedure. We first centre the MD data to ideal arc length for given $\theta_0$. Then the agreement with the given $\theta_0$ is verified by extrapolating the best linear fit of the MD data to wall locations. Finally, the next $\theta_0$ is set as the averaged of the previous estimate and the current estimate.
The final obtained equilibrium angle in figure~\ref{fig:q2-molEff-eqCA}(c) is presented with black dotted lines. 
Inset (figure~\ref{fig:q2-molEff-eqCA}d) shows the angle deviation magnitude from the linear expression near the wall. The deviation is much larger than the noise observed in the bulk of the MD results. Consequently, the deviation near the wall can not be explained by the thermal fluctuations of MD. Therefore some other molecular effects are in play.

The process of removing the unreliable MD interface points and determining the equilibrium contact angle $\theta_0$ is repeated for all considered MD surface charges. Obtained first reliable interface extraction bin locations for comparison with CFM are summarized in the main paper figure~\ref{fig:wall-allQ-summary}(a-d) and the measured contact angles $\theta_0$ are $127^\circ$, $95^\circ$, $69^\circ$ and $38^\circ$ as stated in the main paper text.

\subsection{Polynomial extrapolation of MD interface shape} \label{app:MD-poly-fit}

\begin{figure}
\centering
\includegraphics[width=1.0\linewidth]{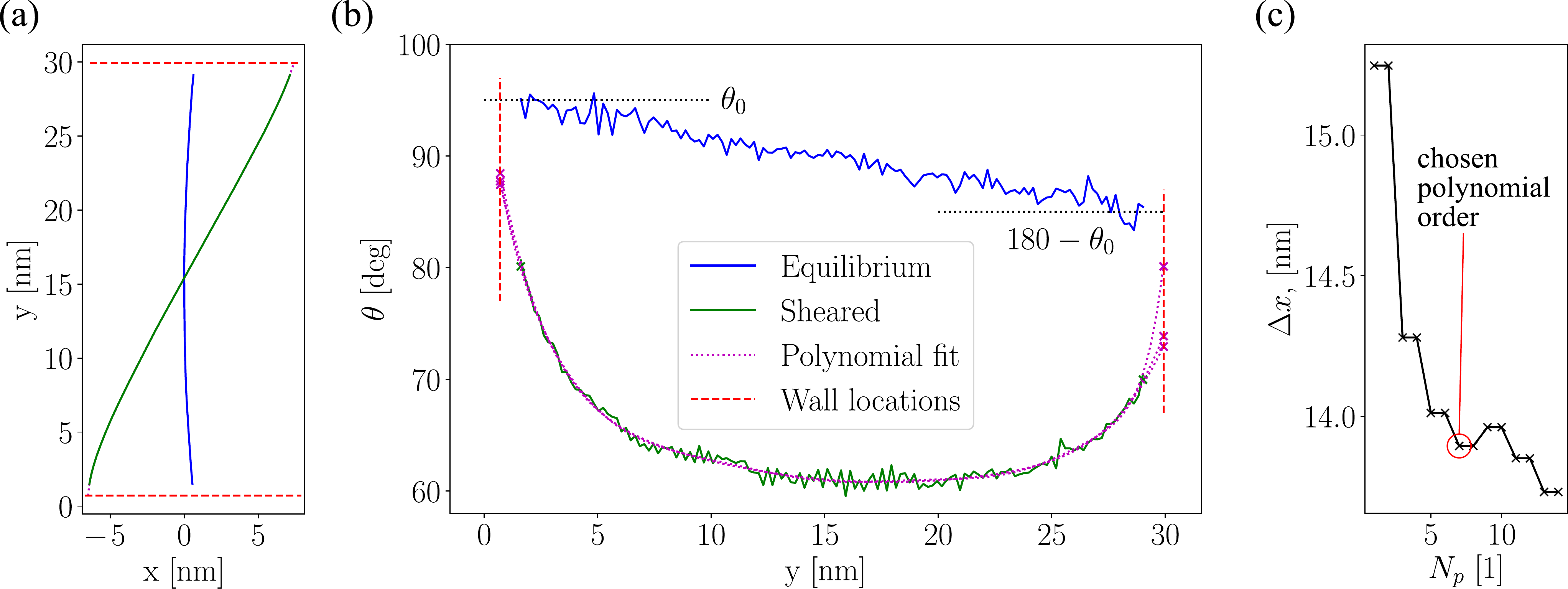}
\caption{MD interface shape (a) and interface angle (b) for $\theta_0 = 95^\circ$ system at equilibrium and sheared (Ca $= 0.20$) configurations. Convergence of $\Delta x$ with respect to polynomial fit order $N_p$ (c). Wall locations are shown with red dashed lines (a,b). The polynomial fit used to obtain the drop displacement estimate is shown with a purple dotted line (a). The equilibrium angle is shown with a black dotted line (b). The purple crosses (b) represent possible contact angle measurements for a few $N_p$ values.}
\label{fig:q2-Ca0.20-sheared}
\end{figure}

In this appendix, we describe how a polynomial fit is used to determine the final drop displacement for comparison with CFM models. In addition, the usefulness of the fit for extracting dynamic contact angle is assessed. Recall that through the extraction of $\theta_0$ values (appendix~\ref{app:md-itf-extr-th0}) the first reliable bins for comparison with CFM have been identified (shown with green in figure~\ref{fig:wall-allQ-summary}a-d). Consequently, there is a gap in interface data near the walls. To illustrate this, we show the equilibrium and sheared ($\Ca = 0.20$) interface shapes for $\theta_0 = 95^\circ$ in figure~\ref{fig:q2-Ca0.20-sheared}(a). With a red dashed line, we show the assumed wall position. In CFM models, however, the interface shape continues to evolve smoothly until meeting the wall. Consequently, when comparing MD with CFM models, the empty data region near the wall is undesirable.

To remedy this issue, we introduce a fit of the interface shape. Unlike for the equilibration run -- for which the function corresponding to the interface shape was known --, the functional form of the dynamic interface is not known. Therefore we choose a polynomial with some order $N_p$ that is not yet known. To fix the $N_p$ parameter, a convergence study of extrapolated drop displacement by varying the $N_p$ parameter is carried out. The example result of the convergence study is shown in figure~\ref{fig:q2-Ca0.20-sheared}(c) for $\theta_0 = 95^\circ$ and $\Ca = 0.20$. We typically observe that initially there are large changes of the extrapolated drop displacement. However, after some order, the magnitude of difference reduces. To settle on the final polynomial order, we use the following guidelines. As a rule of thumb, we choose the order after which the drop displacement $\Delta x$ visually seems to oscillate around some value. In addition, we introduce an arbitrary limit $N_p < 12$ to avoid over-fitting the MD data. Finally, we evaluate the agreement with the interface angle (figure~\ref{fig:q2-Ca0.20-sheared}b) and increase the order if the agreement is not satisfactory. The final chose polynomial order for $\Ca = 0.20$ is $N_p = 7$. The final obtained steady displacement from MD is $\Delta x = 13.89$ nm. The fitted polynomial for the interface, the shape is shown in figure~\ref{fig:q2-Ca0.20-sheared}(a,b) with a magenta dotted line.

While the $\Delta x$ can be obtained by extrapolating the polynomial fit, this approach does not give reliable measurements of the dynamic contact angle at the wall. To illustrate this, in figure~\ref{fig:q2-Ca0.20-sheared}(b) we add two more polynomial fits with $N_p = 8$ and $N_p = 11$. All three polynomial orders give very similar $\Delta x$ values (\ref{fig:q2-Ca0.20-sheared}c). However, the predicted advancing dynamic contact angles at the wall (magenta crosses in figure~\ref{fig:q2-Ca0.20-sheared}b) are significantly different. Similar differences have also been observed for receding contact lines for other simulations. Consequently, the first reliable measurement of the dynamic contact angle can be taken only at some distance from the wall (green crosses in figure~\ref{fig:q2-Ca0.20-sheared}b). The polynomial extrapolation in the main paper is used to get a more accurate \emph{a posteori} steady $\Delta x$ measurement and to read off the dynamic contact angle at the reliable bin location. The convergence of $\Delta x$ is rechecked for each unsteady MD simulation by following the guideline explained above.

\section{Choice of vapour properties} \label{app:vap-prop}

%

According
to empirical formulae~\citep{engtoolb_watvapor2004}, the water vapour
saturation pressure (which is the same as the gas pressure in the
absence of other gasses) at $T = 300\,K$ is
\begin{equation}
p_{wv} = \frac{\exp(77.345 + 0.0057\,T - 7235/T)}{T^{8.2}} 
= 3523.88\,\mathrm{Pa}.
\end{equation}
The water vapour density is
\begin{equation}
\rho_{wv} = 0.0022\,\frac{p_{wv}}{T} = 0.0258\,\mathrm{kg/m}^3.
\end{equation}
The viscosity of water vapour (also called steam) can be looked
up in tables~\citep{engtoolb_gasviscos2014}. Linear interpolation
between two given values closest to $T = 300\,K$ gives us
\begin{equation}
\mu_{wv} = \mu_{20^\circ} + \left(\mu_{50^\circ} - \mu_{20^\circ}\right)
\frac{323.15\,K - T}{30\,K} = 1.04 \cdot 10^{-5}\,\mathrm{Pa} \cdot
\mathrm{s}.
\end{equation}
These are the parameters reported in the main
paper, figure~\ref{fig:intro0}. It was also checked that the results are only weakly sensitive to the exact value of the vapour viscosity.

\section{Wall location and no-slip condition} \label{app:wall-loc}

\begin{figure}
\centering
\includegraphics[width=1.0\linewidth]{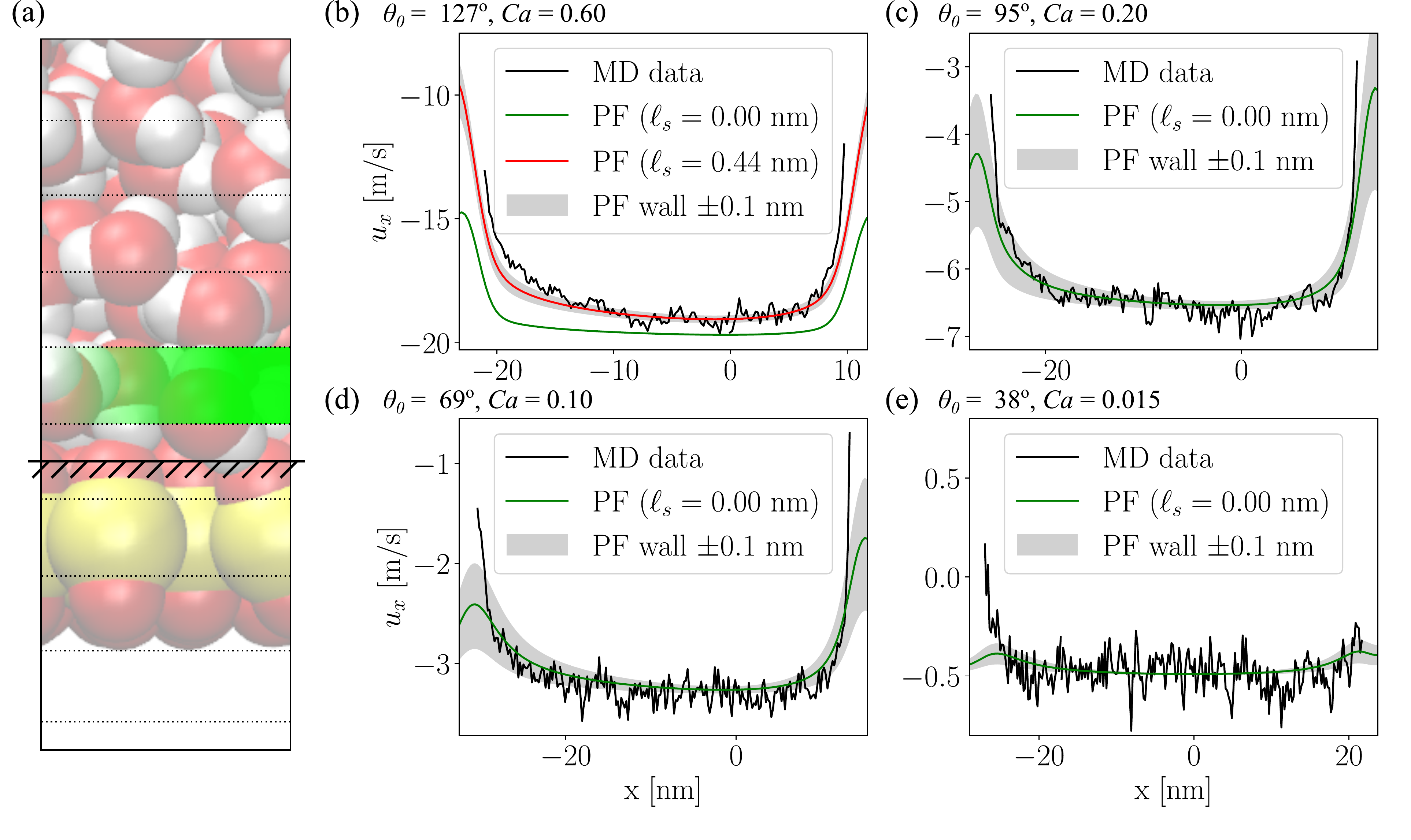}
\vspace*{-20pt}
\caption{Assumed hydrodynamic wall position for the CFM simulations with respect to the molecular picture (a). Horizontal velocity $u_x$ (b-e) over horizontal slice at the bin roughly corresponding to liquid density peak (a, coloured in green).  Results for $\theta_0 = 127^\circ$ (b), $95^\circ$ (c), $69^\circ$ (d) and $38^\circ$ (e) are reported. Calibration $\Ca$ number is stated in panel titles. Expected sensitivity to a slight shift in wall location is given as a grey shaded region.}
\label{fig:allQ-wallLoc-2ndBin-PF}
\end{figure}

In this appendix, we motivate the choice of the hydrodynamic wall position and the  applicability of the no-slip condition. We also how that small changes in wall location can influence results near the contact line significantly.
Recall that the wall position in CFM is set at the centre of the bin with coordinate $y = 0.7$ nm (black line in figure~\ref{fig:allQ-wallLoc-2ndBin-PF}a). If the chosen wall location is appropriate, the CFM should predict flow velocity accurately even a very small distance above the wall. Therefore we select the next bin with coordinate $y = 0.9$ nm (green bin in figure~\ref{fig:allQ-wallLoc-2ndBin-PF}a) to compare velocity distribution between the MD and PF models. We omit VOF from the comparison for clarity and we look at all calibration simulations. 

We extract $u_x$ from MD at the green bin. The MD velocity data is obtained with help of two averaging approaches. The first approach is the MD frame average over the steady regime together with correction for the COM, the same way as done in appendix~\ref{app:md-equil}. This provides the global flow field data and interface shape. Locally, as we approach the two-phase interface, the stochastic interface oscillations become present and can influence the measurement of the hydrodynamic variables. To reduce this influence, we repeat the averaging procedure over all frames in a steady regime, but instead of centring those around the COM, we centre them around the instantaneous interface positions at the left and the right side of the drop. Using this approach, we obtain a cleaner signal from MD closer to the interface. To obtain a single velocity curve, we move the interface averaged results to the global interface location. Then, we replace the velocity data from the COM averaged data with the interface average data until $\approx 10$ nm away from the contact line. The noisy data on the vapour side is neglected. Finally, to further reduce the noise in MD, we make use of the symmetry in the system and take the mean between profiles obtained at the bottom and top walls. The obtained MD stream-wise velocity distribution is shown in figure~\ref{fig:allQ-wallLoc-2ndBin-PF}(b-e) with solid black lines. Contact angles and $\Ca$ numbers are presented in the title of individual panels.

For comparison, we extract the $u_x$ distribution along $x$ coordinate at the same $y$ location from all calibration no-slip PF simulations. The PF results for $\theta_0 = 127^\circ$ are shown in figure~\ref{fig:allQ-wallLoc-2ndBin-PF}(a) with a green solid line. We observe that PF results do not agree with MD results over the full span of $x$ coordinate. We repeat the PF simulation, by gradually increasing the $\ell_s$ value until a good match is obtained. Through this, we obtain $\ell_s = 0.44$ nm. The PF velocity results with $\ell_s = 0.44$ nm are shown in figure~\ref{fig:allQ-wallLoc-2ndBin-PF}(a) with a solid red line. Now, a good agreement between MD and PF is obtained.

For $\theta_0 \leq 95^\circ$ (figure~\ref{fig:allQ-wallLoc-2ndBin-PF}c-e) we observe that agreement between PF velocity predictions and MD results is good at $\ell_s = 0$. Very good correspondence is obtained at the centre of the drop. As contact line regions are approached, the agreement deteriorates. However, the agreement below the liquid bulk is sufficient to conclude that the no-slip condition ($\ell_s = 0$) is appropriate for these contact angles.

The hydrodynamic wall position in the current work is essentially an assumption. Therefore we have also investigated the sensitivity of the velocity profile obtained from PF to small perturbations in wall location. With grey area (figure~\ref{fig:allQ-wallLoc-2ndBin-PF}b-e) we show the region in which the PF results would fall if the wall location would be moved up or down by $0.1$ nm. These results are obtained by sampling by 0.1 nm closer or further away from the solid wall. It was verified that this is equivalent to actually changing the wall position and also the channel height. By looking at the grey region, we observe that the variations are very narrow for all $\theta_0$ values in the centre of the drop. However, approaching the contact line region for $\theta_0 = 95^\circ$ and $69^\circ$ the variation grows. This suggests that for the description of the processes near the contact line the exact location of the solid wall could play a significant role.

\rev{It has to be recognised that the slip length of MD systems has been studied extensively before, see for example works by \cite{thompson1997general} and \cite{huang2008water}. In particular, \cite{huang2008water} investigated the slip length of SPC/E water over a range of surfaces, from silane monolayers to more common Lennards-Jones models. They found that, up to a good accuracy, $\ell_s \sim \left( 1 + \cos\theta_0 \right)^{-2}$. It has to be marked that, to the best of authors knowledge, similar study in MD with surfaces that form hydrogen bonds with water has not been carried out before and is out of scope of the current study. Nevertheless, the slip length values obtained in current work qualitatively agree with results of \cite{huang2008water} -- as $\theta_0$ increases, the slip length grows as well. However, the current inaccuracy of wall location and selected binning resolution ($0.2$ nm) prohibits determination of the exact slip length variation for contact angles $\theta_0 = 38^\circ - 95^\circ$.}

Finally, it is interesting to note that the liquid density variations (\S\ref{sec:md-sheared}) do not impede reliable flow velocity measurements. Reliable velocity measurements can be taken closer to the wall if compared to reliable interface angle measurements (compare red and green bin locations in figure~\ref{fig:wall-allQ-summary}a-d). In addition, through determining the validity of the no-slip condition, we have demonstrated that PF can accurately predict the liquid velocity distribution as close as $0.2$ nm above the last oxygen atom of the solid substrate.

\section{Streamlines from PF near the contact line} \label{app:pf-stream}

\begin{figure}
\centering
\includegraphics[width=0.95\linewidth]{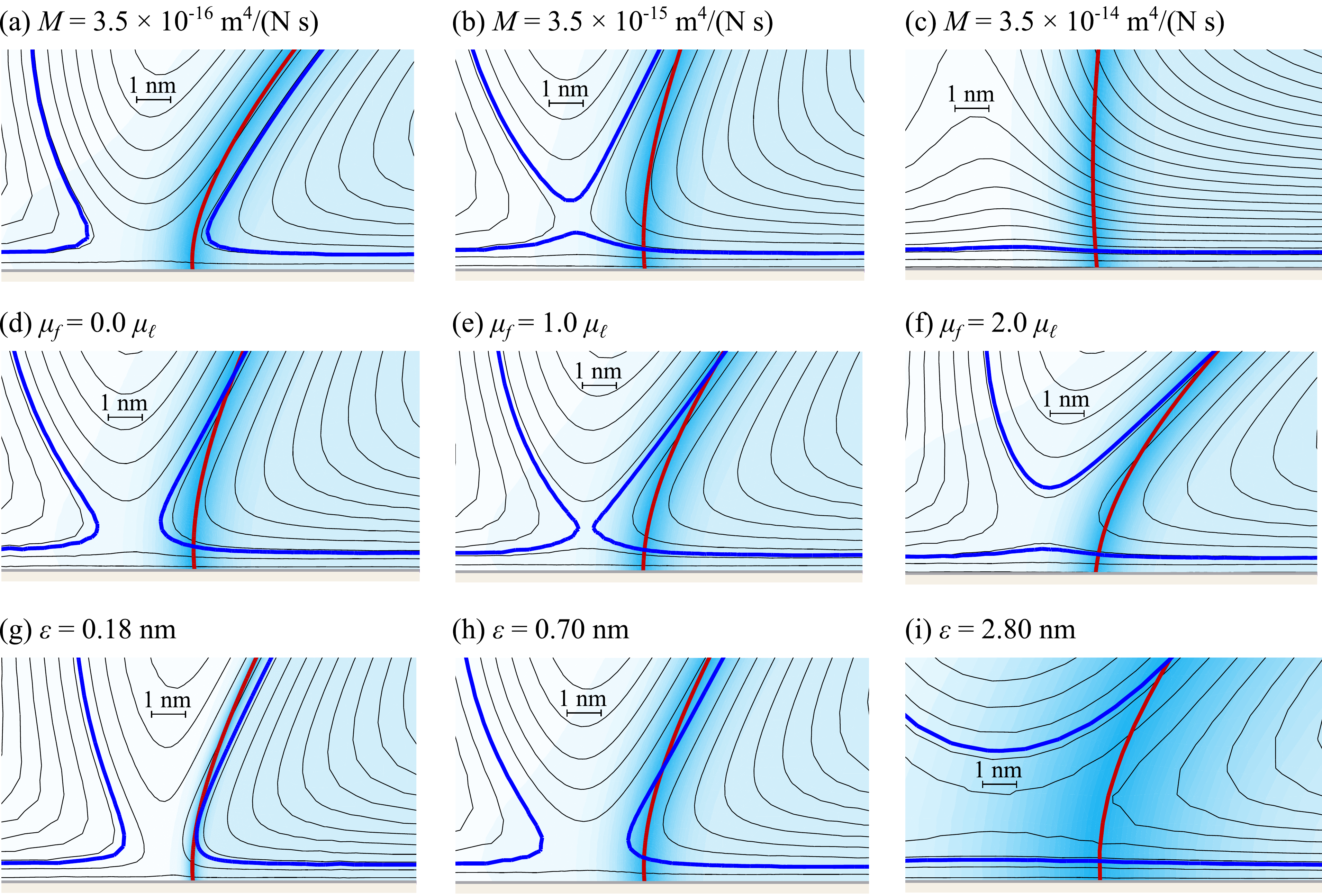}
\caption{Streamlines from PF simulations near the bottom left receding contact line. In all simulations, equilibrium contact angle $\theta_0 = 95^\circ$ and Capillary number $\Ca = 0.20$. The blue streamline describes fluid parcel originating from within the liquid drop at $0.5$ nm distance from the wall (second blue streamline has the same stream function value). The red isoline corresponds to two-phase interface, defined as $C = 0$. With light blue colour we show the variations of the C function. PF mobility (a-c), contact line friction (d-f) and interface thickness (g-i) are varied. Other parameters in the corresponding row are kept constant. In (a-c), $\mu_f = 0$ and $\epsilon = 0.7$ nm. Then $\epsilon = 0.7$ nm and $M = 1.75 \times 10^{-15}$ m$^4$/(N s) are used in (d-f). In (g-i), we have $\mu_f = 0$ and $M = 1.08 \times 10^{-15}$ m$^4$/(N s).}
\label{fig:streamL-PF-param-var}
\end{figure}

To deepen the understanding of the parameter $M$, $\mu_f$ and $\epsilon$ influence on the PF results, the steady flow field in the vicinity of the receding contact line is investigated. For this, the PF simulation with $\theta_0 = 95^\circ$ and $\Ca = 0.20$ is run until the steady $\Delta x$ is reached. The flow field at the last time instant is used to compute the streamlines. With black lines in figure~\ref{fig:streamL-PF-param-var} we present streamlines near the bottom left receding contact line in a zoomed-in window of roughly $12$ nm $\times \ 7$ nm. The two-phase interface, defined as $C = 0$, is presented with a thick red line. The thick blue line identifies a streamline that originates from within the liquid drop at a $0.5$ nm distance from the bottom wall. This particular streamline can be leveraged to compare the amount of streamline crossing and the amount of overshoot at the two-phase interface.

In the first row (figure~\ref{fig:streamL-PF-param-var}a-c) the influence of PF mobility is shown. First, we investigate the flow field with the smallest $M$ (figure~\ref{fig:streamL-PF-param-var}a). Overall, the streamlines are similar to the wedge solution derived and presented by~\cite{huh1971hydrodynamic} and the PF solution thoroughly analysed by~\cite{seppecher1996moving}. On the large viscosity side (in the liquid part), there is only one vortex, while on the small viscosity side (in the vapour part) there are two adjacent vortices. Due to the diffuse nature of the model, the stagnation point is displaced slightly to the left and above the contact line at the wall defined by $C = 0$. In addition, it can be observed that the blue streamline approaches the two-phase interface, then turns and follows the two-phase interface tangentially. By setting $M$ ten times larger (figure~\ref{fig:streamL-PF-param-var}b), it can be observed that the streamline crossing over the interface is increased. The blue streamline now crosses the interface (red line) and continues in the vapour phase. Whereas streamline originating at a slightly larger distance from the wall (see a black line that partially overlaps with the blue line) turns and follows the interface tangentially in the vapour side. For $M$ hundred times larger (figure~\ref{fig:streamL-PF-param-var}c), the streamline crossing is increased even more. The blue streamline proceeds straight into the vapour phase, and so do the streamlines originating up to a distance of roughly $2$ nm above the wall. The original wedge flow pattern can barely be recognised. 

The second row (figure~\ref{fig:streamL-PF-param-var}d-f) illustrates the effect of varying contact line friction $\mu_f$. When there is no friction (figure~\ref{fig:streamL-PF-param-var}d), the contact angle at the wall is equal to $\theta_0$ and the marked streamline overshoots the interface by around $1$ nm and is pulled back within the liquid drop higher above the wall. Increasing friction to $\mu_\ell$ (figure~\ref{fig:streamL-PF-param-var}e) leads to an overshoot of the streamline around $2$ nm. Higher above the wall, the streamline is pulled back and continues parallel to the interface at roughly the same distance from the interface as observed in figure~\ref{fig:streamL-PF-param-var}(d). The contact angle at the wall deviates from $\theta_0$. For the largest contact line friction (figure~\ref{fig:streamL-PF-param-var}f) we observe an even more pronounced contact angle departure from $\theta_0$. In addition, the marked streamline crosses the interface and continues in the vapour phase. Finally, in the third row (figure~\ref{fig:streamL-PF-param-var}g-i) we exemplify the effect of the interface thickness. As $\epsilon$ is reduced, the behaviour of the blue streamline changes from full crossover to vapour phase (figure~\ref{fig:streamL-PF-param-var}i), to a small overshoot of an order of nm (figure~\ref{fig:streamL-PF-param-var}h) and finally to no crossing over the interface (figure~\ref{fig:streamL-PF-param-var}g). The interface shape, on the other hand, remains practically the same for $\epsilon = 0.70$ nm and $0.18$ nm, which is again a signature of the sharp interface limit.

\bibliographystyle{jfm}
\bibliography{references}

\end{document}